\documentclass[aps,prb,10pt,twocolumn,amsfonts,superscriptaddress,showpacs,floatfix]{revtex4-1}

\usepackage{braket,color,verbatim,bm,bbold,amsmath,amssymb,epsfig,feynmf,tikz,float}
\usepackage[colorlinks,bookmarks=false,citecolor=blue,linkcolor=blue,hyperfootnotes=true,urlcolor=blue]{hyperref}

\newcommand{\bw}{\begin{widetext}}
\newcommand{\ew}{\end{widetext}}
\newcommand{\be}{\begin{equation}}
\newcommand{\ee}{\end{equation}}
\newcommand{\bea}{\begin{eqnarray}}
\newcommand{\eea}{\end{eqnarray}}
\def\fr#1{(\ref{#1})}
\def\abs#1{|#1|}

\def\ri{\rm i}

\newcommand{\limb}{\lim\!{}_{\emph{sc}}}

\def\nn{\nonumber\\}

\def\eps{\epsilon}

\def\ri{{\rm i}}

\definecolor{violet}{rgb}{0.62,0,1}
\definecolor{lightblue}{rgb}{0.12,0.56,1}
\definecolor{green}{rgb}{0.13,0.55,0.13}

\begin{document}

\title{Thermalization and light cones in a model with weak
  integrability breaking}  

\author{Bruno Bertini}
\affiliation{The Rudolf Peierls Centre for Theoretical Physics, University of Oxford, Oxford, OX1 3NP, United Kingdom}
\affiliation{SISSA and INFN, Sezione di Trieste, via Bonomea 265, I-34136, Trieste, Italy}
\author{Fabian H.L. Essler}
\affiliation{The Rudolf Peierls Centre for Theoretical Physics, University of Oxford, Oxford, OX1 3NP, United Kingdom}
\author{Stefan Groha}
\affiliation{The Rudolf Peierls Centre for Theoretical Physics, University of Oxford, Oxford, OX1 3NP, United Kingdom}
\author{Neil J. Robinson}
\affiliation{Condensed Matter Physics and Materials Science Division, Brookhaven National Laboratory, Upton, New York 11973, USA}

\date{\today}

\begin{abstract}
We employ equation of motion techniques to study the non-equilibrium
dynamics in a lattice model of weakly interacting spinless
fermions. Our model provides a simple setting for analyzing the
effects of weak integrability breaking perturbations on the time
evolution after a quantum quench. We establish the accuracy of the
method by comparing results at short and intermediate times to
time-dependent density matrix renormalization group computations. For 
sufficiently weak integrability-breaking interactions we always observe
\textit{prethermalization plateaux}, where local observables relax to 
non-thermal values at intermediate time scales. At later times a
crossover towards thermal behaviour sets in. We determine the
associated time scale, which depends on the initial state, the band
structure of the non-interacting theory, and the strength of the
integrability breaking perturbation. Our method allows us to analyze
in some detail the spreading of correlations and in particular the
structure of the associated light cones in our model. We find that the
interior and exterior of the light cone are separated by an
intermediate region, the temporal width of which appears to scale with
a universal power-law $t^{1/3}$.
\end{abstract}

\pacs{71.10.Fd, 05.70.Ln, 71.10.Pm, 03.75.Ss}

\maketitle

\section{Introduction}

It is now well established that there is a significant difference
between the nonequilibrium dynamics of integrable and nonintegrable
quantum systems after a quantum quench.~\cite{PolkonikovRMP11,
GE15,DKPR15,EF16,CC16,C16,CaCh16,DeLucaArxiv16,BD:review, VM:review, LangenReview, 
RigolPRL16, VR:review} Generic (nonintegrable) systems thermalize: local observables relax towards
stationary values described by a Gibbs ensemble with an effective temperature set 
by the initial state.~\cite{DeutschPRA91,SrednickiPRE94,RigolNature08,RigolPRL09,
RS10,BiroliPRL10,BanulsPRL11,aditi,rigol14} Integrable systems, however, relax towards a 
generalized Gibbs ensemble (GGE),\cite{RigolPRL07, RigolPRA06,CazalillaPRL06,CalabreseJStatMech07,
CramerPRL08,BarthelPRL08,FiorettoNJP10,P:meanvalues,CEF,DBZ:work,FE,CauxPRL12,EsslerPRL12,ColluraPRL13, QAPRL,
MussardoPRL13,PozsgayJStatMech13,FagottiJStatMech13,KSCCI,FagottiPRB14,WoutersPRL14,
PozsgayPRL14,BKCexcited, KormosPRA14,DeNardisPRA14,SotiriadisJStatMech14,GoldsteinArxiv14,
EsslerArxiv14, Betal:LL, qbosons, MPTW:XXZ, IlievskiPRL15, S:memory, Ilievskietal, Pirolibound, BPC:sinhG, 
PVC:XXZ, BS:qlGGE}, which retains an infinite amount of information on the initial state.

This raises interesting questions: how does adding a weak
perturbation to an integrable model affect its non-equilibrium dynamics?
Does the proximity to an integrable theory influence the dynamics at
finite times? In classical few-particle systems, this is a long-studied
problem: a weak integrability breaking perturbation induces a
fascinating crossover between nonergodic and chaotic motion, where the
system retains aspects of the nonergodic integrable motion on
intermediate time scales.\cite{KAM} It has recently been
understood that something analogous occurs in quantum many-body systems.
Models with weak integrability breaking perturbations have been found
to exhibit transient behaviour, in which observables relax to
non-thermal values at intermediate times. This phenomenon
was termed \textit{prethermalization} (PT)\cite{MK:prethermalization} 
and has been observed in a number of models\cite{MK:prethermalization,
RoschPRL08,KollarPRB11,worm13,MarcuzziPRL13,EsslerPRB14,NIC14,
Fagotti14,konik14,BF15,CTGM:pret,knap15,SmacchiaPRB15,BEGR:PRL,FC15,
MenegozJStatMech15,KaminishiNatPhys15, D-14} as well 
as in experiments on ultra-cold bosonic gases.\cite{KitagawaNewJPhys11,
Getal1,Getal2,Getal3,LangenReview} 

The general expectation is that PT is a transient
phenomenon, and that at late times thermalization sets in.
This is a natural assumption, but the evidence in its favor is rather
scant; most of the available numerical~\cite{SWreview} and
analytical~\cite{MK:prethermalization,EsslerPRB14} techniques are not
able to reach sufficiently late times. Recently, progress has been
achieved by means of equations of motion (EOM) techniques. These methods
were used to study a weakly nonintegrable model in infinitely many
dimension, which was observed to thermalize.\cite{SK:EOM} 
Reference~[\onlinecite{BEGR:PRL}] considered a one dimensional
weakly non-integrable model and showed that the single-particle
Green's function exhibits PT at intermediate time scales, while at long
times it eventually evolves towards a thermal stationary state.
In Ref.~[\onlinecite{BEGR:PRL}] the EOM were benchmarked against
time-dependent density renormalization group (t-DMRG) computations and
found to be in excellent agreement for all times accessible with t-DMRG. In
both of these works, at long times expectation values approach their
thermal values with corrections which are exponentially small in time. 

In this work, we expand on the results of Ref.~[\onlinecite{BEGR:PRL}]
and study the PT--thermalization crossover in detail using EOM. In particular, 
we focus on studying how the time scale for thermalization depends on 
initial state properties, the band structure and interaction strength of the post quench
Hamiltonian that governs the time-evolution. Using EOM, and their long-time 
simplification to a quantum Boltzmann-like equation, we can study the time-evolution 
from short times to the PT plateau and beyond. 

This paper is organized as follows. In Sec.~\ref{sec:model} we
introduce the class of interacting lattice fermion models considered
in the following, discuss the limits in which it describes integrable
models, and review some important symmetries of the Hamiltonian. In
Sec.~\ref{sec:prob} we discuss our protocol for inducing the
non-equilibrium dynamics and the ``initial conditions'' this induces,
and introduce the central object of our study, the single-particle
Green's function. Following this, we derive the equations of motion
for the momentum-space two point functions in Sec.~\ref{sec:eom}. In
Sec.~\ref{sec:gf} we present results for the time evolution of the
Green's function, compare the EOM results to t-DMRG computations and
discuss the roles played by the next-nearest neighbour hopping and the
initial state in the PT-thermalization crossover. In
Sec.~\ref{sec:lightcone} we investigate light cone effects in the 
time-evolving Green's function. In Sec.~\ref{sec:qbe} we  
consider the long-time limit of the EOM and show that under certain
assumptions they can be reduced to a set of quantum
Boltzmann-like equations. We then use these to study the
thermalization time scale. Section~\ref{sec:u1b} reports results on the
dynamics after quantum quenches in a modified Hamiltonian that breaks
the global $\textrm{U}(1)$ symmetry associated with particle number
conservation. PT is seen to persist in this case. We conclude in 
Sec.~\ref{sec:conc} and cover a number of technical points in the 
appendices.    

\section{The model}
\label{sec:model}

We consider a three-parameter family of interacting spinless fermion
models with Hamiltonian
\bw
\begin{align}
H(J_2,\delta,U) = -J_1\sum_{l=1}^L \Big[1+(-1)^l\delta \Big] \Bigl( c^\dag_l
c^{\phantom\dagger}_{l+1}+\textrm{H.c.}\Bigr) -J_2\sum_{l=1}^L  \Bigl[ c^\dag_l c^{\phantom\dagger}_{l+2}+\textrm{H.c.}\Bigr]+{U}\sum_{l=1}^L n_ln_{l+1}~,
\label{Eq:Ham}
\end{align}
\ew
where we impose periodic boundary conditions
$c^{\phantom\dagger}_{L+1}\equiv c^{\phantom\dagger}_{1}$. 
Here $c_i$ and $c^\dagger_i$ are spinless fermion operators on site
$i$, obeying the canonical anticommutation relations
\be
\{c^\dag_i,c^{\phantom{\dag}}_j\}=\delta_{ij}\ ,\qquad
\{c_i,c_j\}=0\ .
\ee
The amplitudes $J_1$ and $J_2$ describe tunneling between
nearest-neighbour and next-nearest-neighbour sites respectively, and we
include a nearest-neighbour dimerization of strength
$0\leq\delta<1$. Finally there is a nearest-neighbour density-density 
interaction of strength $U$. From here on we set $J_1=1$ and measure
energies in units of $J_1$. 

There are several limits in which the Hamiltonian~\fr{Eq:Ham} becomes
integrable: 
\begin{enumerate}
\item{} For $U=0$ we are dealing with a non-interacting theory, which is
a particularly simple example of an integrable model.
\item{} If we set $\delta = J_2 = 0$, the model \fr{Eq:Ham} becomes
equivalent to the anisotropic spin-1/2 Heisenberg model in an external magnetic field.\cite{Orbach} 
\item{} The low-energy description for $J_2=0$ and $\delta,U\ll1$ is
given by the quantum sine-Gordon model.\cite{EKreview} 
\end{enumerate}
We have checked by computing the level spacing statistics that the
model is non-integrable away from these limits. The Hamiltonian
$H(J_2,\delta,U)$ is invariant under the following transformations of
the fermion operators  
\begin{enumerate}
\item Global $\textrm{U}(1)$ transformations: ${\cal U}(\phi)$
\be
c_i\rightarrow {\cal U}(\phi) c_i {\cal U}^\dag(\phi)=e^{\ri\phi}c_i~,\qquad\phi\in[0,2\pi ]~,
\ee   
\item Translation by two sites: ${\cal T}_2$ 
\be
c_i\rightarrow {\cal T}_2 c_i {\cal T}_2^\dag= c_{i+2}~,
\ee
\item Inversion with respect to any bond $j$: ${\cal B}_j$ 
\be
c_i\rightarrow {\cal B}_j c_i {\cal B}_j^\dag= c_{2j-i+1}~.
\label{Eq:inversion}
\ee
\end{enumerate}
In the absence of next nearest neighbour tunneling $J_2=0$, the model
exhibits an additional \emph{particle-hole} symmetry at half filling
(one fermion per two sites)
\be
\label{Eq:ph}
c_i\rightarrow {\cal J} c_i {\cal J}^\dag= (-1)^i c^\dag_{i}~.
\ee 

We are interested in the regime of weak interactions, $U \lesssim
1$. In this case, a convenient basis for analyzing the quench dynamics
is provided by diagonalizing the quadratic (non-interacting) part of
the Hamiltonian. This is done by going to Fourier space (using an
elementary cell with two sites) and then carrying out a Bogoliubov
transformation to momentum space fermion creation and annihilation
operators  with anticommutation relations
$\{\alpha_\mu(k),\alpha^\dagger_\nu(q)\}=\delta_{\mu\nu}\delta_{k,q}$
\be
c_l=\frac{1}{\sqrt{L}}\sum_{k>0}\sum_{\eta=\pm}
\gamma_\eta(l,k|\delta)\alpha_\eta(k)\ .
\label{Eq:BogoTrans}
\ee
Here the coefficients are given by 
\bea
\gamma_{\pm}(2j-1,k|\delta)&=&e^{-i k (2j-1)}~,\nn
\gamma_{\pm}(2j,k|\delta)&=&\pm e^{-i k 2 j} e^{-i \varphi_k(\delta)}~,
\eea
where $\varphi_k(\delta)$ is the Bogoliubov angle 
\be
e^{-i \varphi_k(\delta)}=\frac{-\cos k+ i \delta\sin k}{\sqrt{\cos^2 k+ \delta^2\sin^2k}}~.
\ee
In terms of the two species of Bogoliubov fermions the Hamiltonian reads
\bw
\be
H(J_2,\delta,U) = \sum_{\eta=\pm}\sum_{k > 0} \epsilon_\eta(k) \alpha^\dag_\eta(k) \alpha_\eta(k) 
+ U \sum_{\bm\eta}\sum_{\bm k {>0}} V_{\bm \eta}({\bm k}) \alpha^\dag_{\eta_1}(k_1)\alpha^\dag_{\eta_2}(k_2)\alpha_{\eta_3}(k_3)\alpha_{\eta_4}(k_4)~,
\label{Eq:HFourierTransformed}
\ee
where we introduced notations ${\bm \eta} =
(\eta_1,\eta_2,\eta_3,\eta_4)$, ${\bm k} = (k_1,k_2,k_3,k_4)$ and
${\bm k}>0$ is shorthand for $k_i > 0$ for all $i=1,\ldots,4$. 
The single particle dispersion relation is
\be
\epsilon_\eta(k) = -2J_2\cos(2k) + 2\eta
\sqrt{\delta^2+(1-\delta^2)\cos^2(k)}~\ ,
\label{dispersion}
\ee
while the interaction $V_{\bm\eta}({\bm k})$ can be written in a
convenient antisymmetrized form 
\bea
V_{\bm\eta}({\bm k}) &=& - \frac{1}{4} \sum_{P,Q\in S_2} {\rm sgn}(P) {\rm sgn}(Q) V'_{\eta_{p_1}\eta_{q_1}\eta_{p_2}\eta_{q_2}}(k_{p_1},k_{q_1},k_{p_2},k_{q_2})~.
\eea
Here $P=(p_1,p_2)$ and $Q=(q_1,q_2)$ are permutations of $(1,2)$ and $(3,4)$ respectively and 
\bea
V'_{\bm \eta}(\bm k)=\frac{e^{i(k_3-k_4)}}{2L}
&\Big[& \left(\eta_1\eta_2e^{i \varphi_{k_1}(\delta)-i\varphi_{k_2}(\delta)}
+\eta_3\eta_4e^{i \varphi_{k_3}(\delta)-i\varphi_{k_4}(\delta)}\right)\delta_{k_1-k_2+k_3-k_4,0}\nn
&+& \left(\eta_1\eta_2e^{i
  \varphi_{k_1}(\delta)-i\varphi_{k_2}(\delta)}
-\eta_3\eta_4e^{i \varphi_{k_3}(\delta)-i\varphi_{k_4}(\delta)}\right)\delta_{k_1-k_2+k_3-k_4\pm\pi,0} \Big]~.
\label{Int}
\eea
\ew

\section{Quantum Quench}
\label{sec:prob}
\subsection{Quench Protocol}
Our protocol for inducing and analyzing nonequilibrium dynamics is as
follows. We always prepare the system in an initial density matrix $\rho_0$
that is ``not an eigenstate of $H(J_2,\delta_f,U)$'' for any value of $U$
[\emph{i.e.}, it does not commute with $H(J_2,\delta_f,U)$]. An
important condition we impose is that Wick's theorem holds in the
initial density matrix. A convenient choice we use in the following is
provided by the equilibrium states 
\be
\rho_0(\beta_i,\delta_i)=\frac{e^{-\beta_i H(0,\delta_i,0)}}
{{\rm Tr}[e^{-\beta_i H(0,\delta_i,0)}]}~,
\label{thermal}
\ee  
These states include, as a particular case, the ground state of the
Hamiltonian $H(0,\delta_i,0)$. The rationale for considering finite
temperatures $\beta_i<\infty$ is that this provides us with a simple
way of changing the energy density in the initial state.
\subsubsection{Integrable quench}\label{subsec:integrablequench}
One class of quenches we consider is to the non-interacting theory
with Hamiltonian $H(J_2,\delta_f,0)$. The time evolved density matrix
in this case is
\be
\rho(t)=e^{-i t H(J_2,\delta_f,0)}\rho_0 e^{i t H(J_2,\delta_f,0)}~.
\ee

\subsubsection{Integrability breaking quench}
The second class of quenches we consider is to the non-integrable theory
with Hamiltonian $H(J_2,\delta_f,U)$. The time evolved density matrix
in this case is
\be
\rho(t)=e^{-i t H(J_2,\delta_f,U)}\rho_0 e^{i t H(J_2,\delta_f,U)}~.
\ee
Here the interaction with strength $U$ plays the role of a weak
integrability-breaking perturbation, and our aim is to quantify how
this perturbation changes the dynamics compared to the integrable quench.

We stress that our protocol differs in a very important way from weak
interaction quenches which have been analyzed previously with
equations of motion techniques.\cite{SK:EOM,NessiArxiv15} In these
cases, no dynamics are present for $U=0$; hence quenching the
interaction from zero to a finite value simultaneously induces a time
dependence in the problem \emph{and} breaks the
integrability. Accordingly, the effect of the integrability breaking
on the nonequilibrium dynamics is masked.

\subsection{The single-particle Green's function}
The main object of interest in this work is the single-particle
Green's function  
\be
{\cal G}(j,l;t) = {\rm Tr}\left[\rho(t)c^\dag_j c^{\phantom{\dag}}_l\right].
\ee
From the symmetries of the Hamiltonian (and hence those of the initial 
state~\fr{thermal}), the following properties of the Green's function can be derived
\begin{align}
&{\cal G}(j,l;t)={\cal G}(j+2n,l+2n;t)~, \\
&{\cal G}(j,l;t)={\cal G}(j,l;t)^*~, & & j-l=2n+1~, \label{Eq:oddreal}\\
&{\cal G}(j,l;t)={\cal G}(j,2j-l;t)^*~, & & j-l=2n~,
\end{align}
where $n$ is an integer. The Green's function can be obtained from the
two-point functions of the Bogoliubov fermion operators $ \alpha_{\pm}(k)$ 
of the final Hamiltonian $H(J_2,\delta_f,U)$
\be
{\cal G}(j,l;t) =\frac{1}{L} \sum_{k>0} \sum_{\mu,\nu = \pm} 
\gamma^*_\mu(k,j) \gamma^{\phantom{\dag}}_\nu(k,l) n_{\mu\nu}(k,t)\,,
\label{Eq:GreenFun}
\ee
where we have defined
\be
n_{\mu\nu}(k,t) = {\rm Tr}\left[\rho(t)\alpha^\dag_{\mu}(k) \alpha^{\phantom{\dag}}_{\nu}(k)\right]~.
\label{Eq:Defnab}
\ee
Since $\rho_0$ is noninteracting, we can easily evaluate
\fr{Eq:Defnab} for $t=0$ 
\begin{align}
\!\!\!n_{\mu\mu}(k)&=
\frac{1}{2}-\frac{1}{2}\cos\!\big(\Delta\varphi_k(\delta_f,\delta_i)\big)\!\tanh\!\bigg( \frac{\beta\epsilon_\mu^{(0)}(k)}{2}\bigg)~,\\
\!\!\! n_{\mu\bar\mu}(k)&=
\frac{i}{2}\sin\!\big(\Delta\varphi_k(\delta_f,\delta_i)\big)\!\tanh\!\bigg(\frac{\beta\epsilon_\mu^{(0)}(k)}{2}\bigg)~ .
\label{AB:Occval}
\end{align}
Here
$\Delta\varphi_k(\delta_f,\delta_i)\equiv\varphi_k(\delta_f)-\varphi_k(\delta_i)$
and the dispersion relations $\epsilon_\alpha^{(0)}(k)$ are given by 
\fr{dispersion} with $J_2=0$ and $\delta=\delta_i$.
We note that as a consequence of the inversion
symmetry~\fr{Eq:inversion} $\textrm{Re}[{\cal G}(j,j+2n;t)]$ 
depends only on the occupation numbers~$n_{\mu\mu}(k;t)$.

\section{Equations of motion}
\label{sec:eom}
To study the time-evolution of the Green's function we use EOM
techniques.\cite{SK:EOM,INJPhys,NessiArxiv15,BEGR:PRL,ESY:QBE,LS:QBE}
In this section, for the purpose of completeness, we present the
derivation of the EOM. We will closely follow the derivation given in
the Supplemental Material of Ref.~[\onlinecite{BEGR:PRL}] which is in
turn based on the one given in Ref.~[\onlinecite{ESY:QBE}], where the
EOM are used as an intermediate step to derive a quantum Boltzmann
Equation. Our starting point is the Heisenberg equations for the
time-evolved bilinears $\hat n_{\mu\nu}(k,t)=\alpha^\dag_{\mu}(k,t)
\alpha^{\phantom{\dag}}_{\nu}(k,t)$, which read  
\bw
\begin{align}
\frac{\partial}{\partial  t}\hat{n}_{\mu\nu}(k,t)=&
i\bigl[H,\hat{n}_{\mu\nu}(k,t)\bigr] = i\epsilon_{\mu\nu}(k)\hat{n}_{\mu\nu}(k,t)
  +i U\sum_{{\bm \eta}}\sum_{{\bm{q}{>0}}}{Y}_{\mu\nu}^{\boldsymbol{\eta}}(k,\bm{q})\hat{A}_{\boldsymbol{\eta}}(\bm{q},t)\ ,
\label{Heisenberg}
\end{align}
Here we have defined the functions $\epsilon_{\mu\nu}(k) \equiv\epsilon_{\mu}(k)-\epsilon_{\nu}(k)$,
\begin{align}
{Y}_{\mu\nu}^{\boldsymbol{\eta}}(k,\bm{q})\equiv& \delta_{\nu,\eta_4}\delta_{k,q_4}{V}_{\eta_1\eta_2\eta_3\mu}(\bm{q}) 
+\delta_{\nu,\eta_3}\delta_{k,q_3}{V}_{\eta_1\eta_2\mu\eta_4}(\bm{q})
-\delta_{\mu,\eta_2}\delta_{k,q_2}{V}_{\eta_1\nu\eta_3\eta_4}(\bm{q})
-\delta_{\mu,\eta_1}\delta_{k,q_1}{V}_{\nu\eta_2\eta_3\eta_4}(\bm{q})~,
\end{align}
and the operators
\be
\hat{A}_{\boldsymbol{\eta}}(\bm{q},t)\equiv\alpha^{\dag}_{\eta_1}(q_1,t)\alpha^\dag_{\eta_2}(q_2,t)\alpha_{\eta_3}(q_3,t)\alpha_{\eta_4}(q_4,t)~.
\ee
The quartic operators in \fr{Heisenberg} evolve according to
the following Heisenberg equations of motion
\be
\frac{\partial}{\partial t}\hat{A}_{\boldsymbol{\eta}}(\bm{q},t) =
 i {E}_{\boldsymbol{\eta}}(\bm{q})\hat{A}_{\boldsymbol{\eta}}(\bm{q},t)
+i U\sum_{\boldsymbol{\gamma}}\sum_{\bm{p}{>0}} {V}_{\boldsymbol{\gamma}}(\bm{p})
\left[\hat{A}_{\boldsymbol{\gamma}}(\bm{p},t),\hat{A}_{\boldsymbol{\eta}}(\bm{q},t)\right]~,
\label{Heisenberg4point}
\ee
where $
{E}_{\boldsymbol{\eta}}(\bm{q})\equiv{\epsilon}_{\eta_1}(q_1)+{\epsilon}_{\eta_2}(q_2)-{\epsilon}_{\eta_3}(q_3)-{\epsilon}_{\eta_4}(q_4)$. The
commutator on the right-hand side produces operators involving six
fermion creation and annihilation operators. Continuing this procedure
leads to an infinite hierarchy of coupled equations. 
This hierarchy is closely related to the equations of motion for the
reduced density operator obtained in the BBGKY approach, see e.g.
Ref.~[\onlinecite{Bonitzbook}]. The relation between the two follows
directly from the Fock space representation for the reduced density
operators  
\be
\!\!\braket{k'_n\eta'_n,\dots, k'_1 \eta'_1|F_n|k_1\eta_1,\dots, k_n \eta_n}
=\frac{\textrm{Tr}\left[\rho(t)\alpha^\dag_{\eta_1}(k_1)\ldots\alpha^\dag_{\eta_n}(k_n)\alpha^{\phantom{\dag}}_{\eta'_n}(k'_n)\ldots
\alpha^{\phantom{\dag}}_{\eta'_1}(k'_1)\right]}{n! L^n \braket{N}(\braket{N}-1)\cdots(\braket{N}-n)}\, .
\label{eq:relation}
\ee  
Here $F_n$ is the $n$-particle reduced density operators, obtained by tracing out the degrees of freedom associated with all but $n$ particles from the density matrix $\rho(t)$, and $\braket{N}\equiv\textrm{Tr}[\rho(t) N]=\textrm{Tr}[\rho(0) N]$ where $ N$ is the particle-number operator. 
We also defined 
\be
\ket{k_1 \eta_1, \ldots, k_n \eta_n}=\alpha^\dag_{\eta_1}(k_1)\ldots\alpha^\dag_{\eta_n}(k_n)\ket{0}\,,
\ee
where $\ket{0}$ the vacuum state, satisfying
$\alpha_{\eta}(k)\ket{0}=0$ for all $k$ and $\eta$. From
Eq.~\fr{eq:relation} we see that the expectation values of strings of
fermionic operators that appear in our hierarchy are essentially the
matrix elements of the reduced density operators. In particular, by
taking the expectation value of~\fr{Heisenberg}
and~\fr{Heisenberg4point} we recover the first two equations of the
BBGKY hierarchy.  

In order to proceed we integrate \fr{Heisenberg4point} in time and
then take the expectation value in our initial state $\rho_0$. This gives
\be
\braket{\hat{A}_{\boldsymbol{\eta}}(\bm{q},t)}=\braket{\hat{A}_{\boldsymbol{\eta}}(\bm{q},0)}e^{i  t {E}_{\boldsymbol{\eta}}(\bm{q})}
+i U \sum_{\boldsymbol{\gamma}}\sum_{\bm{p}{>0}} \int_{0}^{t}\!{\rm d}s~ e^{i  (t-s){E}_{\boldsymbol{\eta}}(\bm{q})}
{V}_{\boldsymbol{\gamma}}(\bm{p})\left\langle \left[\hat{A}_{\boldsymbol{\gamma}}(\bm{p},s),\hat{A}_{\boldsymbol{\eta}}(\bm{q},s)\right] \right\rangle~,
\label{Eq:fourparticlecumulant}
\ee
where we have defined $\braket{\hat O}\equiv \textrm{Tr}[\rho_0 \hat
  O]$. Taking the expectation value of Eq.~\fr{Heisenberg} in $\rho_0$
and then using \fr{Eq:fourparticlecumulant}, we obtain
a set of exact integro-differential equations for the expectation values
$n_{\mu\nu}(k,t)={\rm Tr}[\rho_0\hat{n}_{\mu\nu}(k,t)]$ (see
Eq.~\fr{Eq:Defnab})  
\begin{align}
\frac{\partial}{\partial  t}{n}_{\mu\nu}(k,t)=&i\epsilon_{\mu\nu}(k) n_{\mu\nu}(k,t)
+i U\sum_{{\bm \eta}}\sum_{{\bm{q}{>0}}}{Y}_{\mu\nu}^{\boldsymbol{\eta}}(k,\bm{q})\braket{\hat{A}_{\boldsymbol{\eta}}(\bm{q},0)} e^{i  t {E}_{\boldsymbol{\eta}}(\bm{q})} \nn
&-U^2\sum_{\boldsymbol{\eta},\boldsymbol{\gamma}}\sum_{\bm{q},\bm{p}{>0}}  \int_{0}^{t}\!{\rm d}s~{Y}_{\mu\nu}^{\boldsymbol{\eta}}(k,\bm{q}) 
e^{i  (t-s){E}_{\boldsymbol{\eta}}(\bm{q})}{V}_{\boldsymbol{\gamma}}(\bm{p})\braket{\hat{A}_{\boldsymbol{\gamma}}(\bm{p},s)\hat{A}_{\boldsymbol{\eta}}(\bm{q},s)}\nn
&+U^2\sum_{\boldsymbol{\eta},\boldsymbol{\gamma}}\sum_{\bm{q},\bm{p}{>0}}  \int_{0}^{t}\!{\rm d}s~{Y}_{\mu\nu}^{\boldsymbol{\gamma}}(k,\bm{p}) 
e^{i  (t-s){E}_{\boldsymbol{\gamma}}(\bm{p})}{V}_{\boldsymbol{\eta}}(\bm{q})\braket{\hat{A}_{\boldsymbol{\gamma}}(\bm{p},s)\hat{A}_{\boldsymbol{\eta}}(\bm{q},s)}.\label{EOM}
\end{align}

Since we focus on cases where Wick's theorem holds for the initial
density matrix $\rho_0$, see Eq.~\fr{thermal}, the expectation value
$\braket{\hat{A}_{\boldsymbol{\alpha}}(\bm{q},0)}$ can be written in
terms of the initial values $n_{\mu\nu}(k,0)$. The
time-dependent eight-point average present in Eq.~\fr{EOM} can be
decomposed into the form
\be
\braket{\hat{A}_{\boldsymbol{\gamma}}(\bm{p},t)\hat{A}_{\boldsymbol{\alpha}}(\bm{q},t)}=f(\{n_{\mu\nu}(k,t)\}) 
+\mathcal C[\braket{\hat{A}_{\boldsymbol{\gamma}}(\bm{p},t)\hat{A}_{\boldsymbol{\alpha}}(\bm{q},t)}]~,
\label{Eq:truncation}
\ee
where $f$ represents the fully disconnected part (which is obtained by
applying Wick's theorem), and $\mathcal C\left[\cdots\right]$ denotes
terms involving the four, six and eight-particle cumulants
[the eight particle cumulant does not contribute to Eq.~\fr{EOM} due to
the antisymmetric structure of the accompanying term].
In order for Eq.~\fr{EOM} to reduce to
a closed set of integro-differential equations, we now assume that the
four and six particle cumulants are negligible at all times. This
assumption is uncontrolled -- we check it by comparison
of our results to those obtained using t-DMRG (it will be
apparent that this assumption is valid for the model and initial conditions
under consideration). This truncation leads a closed system of equations for 
the expectation values
\begin{align}
\frac{\partial}{\partial  t}{n}_{\mu\nu}(k,t) =& 
i {\epsilon}_{\mu\nu}(k) n_{\mu\nu}(k,t)
+4iU\sum_{\gamma_1\gamma_2\gamma_3}\sum_{q{>0}} V_{\gamma_1 \gamma_2 \gamma_3 \mu}(k,q,q,k) 
e^{i \epsilon_{\gamma_1\nu}(k) t} e^{i \epsilon_{\gamma_2\gamma_3}(q) t} n_{\gamma_1\nu}(k,0) n_{\gamma_2\gamma_3}(q,0)~\nn
&-4iU\sum_{\gamma_1\gamma_2\gamma_3}\sum_{q{>0}} V_{\nu \gamma_2 \gamma_3 \gamma_1}(k,q,q,k) 
e^{i \epsilon_{\mu\gamma_1}(k) t} e^{i \epsilon_{\gamma_2\gamma_3}(q) t} n_{\mu\gamma_1}(k,0) n_{\gamma_2\gamma_3}(q,0)~\nn
& -U^2  \int_0^t \!\textrm{d} t'  \sum_{\vec{\gamma}}\sum_{k_1,k_2,k_3{>0}} \!\!\!\!\!L^{\vec{\gamma}}_{\mu\nu}(k_1,k_2,k_3; k; t-t') 
n_{\gamma_1\gamma_2}(k_1,t') n_{\gamma_3\gamma_4}(k_2,t')n_{\gamma_5\gamma_6}(k_3,t')\nn
& - U^2  \int_0^t \!\textrm{d}t'  \sum_{\bm \gamma}\sum_{k_1,k_2{>0}} \!\!\!\!K^{\bm\gamma}_{\mu\nu}(k_1,k_2; k; t-t') 
n_{\gamma_1\gamma_2}(k_1,t') n_{\gamma_3\gamma_4}(k_2,t')~.
\label{Eq:EOM}
\end{align}
Here we have introduced notations such that $\vec
\gamma=(\gamma_1\ldots\gamma_6)$ and the kernels are given by 
\begin{align}
&K^{\bm\gamma}_{\mu\nu}(k_1,k_2;k;t) = 4 \sum_{k_3,k_4{>0}} \sum_{\eta,\eta'} 
X^{\gamma_1\gamma_3\eta\eta';\eta\eta'\gamma_4\gamma_2}_{{\bm k};{\bm k}'} (\mu,\nu;k;t),\nn
&L^{\vec\gamma}_{\mu\nu}(k_1,k_2,k_3;k;t) = 
8 \sum_{\eta}\sum_{k_4{>0}} X^{\gamma_1\gamma_3\gamma_6\eta;\eta\gamma_5\gamma_4\gamma_2}_{{\bm k};{\bm k}'}(\mu,\nu;k;t)
- 16\sum_{\eta} X^{\gamma_1\gamma_3\eta\gamma_4;\gamma_5\eta\gamma_6\gamma_2}_{k_1k_2k_1k_2;k_3k_1k_3k_1}(\mu,\nu;k;t)~,\nn
&X^{{\bm\gamma};{\bm\eta}}_{{\bm k};{\bm q}}(\mu,\nu;q;t) =
Y^{\bm\gamma}_{\mu\nu}({\bm k},q)V_{\bm\eta}({\bm q}) e^{i
  E_{\bm\gamma}({\bm k})t} - ({\bm \gamma},{\bm k})\leftrightarrow({\bm
  \eta},{\bm q}).\label{Eq:kernelsEOM}
\end{align}

In the framework of the BBGKY approach, our truncation scheme is
sometimes referred to as the \textit{second Born
  approximation}.\cite{Bonitzbook} We note that the same result can be
obtained in the non-equilibrium Green's function
approach,\cite{KBbook} as discussed in Ref.~[\onlinecite{HBBcomparison}].  

It is useful to note that our truncation scheme and Eqs. \fr{Eq:EOM} conserve the total energy
\begin{align}
E=&\sum_{\eta=\pm}\sum_{k > 0} \epsilon_\eta(k) n_{\eta\eta}(k,t) 
+ U \sum_{\bm\eta}\sum_{\bm k {>0}} V_{\bm \eta}({\bm k}) \braket{\hat{A}_{\boldsymbol{\eta}}(\bm{k},0)}e^{i  t {E}_{\boldsymbol{\eta}}(\bm{k})}\notag\\
&+i U\sum_{\boldsymbol{\gamma},\boldsymbol{\eta}}\sum_{\bm{k},\bm{p}{>0}} \int_{0}^{t}\!{\rm d}s~ e^{i  (t-s){E}_{\boldsymbol{\eta}}(\bm{q})}
V_{\bm \eta}({\bm k}) {V}_{\boldsymbol{\gamma}}(\bm{p})f(\{n_{\mu\nu}(k,t)\}) ~,
\end{align}
where $f(\{n_{\mu\nu}(k,t)\})$ is the Wick's theorem part of $\braket{\hat{A}_{\boldsymbol{\gamma}}(\bm{p},t)\hat{A}_{\boldsymbol{\alpha}}(\bm{q},t)}$ appearing in \fr{Eq:truncation}. 
\ew

Solving the system of integro-differential equations~\fr{Eq:EOM} is computationally demanding; we designed an algorithm which scales as $L^3 \times T$, where $T$ is the number of time steps and $L$ the number of lattice sites. Our algorithm is based on the following idea: we store the values of the integrals in variables of the form
\be
I_{\alpha_1\ldots\alpha_j}(k_1,k_2,k_3; t)=\int_0^t {\rm d}s~ F_{\alpha_1\ldots\alpha_j}(k_1,k_2,k_3; s),
\ee
where $F(k_1,k_2,k_3; s)$ contains products of $n_{\mu\nu}(k_j,t)$,
vertex functions and oscillating phases $e^{i\epsilon_{\mu\nu}(k) t}$;
then we solve the extended system of equations for
$\{{n}_{\mu\nu}(k,t)\}$ and $\{I(k_1,k_2,k_3; t)\}$ by means of a
fourth-order Runge-Kutta method.\cite{numericalrecipes} Using this
procedure we can reach long times $J_1t\sim 100$ on large systems
$L\sim 400$ (an algorithm with similar scaling was proposed in
Ref.~[\onlinecite{NessiArxiv15}]). The maximum times that we consider
are controlled by the appearance of finite-size related traversals.\cite{EF16}

\subsection{Window of applicability}
Truncating the Green's function hierarchy is an uncontrolled
approximation and an important question is: in what parameter regime we
may expect it to be accurate? A crucial aspect of our work is that we
always initialize the system in states where the neglected cumulants
vanish. This means that for small $U$ and short times the EOMs will
provide a good approximation. We have verified this by independent
checks, \emph{cf.} Sec.~\ref{sec:gf}. Over time the higher cumulants
may grow and eventually become important. If this happens, our
approach will cease to be quantitatively accurate. The basic premise
of our work is to apply the EOM approach to some intermediate time
window. The behaviour at asymptotically late times may well show
features not captured by our method.\cite{Luxetal,kim}

\subsection{Leading order approximation}

The EOM~\fr{Eq:EOM} that we have derived are the result of a second
order expansion in the interaction parameter $U$: we approximately
take into account the effect of the four-particle connected cumulants
in the expectation value of Eq.~\fr{Heisenberg} with respect to the
initial density matrix $\rho_0$ by means of
Eqs.~\fr{Eq:fourparticlecumulant}. A less accurate `leading order'
approximation would be to neglect all four-particle cumulants from the
expectation value of \fr{Heisenberg}: this gives rise to a 
simpler system of equations, which read 
\bw
\begin{align}
\frac{\partial}{\partial  t}{n}_{\mu\nu}(k,t) =& i {\epsilon}_{\mu\nu}(k) n_{\mu\nu}(k,t)\nn
&+4iU\sum_{\gamma_1\gamma_2\gamma_3}\sum_{q{>0}} \Big[ V_{\gamma_1 \gamma_2 \gamma_3 \mu}(k,q,q,k) n_{\gamma_1\nu}(k,t) n_{\gamma_2\gamma_3}(q,t)
- V_{\nu \gamma_2 \gamma_3 \gamma_1}(k,q,q,k) n_{\mu\gamma_1}(k,t) n_{\gamma_2\gamma_3}(q,t)\Big]~.\label{Eq:firstorderEOM1}
\end{align}
\ew
For short times, the right hand sides of Eqs.~\fr{Eq:firstorderEOM1} and 
Eqs.~\fr{Eq:EOM} coincide with the perturbative expansion of  
$i\textrm{Tr}\bigl[\rho_0\bigl[H,\hat{n}_{\mu\nu}(k,t)\bigr]\bigr]$ to first and 
second order in $U$, respectively.

Equations~\fr{Eq:firstorderEOM1} give results which are equivalent to
those found by means of the first-order continuous unitary
transformation (CUT)\cite{W:CUT, UhrigCUT, kehreinbook,
  MK:prethermalization} approach used in
Ref.~[\onlinecite{EsslerPRB14}]. 
Solving the equations up to $O(U^2)$ corrections, it is 
easy to extract the expectation values of $\hat n_{\mu\nu}(k)$ in the 
``deformed GGE'' of Ref.~[\onlinecite{EsslerPRB14}].

The EOM~\fr{Eq:EOM} at short times refine the leading order
description (obtained by either the leading order EOM or CUT approach)
by going to next order in perturbation theory. However, at later times
we will see that non-perturbative feedback mechanisms present in the
second order EOM cause a drifting away from the PT plateau observed
in the leading order approximations.

\section{The Green's function from the equations of motion}
\label{sec:gf}

We now turn our attention to computation of the time-evolution of the
Green's function~\fr{Eq:GreenFun} by means of the EOM. We first begin
by providing a crucial check of the validity of the approximations
underlying the EOM by direct comparison to t-DMRG computations.  
\begin{figure}[ht]
\begin{tabular}{l}
\includegraphics[width=0.425 \textwidth]{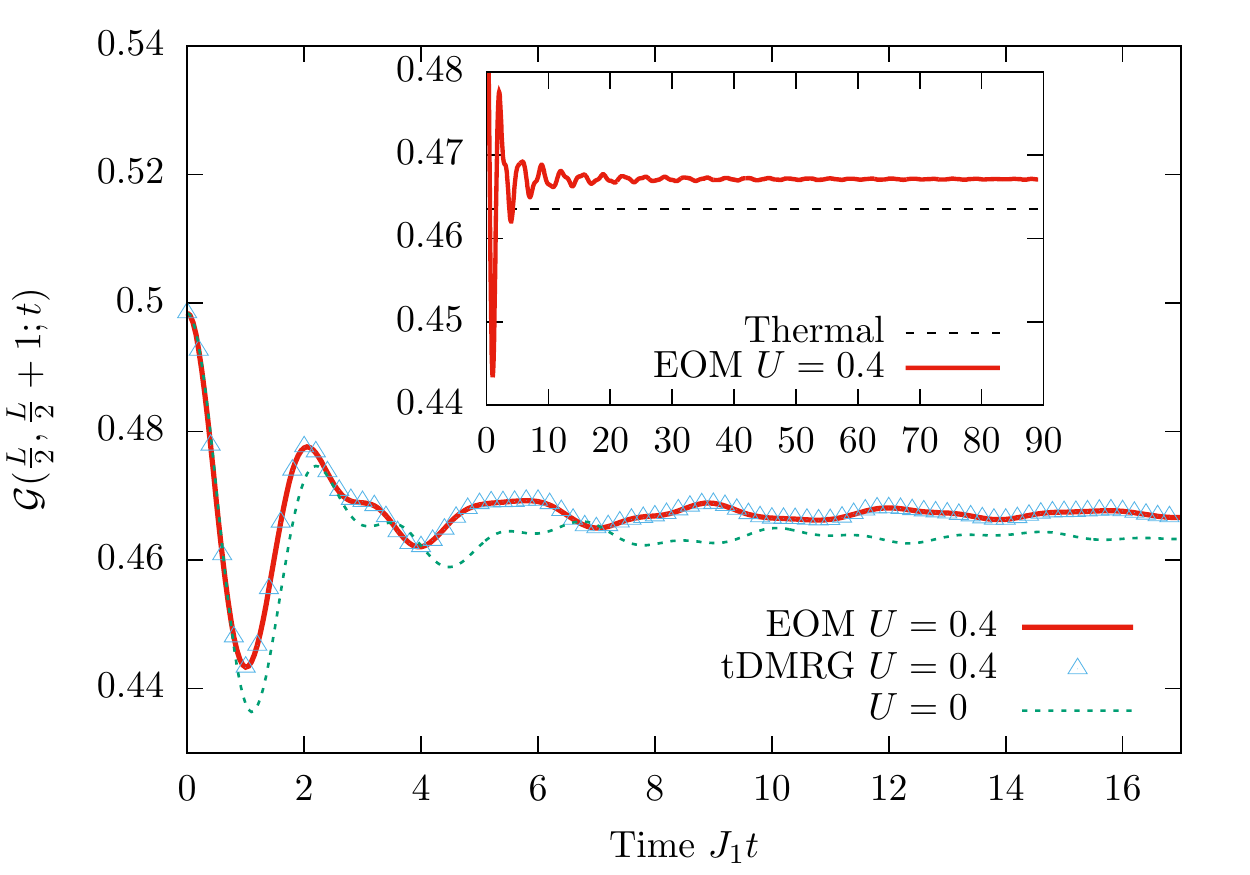} \\
\end{tabular}
\caption{
Green's function ${\cal G}({L}/{2},{L}/{2}+1;t)$ on a $L=256$ site
system for a quench where the system is prepared in the ground state
of $H(0,0.8,0)$ and time evolved with $H(0,0.4,0.4)$. EOM results (red
line) are in excellent agreement with t-DMRG
computations~\cite{EsslerPRB14} (triangles); the green dotted line
shows the result of the ``integrable quench'' (\emph{i.e.} time
evolution generated by $H(0,0.4,0)$, \emph{cf.}
Sec.~\ref{subsec:integrablequench}). Inset: behaviour on a larger time
interval.  }
\label{Fig:G1}
\end{figure}
In Fig.~\ref{Fig:G1} we report the time-evolution of ${\cal
  G}(L/2,L/2+1)$ computed by means of the EOM and t-DMRG for a 
quench in which the system is prepared in the ground state of
$H(0,0.8,0)$ and time-evolved with $H(0,0.4,0.4)$. We see that,
despite $U=0.4$ being relatively large, there is remarkably good
agreement between the two methods for all times accessible to the
t-DMRG computations. Importantly, t-DMRG results \cite{EsslerPRB14} for the Green's
functions at larger distances are similarly well reproduced. This
agreement confirms that the EOM method is accurate for small values of
$U$ on short and intermediate time scales. The main advantage of the EOM method
compared to t-DMRG is that it allows us to access larger systems and
longer times than those previously reported.\cite{EsslerPRB14} 

We observe very long-lived PT plateaux, as is
exemplified in the inset of Fig.~\ref{Fig:G1}. There is an
intermediate time window during which the Green's functions appear to
settle to quasi-stationary values. These are well-separated from the
thermal values, computed via exact diagonalization (ED) on a system of
$L=16$ sites. In computing the thermal values, we adopt the following
procedure: we first compute the energy density in our system, given by  
\be
e=\frac{1}{L}{\rm Tr}\left[\rho_0 H(J_{2},\delta_f,U)\right]~.
\ee
We then determine the effective temperature $1/\beta_{\rm eff}$ of the
thermal ensemble for the post-quench Hamiltonian $H(J_{2},\delta,U)$
through  
\be
e\overset{!}{=}\frac{1}{L}{\rm Tr}\left[\frac{1}{Z}e^{-\beta_{\rm eff}H(J_{2},\delta_f,U)}
H(J_{2},\delta_f,U)\right]\Bigr|_{\textrm{fixed}\, n}~.
\label{thermalens}
\ee  
In practice we compute \fr{thermalens} by ED, where the trace is
performed over states with a fixed particle number density $n={\rm
  Tr}\left[\rho_0  N\right]/L$. We then compute the single-particle
Green's function in thermal equilibrium at temperature $1/\beta_{\rm
  eff}$ using the same method. Our ED results for the thermal value of
${\cal G}(L/2,L/2+1)$ are consistent with the quantum Monte Carlo
results reported in Ref.~[\onlinecite{EsslerPRB14}]. 

The quasi-stationary values to which the Green's functions relax are
compatible with the CUT results of Ref.~[\onlinecite{EsslerPRB14}] up
to order $U^2$ corrections. This means that (up to the $O(U^2)$ 
corrections) the stationary values can be described the ``deformed GGE'' 
ensemble,\cite{EsslerPRB14} which corrects the stationary values of 
the non-interacting GGE to $O(U)$.

\subsection{Effects of next-nearest-neighbour hopping and finite
temperature initial states.}
\label{subsec:nnnhopping}

To investigate whether the prethermalized regime eventually evolves toward
thermal equilibrium, it is convenient to both invoke a non-zero
next-nearest-neighbour hopping amplitude $J_2$, and to initialize the
system in a thermal density matrix rather than a ground state. 
Here we focus on the thermal initial state \fr{thermal} with inverse 
temperature $\beta_i=2$ and $\delta_i=0$. A detailed analysis of the
dependence of the time evolution on $\beta_i$ and $\delta_i$ is carried out in 
Sec.~\ref{sec:initialstate}. The dependence of the dynamics of ${\cal
  G}(i,j;t)$ on the final dimerization $\delta_f$, the sign of
interaction $U$, and the presence of particle-hole symmetry is discussed in
Appendix~\ref{app:details}. 

\begin{figure}[b!]
\includegraphics[width=0.425\textwidth]{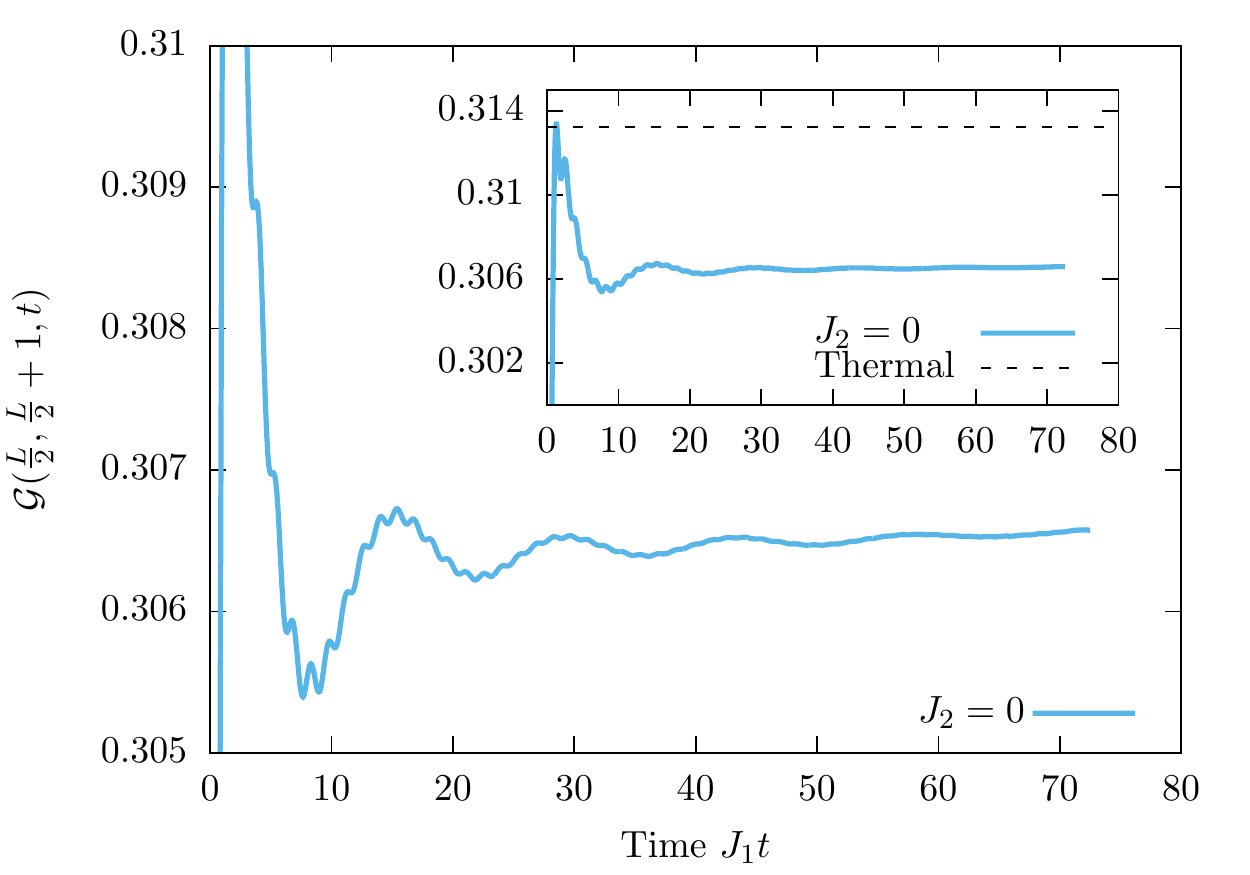}
\caption{
Green's function ${\cal G}({L}/{2},{L}/{2}+1;t)$ for a system
of size $L=320$ prepared in state with density matrix
$\rho_0(2,0)$\fr{thermal} and time-evolved with $H(0,0.1,0.4)$. 
The insets show the position of the ED thermal values.
}
\label{Fig:J20zoomed}
\end{figure}

We start by investigating the effects of a finite temperature initial
state on the dynamics with $J_2=0$. In Fig.~\ref{Fig:J20zoomed} we
show results for the Green's functions ${\cal
  G}({L}/{2},{L}/{2}\pm1;t)$ for a $L=320$ site system time-evolved
with the Hamiltonian $H(0,0.1,0.4)$ and initially 
prepared in a thermal state \fr{thermal} with density matrix
$\rho_0(2,0)$. We observe a very slow drift towards the thermal
value (note the scale on the y-axis). 
This should be contrasted to starting from the ground state and
$J_2=0$, \emph{cf.} Fig.~\ref{Fig:G1} and
Ref.~[\onlinecite{EsslerPRB14}], where no drift is observed on the
time scales accessible to us. 

\begin{figure}[t!]
\includegraphics[width=0.45\textwidth]{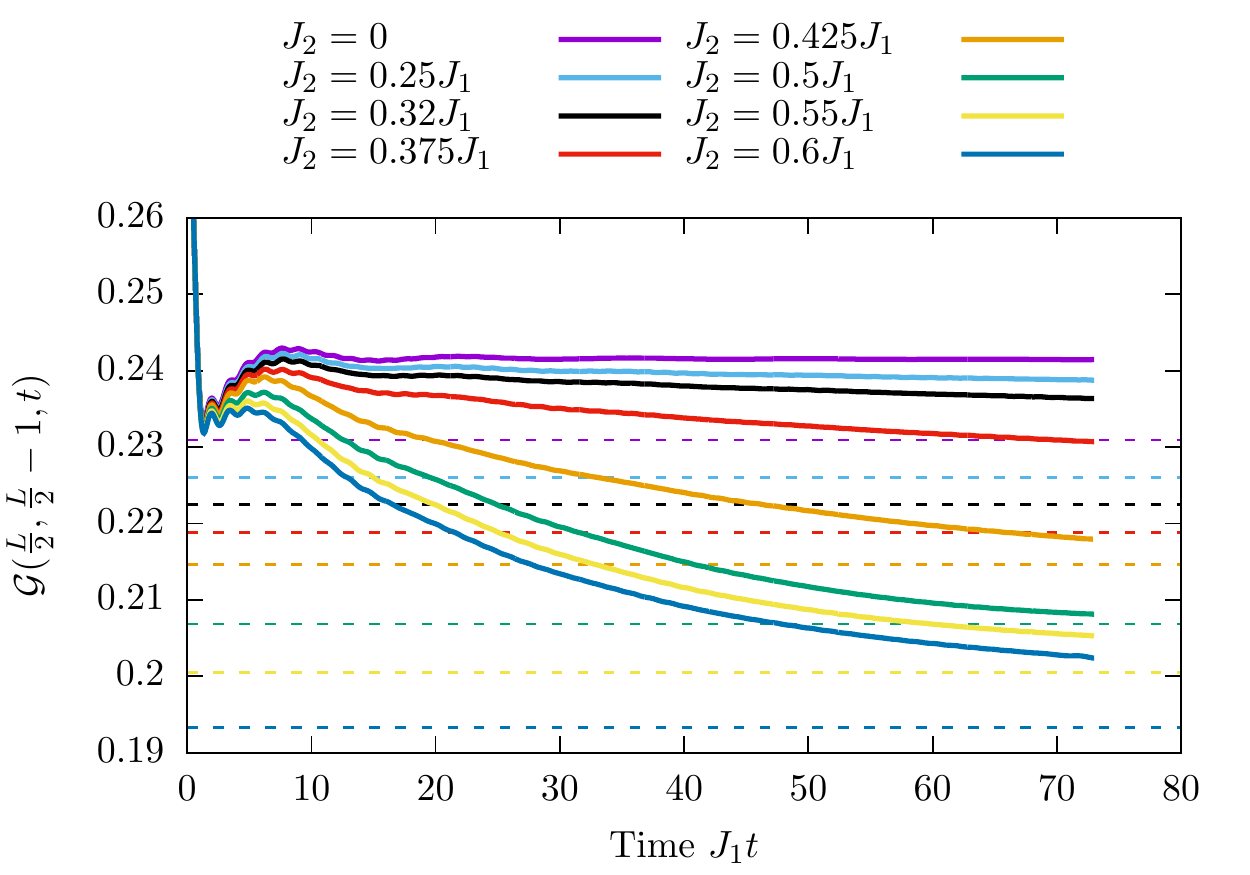}
\caption{Green's function ${\cal G}({L}/{2},{L}/{2}-1;t)$ for a system
of size $L=320$ prepared in state with density matrix
$\rho_0(2,0)$\fr{thermal} and time-evolved with $H(J_2,0.1,0.4)$. 
Dashed lines indicate the thermal values computed by ED of $L=16$ sites.
Different colours correspond to different values of 
$J_2=0,0.25,0.32, 0.375,0.425,0.5,0.55,0.6$ (top to bottom). } 
\label{Fig:J2dep}
\end{figure}

We now turn our attention to the effects of including a next-nearest-neighbour
tunneling $J_2$ at fixed $U$. In Fig.~\ref{Fig:J2dep} we
report the time-evolution of the Green's function ${\cal
  G}(L/2,L/2-1)$ for the system prepared in the thermal density matrix
$\rho_0(2,0)$ defined in Eq.~\fr{thermal} and subsequently
time-evolved with the Hamiltonian $H(J_2,0.1,0.4)$ for
$J_2=0,\ 0.25,\ 0.375,\ 0.425,\ 0.5,\ 0.55,\ 0.6$. 
We see that the main effect of increasing $J_2$ at fixed $U$ is to
induce a drift off the PT plateau towards the thermal
values.

\begin{figure}[ht]
\includegraphics[width=0.45\textwidth]{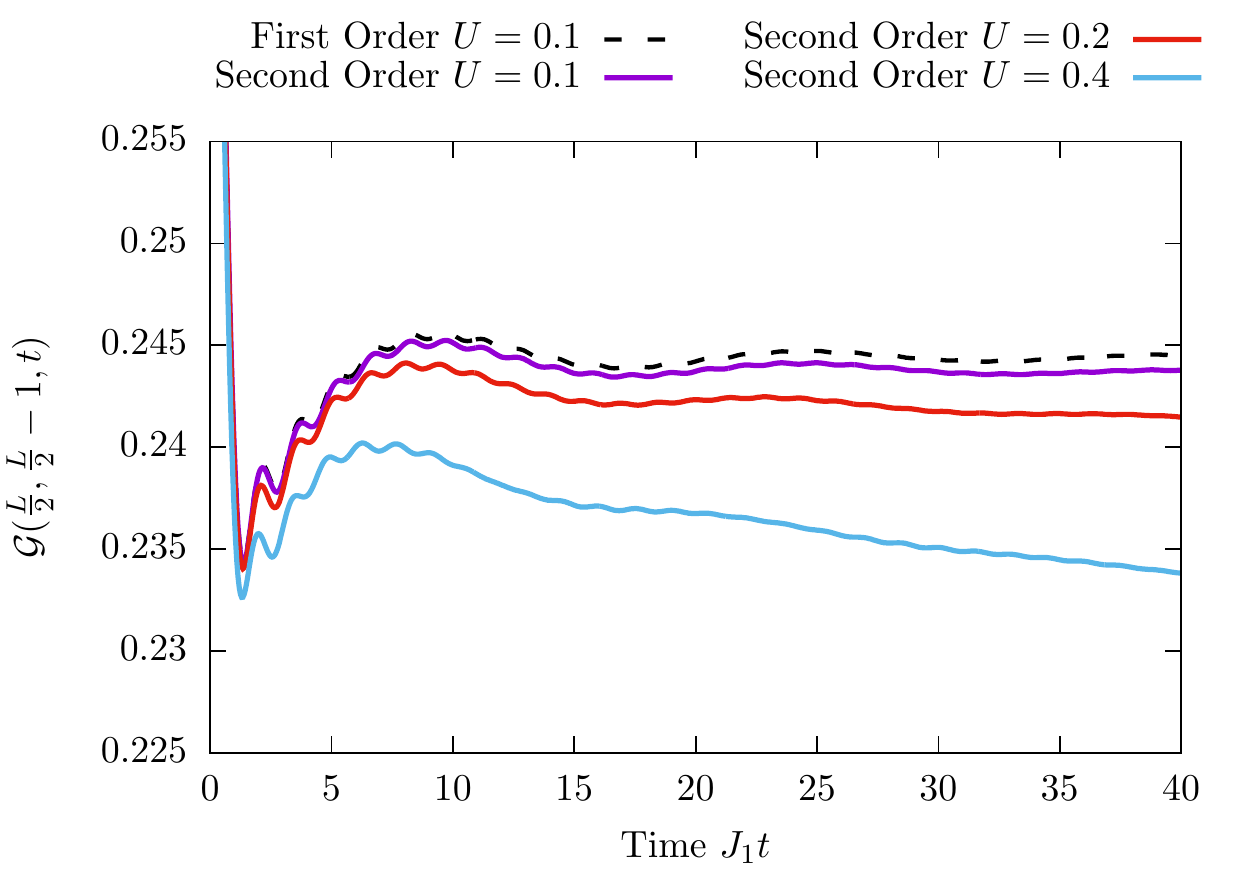}
\caption{Green's function ${\cal G}({L}/{2},{L}/{2}-1;t)$ for a system
of size $L=320$ prepared in the density matrix
$\rho_0(2,0)$~\fr{thermal} and time-evolved  with $H(0.375,0.1,U)$ for
several values of $U$. Solid lines show results from the second order  
EOM~\fr{Eq:EOM} for $U=0.1,0.2,0.4$ (top to bottom). For $U=0.1$ we compare with 
the first order EOM (dashed).}
\label{Fig:VaryingU}
\end{figure}

For weak next-nearest-neighbour hopping the system is close to the
prethermalized quasi-stationary state over a large time interval as is
illustrated in Fig.\,\ref{Fig:J2dep}. Increasing the value of the
$J_2$ causes this time window to significantly reduce and for large
values of $J_2$ expectation values rapidly approach their thermal
values. Importantly, the first order EOM remain prethermalized
\textit{for all times} and for all strengths of $J_2$. 
We stress that this does not imply that the first order EOM ``do not
work'' for large $J_2$: for any given value of $J_2$
\emph{we always observe a PT plateau} as long as $U$ is
sufficiently small. In this regime the first order EOM are in good
agreement with those at second order. This point is illustrated in
Fig.~\ref{Fig:VaryingU}.

\begin{figure}[t!]
\includegraphics[width=0.5\textwidth]{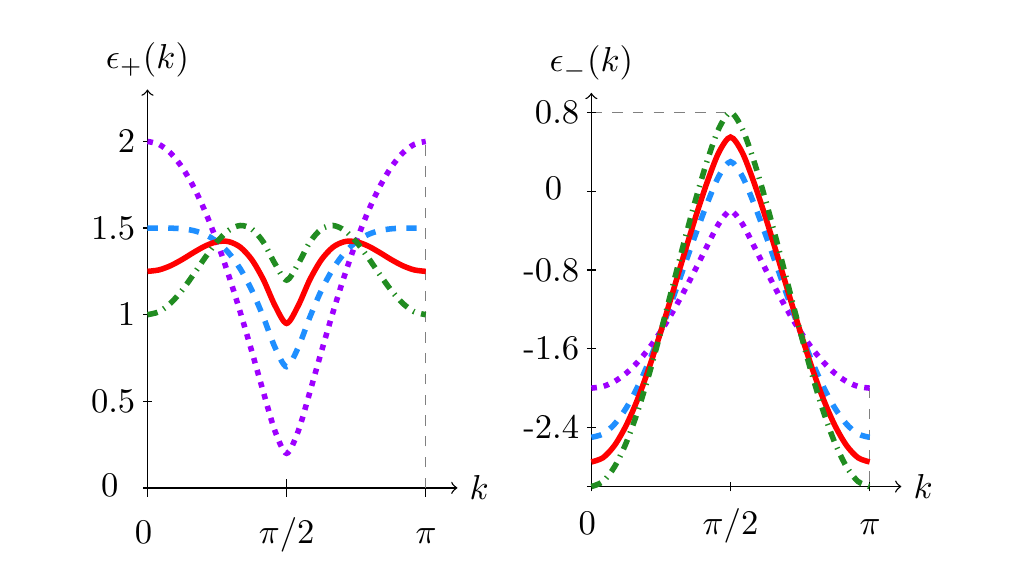}
\caption{Dispersion relations for the two bands of Bogoliubov fermions
in the non-interacting model with $J_1=1$, $\delta=0.1$ and 
$J_2=0$ (violet, dotted), $J_2=0.25$ (blue, dashed), $J_2=0.375$ (red,
solid), $J_2=0.5$ (green, dot-dashed). Increasing $J_2$ leads to
additional crossings at a fixed energy for $\epsilon_+(k)$. For $J_2 >
0.6$ additional crossings at fixed energy appear as the ranges of
$\epsilon_+(k)$ and $\epsilon_-(k)$ overlap.} 
\label{fig:disp}
\end{figure}

In summary, \emph{at fixed $U$} the addition of $J_2$ allows us to 
tune the crossover timescale between the prethermalized and
thermalized regimes. Some understanding of the strong dependence on
$J_2$ can be gained by considering $J_2 > 0.25$, where additional
scattering channels open due to crossings at a fixed energy (see
Fig.~\ref{fig:disp}), which promotes relaxation.
Figure~\ref{fig:disp} exhibits a second important effect of $J_2$: it
changes the bandwidths $W_{1,2}$ of both bands and leads to a
reduction of $W_{\rm min}={\rm min}(W_1,W_2)$. This in turn leads to
a larger value $U/W_{\rm min}$ of the dimensionless interaction strength.
However, as is shown in Appendix~\ref{sec:phsymmetry}, even in cases
where $W_{\rm min}$  is unchanged, the opening of additional scattering
channels (\emph{i.e.} increasing the number of crossings at fixed
energy) is sufficient to speed up the relaxation. These findings are
in accord with recent work on the relaxational dynamics in the
Hubbard model\cite{FMS:NIQBE, BK:rates} by means of quantum Boltzmann
equation methods. In these works it was observed that adding a 
next-nearest neighbour hopping leads to thermalization.

\subsection{Beyond the prethermalization plateaux}\label{sec:beyondPT}
\begin{figure}[t!]
\begin{tabular}{l}
\includegraphics[width=0.425\textwidth]{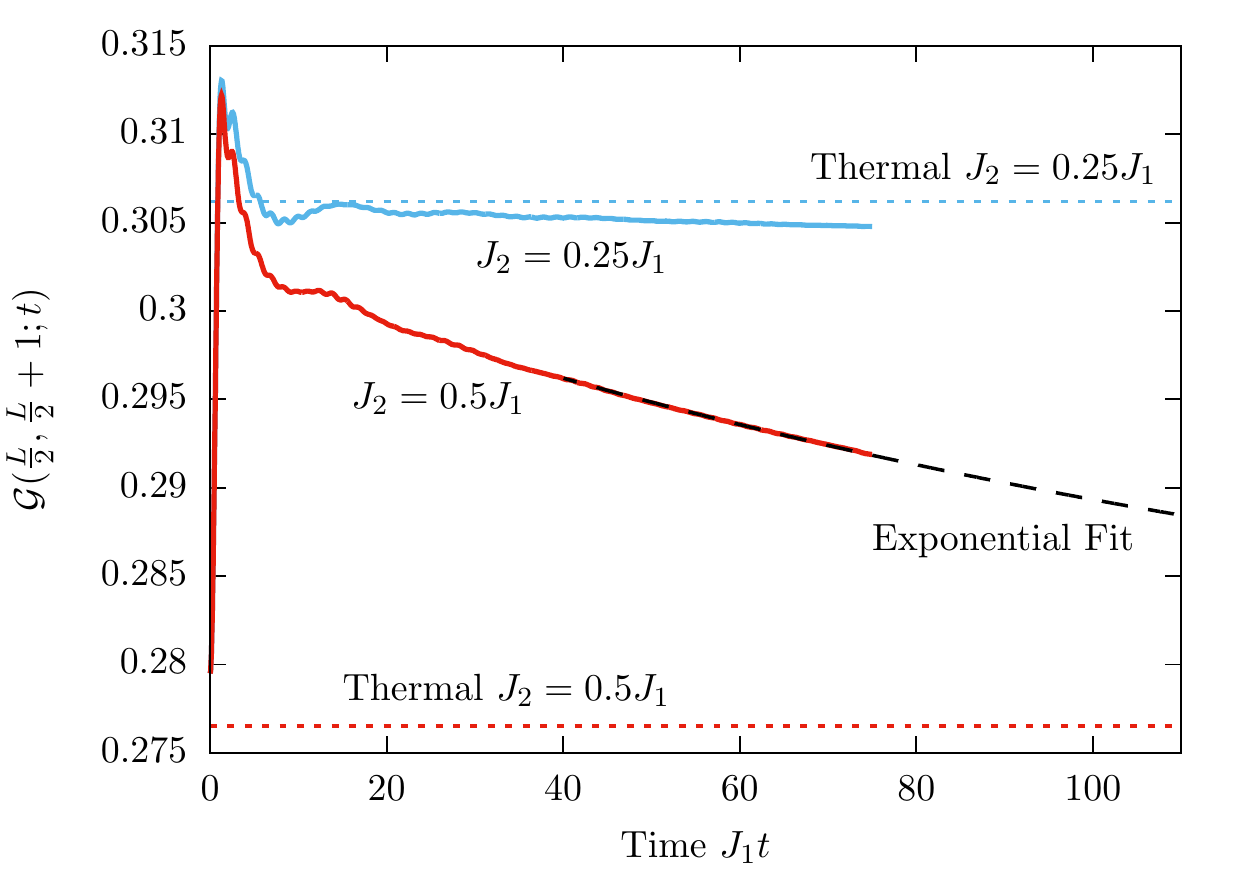}\\
\includegraphics[width=0.425\textwidth]{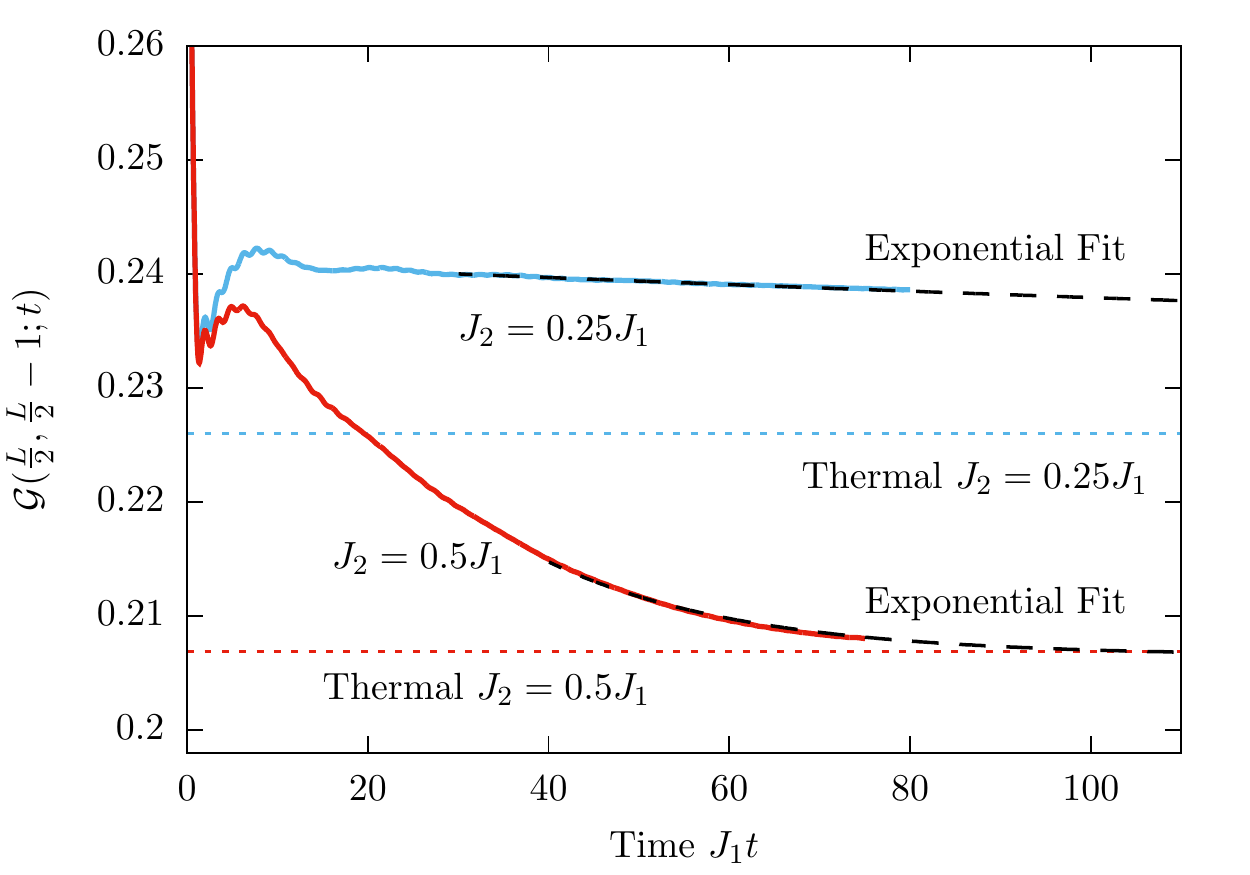}
\end{tabular}
\caption{
Green's function ${\cal G}({L}/{2},{L}/{2}\pm1;t)$ for a
system of size $L=320$ prepared in state with density matrix
$\rho_0(2,0)$ and time-evolved with
$H(J_2,0.1,0.4)$. Expected steady state thermal values are shown
as dotted lines, whilst the black dashed 
lines are exponential fits to Eq.~\fr{Eq:ExpFit}. Data in the lower
panel have been previously reported in Ref.~[\onlinecite{BEGR:PRL}].
}  
\label{fig:TPG1a}
\end{figure}

\begin{figure}[ht]
\begin{tabular}{l}
\includegraphics[width=0.425\textwidth]{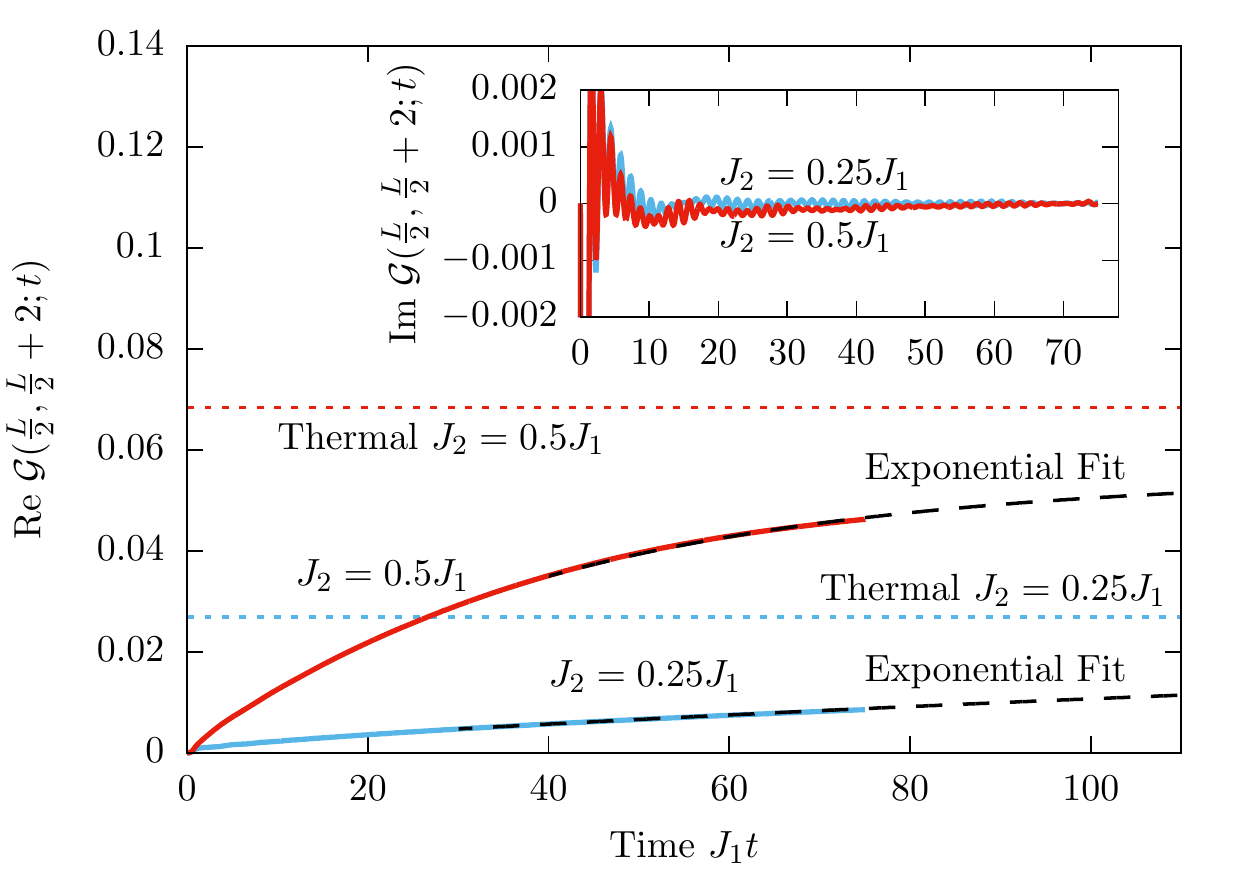}
\end{tabular} 
\caption{
Real and imaginary (inset) parts of the Green's function ${\cal G}({L}/{2},{L}/{2}+2;t)$ for a system of size
$L=320$ prepared in state with density matrix
$\rho_0(2,0)$ and time-evolved with
$H(J_2,0.1,0.4)$. Expected steady state thermal values, are shown by dotted lines
while the black dashed lines are exponential fits to
\fr{Eq:ExpFit}. These data have been previously reported in
Ref.~[\onlinecite{BEGR:PRL}].}
\label{fig:TPG2a}
\end{figure}

\begin{figure}[ht]
\begin{tabular}{l}
\includegraphics[width=0.425\textwidth]{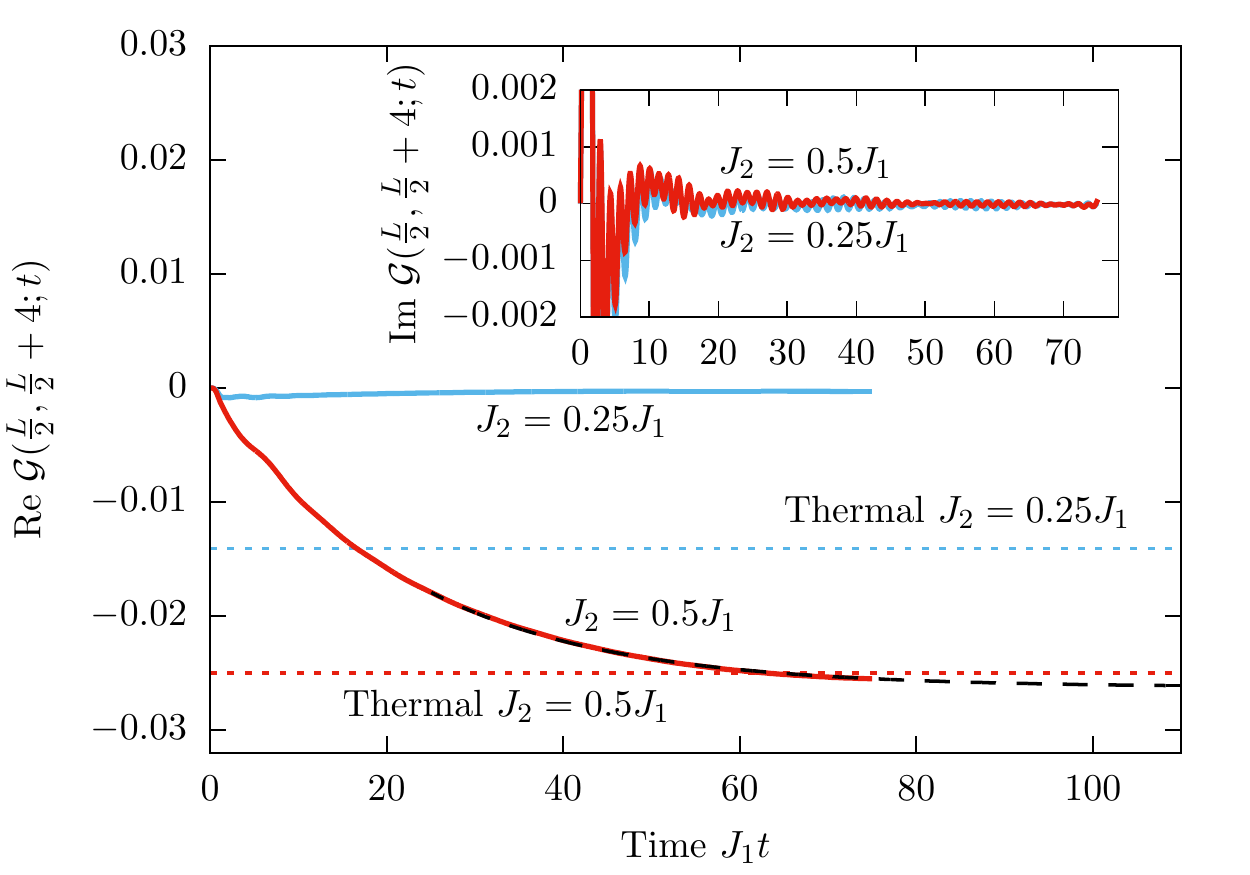}  
\end{tabular} 
\caption{
Same as Fig.~\ref{fig:TPG2a} for ${\cal G}({L}/{2},{L}/{2}+4;t)$.
}
\label{fig:TPG2b}
\end{figure}

As we have stressed before, for sufficiently small integrability
breaking parameter $U$ we always observe a PT
plateau. On the other hand, by keeping $U$ fixed and increasing $J_2$ we
can access a regime beyond
PT. Figures~\ref{fig:TPG1a}--\ref{fig:TPG2b} show the 
time-evolution of the Green's function for different separations and
two values of the next-nearest-neighbour hopping amplitude $J_2$ which
generate qualitatively different evolution of the local
observables. With $J_2 = 0.25$, the Green's function remains close to
the value in the prethermalized state for long times, whilst for $J_2=0.5$ the
system rapidly thermalizes. The thermal values shown in the figures
are computed by ED of small systems of $L=16$ sites. Note that the
Green's function for even separations are complex and their real parts
always show a smooth behaviour in time, see for example Fig.~\ref{fig:TPG2a}. 
This is because they depend only on the occupation numbers $n_{\mu\mu}(k,t)$ 
which are slowly varying functions of time, \emph{cf}. Section~\ref{sec:qbe}.


We now focus on the Green's function between sites at distances such
that we observe a clear drift towards the thermal values on the
time scales accessible to us, see Figs.~\ref{fig:TPG1a}, \ref{fig:TPG2a} and \ref{fig:TPG2b}. The observed
behaviour is compatible with an exponential decay towards the thermal
value  
\be
{\cal G}(i,j;t) \sim {\cal G}(i,j)_{\text{th}} + A_{ij}(J_2, \delta, U) e^{-t/\tau_{ij}(J_2, \delta_f, U)}~.
\label{Eq:ExpFit}
\ee
Here ${\cal G}(i,j)_{\text{th}}$ is the thermal Green's function at
temperature $1/\beta_{\rm eff}$. In general the relaxation times for
the real and imaginary parts of the Green's function between evenly
separated sites are different, and we denote them by $\tau_{ij}(J_2,
\delta_f, U)_{\textrm{r,i}}$ in the following.  In some cases, for
example the $J_2=0.5$ case of Figs.~\ref{fig:TPG2a}~and~\ref{fig:TPG2b},
to obtain a better 
fit we have to allow the thermal value ${\cal G}(i,j)_{\text{th}}$ to
deviate from the ED result by a small amount. We believe that this
(tiny) discrepancy can be explained by a combination of errors in the
EOM and finite-size effects on the ED result. 

\begin{figure}[ht]
\includegraphics[width=0.45\textwidth]{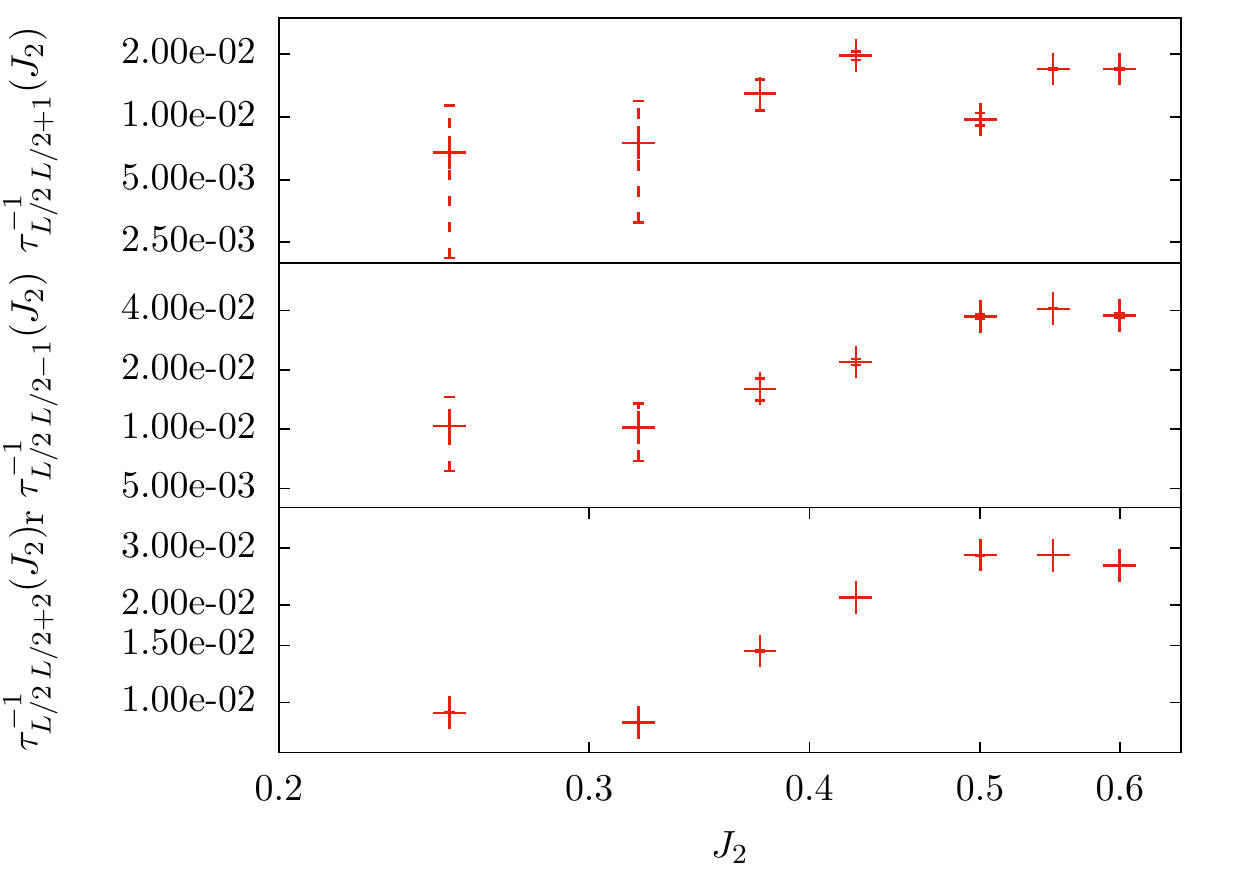}
\caption{Inverse relaxation times $\tau^{-1}_{{L}/{2}\,{L}/{2}+ j}(J_2)$ with $j=1,-1,2$
(top to bottom) characterizing the late time behaviour of the Green's
function (\emph{cf.} Fig.~\ref{Eq:ExpFit}). The system is prepared in
the initial density matrix $\rho_0(2,0)$ \fr{thermal} and
time-evolved under the Hamiltonian $H(J_2,0.1,0.4)$. The system size
is $L=320$ (\emph{cf.} Fig.~\ref{Fig:J2dep}). Errors are estimated by
varying the initial time at which the exponential fit is applied.} 
\label{Fig:expvsJ2}
\end{figure}

In Fig.~\ref{Fig:expvsJ2} we show the inverse relaxation times determined
by fitting the decay of the Green's function to the form \fr{Eq:ExpFit}
for a system prepared in the thermal state~\fr{thermal} with density
matrix $\rho_0(2,0)$ and time-evolved under the Hamiltonian
$H(J_2,0.1,0.4)$ on a $L=320$ site chain. We see that the relaxation times
$\tau_{ij}(J_2,\delta_f, U)$ are quite sensitive to the value of 
$J_2$, which in turn has a large influence on whether drifting towards 
thermalization can be observed within the time window accessible to us.
Increasing the separation between the two sites leads to an increase
of the relaxation times. This takes the time window in which
\fr{Eq:ExpFit} holds beyond the regime accessible to us by a numerical
solution of \fr{Eq:EOM}. However, we conjecture that the 
relation \fr{Eq:ExpFit} describes the relaxation towards the thermal
value of the Green's function for any value of the separation if one
waits for long enough times. This is in some sense a ``minimal''
assumption: it is reasonable to think that the relaxation behaviour of
the Green's function remains qualitatively the same for any separation
of the two sites, providing $\abs{i-j}\ll L$. In the following, we
will give other arguments in favor of this conjecture by exploring the
dynamics for longer times with a quantum Boltzmann equation, which can
be derived as the scaling limit of the equations \fr{Eq:EOM} for the
occupation numbers, see Sec.~\ref{sec:qbe}.

\subsection{Initial state dependence}\label{sec:initialstate}
We now turn to the dependence of the relaxation of local observables
on properties of the initial state. We note that for integrable models
this question has been the subject of extensive numerical studies, see
for example~[\onlinecite{RigolPRA11,HePRA12,HePRA13,TorresHerreraPRE13}].

\subsubsection{Dependence on the energy density}
\begin{figure}[t!]
\begin{tabular}{l}
\includegraphics[width=0.425\textwidth]{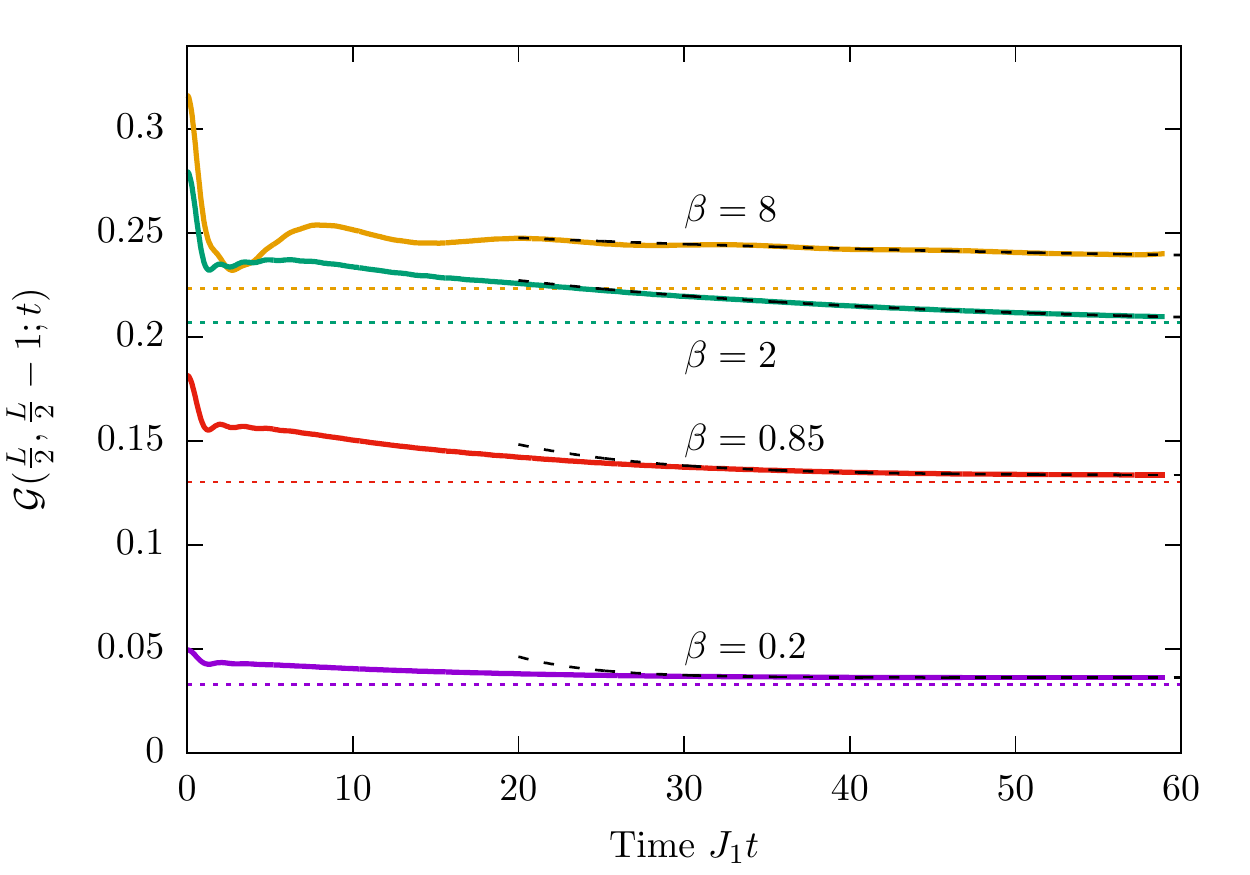}
\end{tabular}
\caption{${\cal G}({L}/{2},{L}/{2}-1;t)$  for a system with
Hamiltonian $H(0.5,0.1,0.4)$ and sizes $L=256, 320$ initially prepared in the density matrix $\rho_0(\beta,0)$ for four different
values of $\beta$. The expected steady state thermal values are
indicated by dotted lines and the exponential fit~\fr{Eq:ExpFit} by
a dashed line.}  
\label{Fig:endep}
\end{figure}

We
first consider the dependence of the relaxational behaviour on the
energy density of the initial state. We note that the energy density
of the various quenches we considered in
Secs.~\ref{subsec:nnnhopping} and \ref{sec:beyondPT} (see also Appendix~\ref{subsec:deltadep}) was, in fact, fixed because the initial state satisfies
\be
\textrm{Tr}\left[\rho_0(\beta,0) H(J_2,\delta,U)\right]
= \textrm{Tr}\left[\rho_0(\beta,0) H(0, 0,U)\right]\,.  
\ee
In Fig.~\ref{Fig:endep} we present the time-evolution of the Green's
function $\mathcal{G}(L/2,L/2-1;t)$ starting from the initial density
matrices $\rho_0(\beta,0)$~\fr{thermal} with $\beta = 0.2,0.85,2,8$. 

\begin{figure}[t!]
\includegraphics[width=0.45\textwidth]{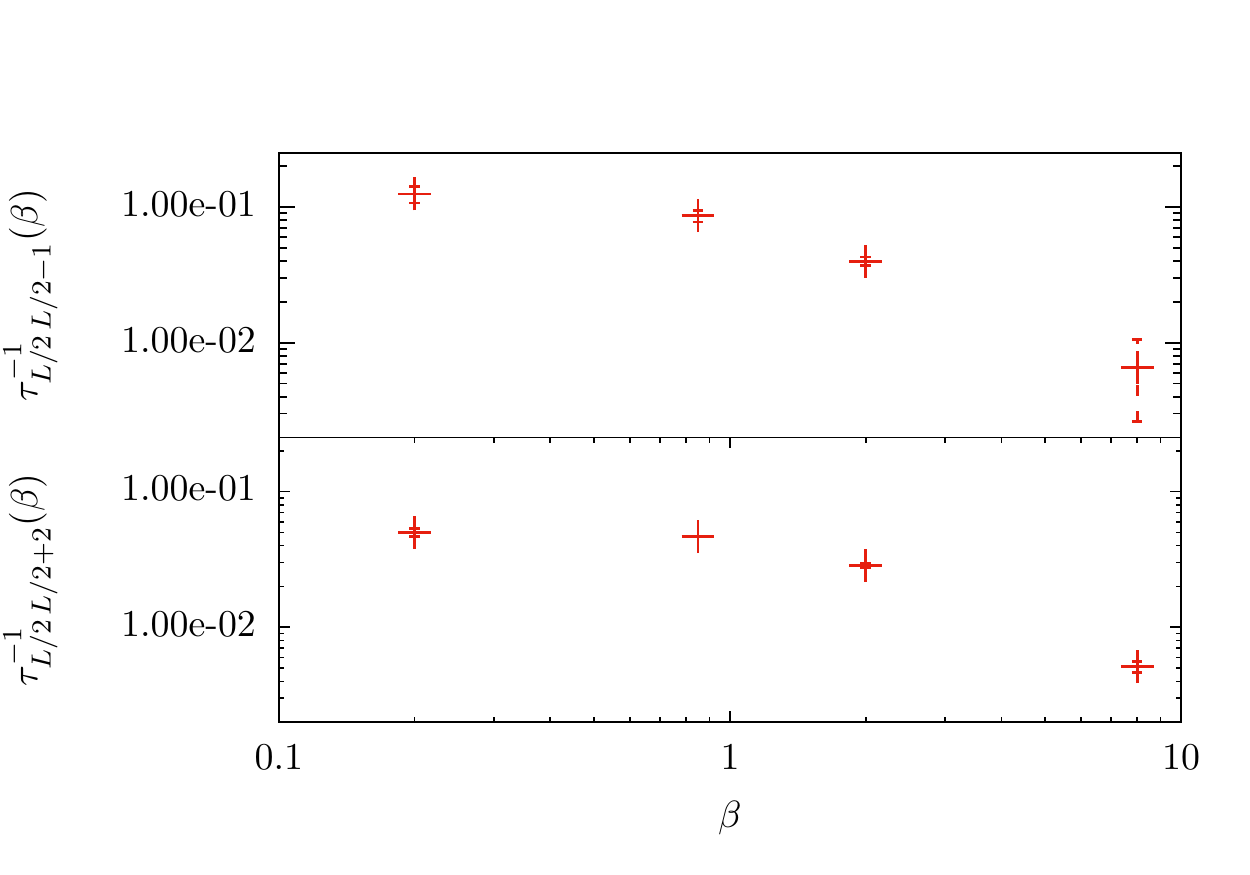}
\caption{Inverse Relaxation times
$\tau^{-1}_{i\,i-1}(J_2,\delta,U,\beta)$ (above) and
$\tau^{-1}_{i\,i+2}(J_2,\delta,U,\beta)$ (below) obtained by fitting
the data in Fig.~\ref{Fig:endep} with Eq.~\fr{Eq:ExpFit}. Error bars are
estimated by varying the initial time at which the exponential fit
is applied.\cite{notefig}} 
\label{Fig:expbeta}
\end{figure}

Our results suggest that, fixing all other parameters, \textit{the
time window for which observables show a prethermalized behaviour
increases when the temperature of the initial state is decreased}
(i.e., with increasing $\beta$). In order to quantify this statement,
we note that the data are well described by the ``exponential
relaxation'' introduced in Eq.~\fr{Eq:ExpFit}. We plot the inverse
exponents $\tau^{-1}_{i,i-1}$ and $\tau^{-1}_{i,i+2}$ as a function of $\beta$ in
Fig.~\ref{Fig:expbeta}; we see that even when there is clear decrease
of both exponents with $\beta$, the
dependence is not of a simple power-law form. In
Ref.~[\onlinecite{SK:EOM}], Stark and Kollar examined the limit of
infinite dimensions $d\rightarrow\infty$ and found that the exponent 
depends on the \emph{final} inverse temperature $\beta_f$ as
$\tau^{-1}_{i,j}\propto \beta_f^{-2}$. 
In the limit of high initial temperature $\beta_i\sim0$, we expect the initial and final temperatures to be comparable $\beta_i \sim \beta_f$. In this limit we find a dependence $\tau^{-1}_{i\,j}\propto\beta_f^{-\alpha}$ with $\alpha \approx 0.1 - 0.2$. The different  exponent compared to Ref.~[\onlinecite{SK:EOM}] has its origin in the distinct quench protocol as well as dimensionality.

We note that for small $\beta$ the range of variation of the Green's
function is very small. This is reasonable: both the initial and
thermal density matrices have the form $\rho \propto \mathbb{1} +
O(\beta)$, where $\mathbb{1}$ is the identity matrix. This means that
to leading order in $\beta$, the expectation values do not time
evolve.  

\subsubsection{Different initial states at fixed energy density}
\begin{figure}[ht]
\begin{tabular}{l}
\includegraphics[width=0.425\textwidth]{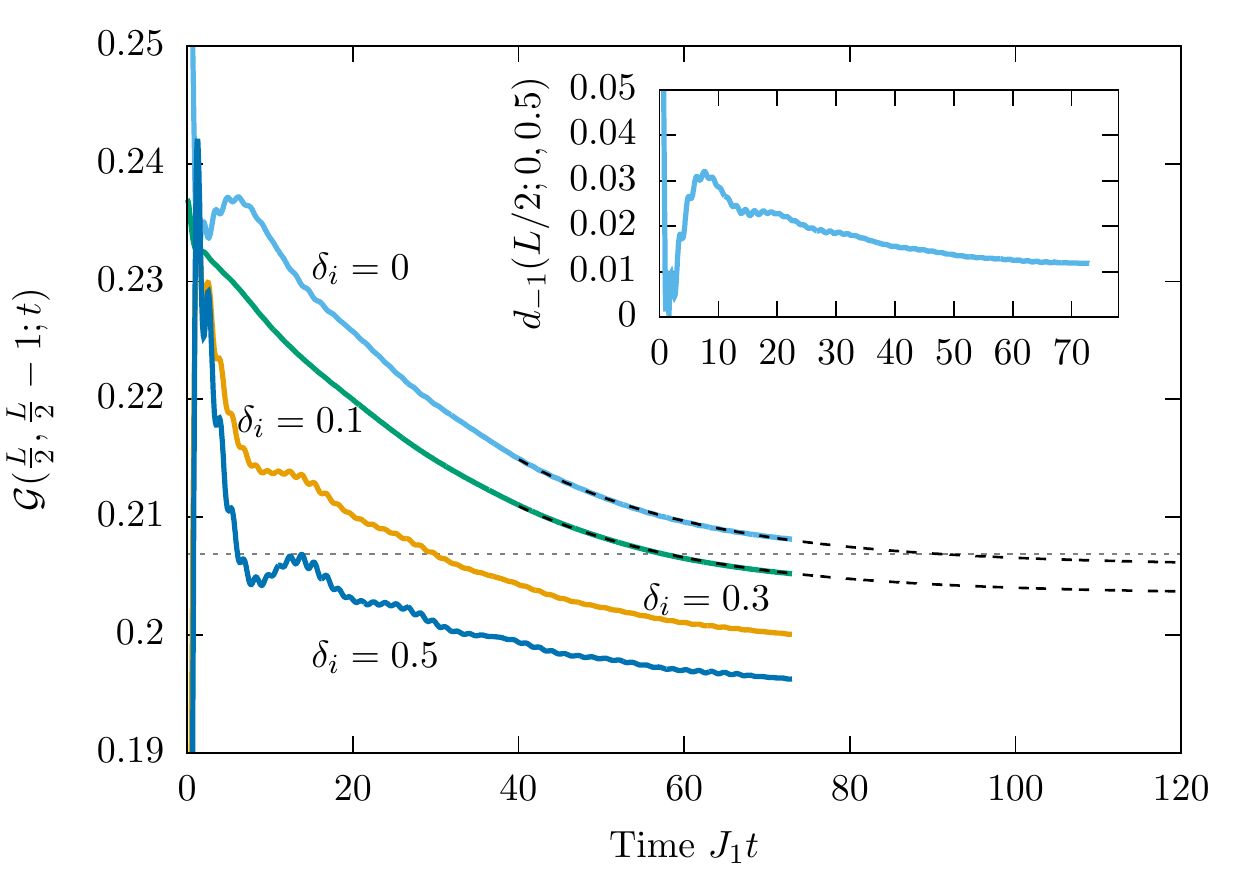} 
\end{tabular}
\caption{${\cal G}({L}/{2},{L}/{2}-1;t)$ for a system with
Hamiltonian $H(0.5,0.1,0.4)$ and size $L=320$ initially prepared in
a density matrix $\rho_0(\beta,\delta_i)$, where $\beta$ is chosen
such that all initial states have the same energy density. Dashed
lines show the exponential fit \fr{Eq:ExpFit}. Grey dotted lines
indicate the thermal values computed by ED. The insets show the behaviour of
  $d_{j}(i;0,0.5)$ [\emph{cf.} Eq.~\fr{Eq:dist}]. }  
\label{Fig:diffinitialstate}
\end{figure}

As our model is non-integrable, at late times we expect that the 
only relevant information contained in the initial conditions should
be the energy density. 
Concomitantly, we expect that the Green's functions evolve 
towards the same limiting values when starting from macroscopically different
initial states which have the same energy density.
 In order to investigate this point, we compare the
time-evolution of the Green's function starting from the density
matrix $\rho_0(\beta,\delta_i)$ for four different sets of
$(\beta,\delta_i)$, chosen such that the energy density in each case 
is identical (the number density is $1/2$ in $\rho_0$ with $J_{2i}=0$). 
By construction, these initial states are macroscopically different. As shown in
Fig.~\ref{Fig:diffinitialstate}, in each of the four cases the time evolution
of Green's functions proceeds quite differently at short and
intermediate times. The late time behaviour, however, appears to be
compatible with the same stationary value within the errors associated
with our approximations. This can be seen in the inset of
Fig.~\ref{Fig:diffinitialstate}, which shows the difference between
the two most distant curves
\be
d_{j}(i;0,0.5)=\bigl|\mathcal{G}(i,i+j)|_{\delta_i=0}-\mathcal{G}(i,i+j)|_{\delta_i=0.5}\bigr|\,.
\label{Eq:dist}
\ee
We may push this analysis further by carrying out fits of the EOM
results to the exponential form~\fr{Eq:ExpFit}, and then extrapolating
to late times. In cases where the relaxation is fast, e.g. $\delta_i =
0,0.1$), we find that the extrapolated ``stationary values'' are in
good agreement with one another and the expected thermal result
computed by ED on $L=16$ sites (the differences are $\sim 10^{-3}$).
For larger values of $\delta_i$ the relaxation is slower and the
extrapolated values differ significantly from the thermal result. This
is perhaps not surprising, as the quality of the exponential fit is
not expected to be as good in these cases. We note that the case with
$\delta_i = \delta_f = 0.1$ is very similar to the one considered in
Ref.~[\onlinecite{NessiArxiv15}].

\section{Light cone effects}\label{sec:lightcone}
The CUT approach \cite{MK:prethermalization,EsslerPRB14} to quantum
quenches provides a simple intuitive picture of PT. Switching on weak
interactions does not immediately destroy the free (non-interacting)
quasi-particles, but rather ``dresses'' them through particle-hole
excitations. To leading order in $U$ this deforms their dispersion,
and higher orders will generate quasi-particle decay and render their
lifetime finite. This picture propounds the idea that deviations from the
PT plateau and eventual relaxation to a thermal state may be related to
quasi-particle decay at finite energy densities. A very direct probe
of quasi-particle propagation is provided by light cone effects.
\cite{cc-05,dechiara06,CalabreseJStatMech07,laeuchli08,FC08,manmana09,
CEF,cetal-12,langen13,bonnes14,Krutitsky14,EF16,CC16,F:LocalSwitch,
BD:review, VM:review, BF:defect, CADY:hydro, BCDF:transport} We expect
the Green's function to exhibit such an effect, where the propagation 
velocity of the light cone is determined by the maximal quasi-particle
group velocity. In cases where quasi-particles are long lived, we expect to
observe a ``clean'' light cone effect over a large time window. On the
other hand, when there is a substantial decay rate, we expect the
structure of the light cone to be modified. While it is more or less
obvious that this intuition will hold for local quenches\cite{localQ}
it is a priori completely unclear whether it carries over to global
ones. Closely related questions have been recently investigated for
global quenches in the context of perturbed conformal field theories
\cite{C:furtherresults} as well as certain lattice models
\cite{confinementnoneq}.  

\subsection{Non-interacting case}
\begin{figure}[ht]
\includegraphics[width=0.45\textwidth]{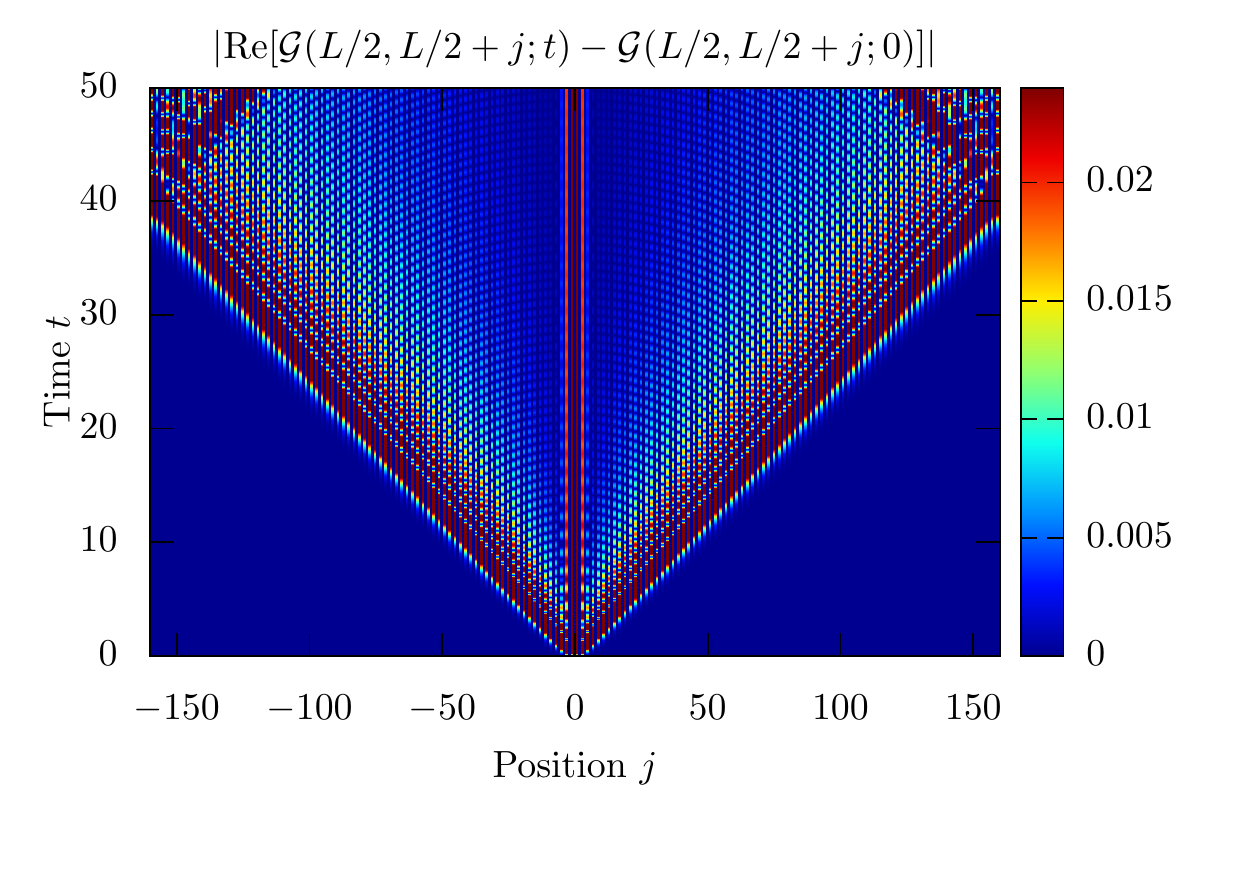} \quad
\includegraphics[width=0.45\textwidth]{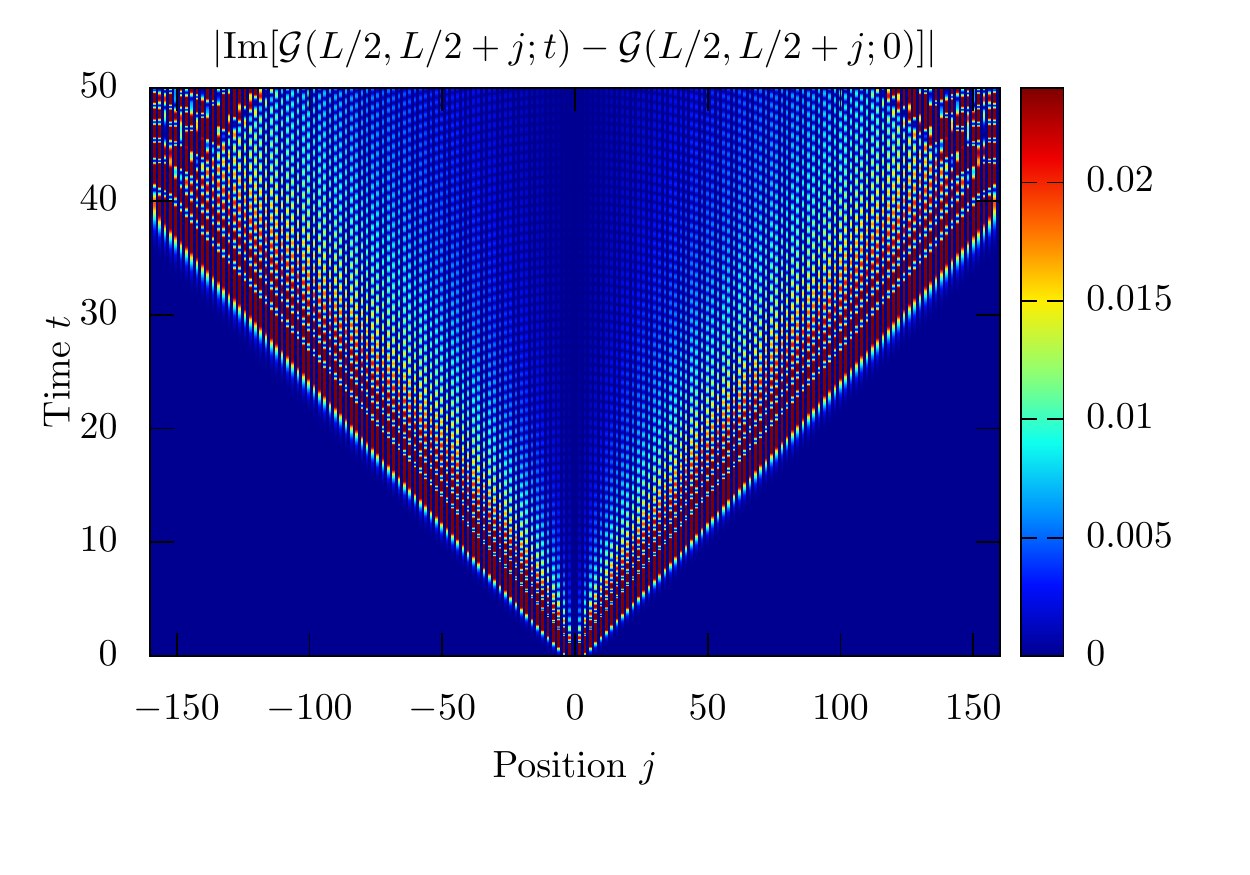}
\caption{Dynamics of the full Green's function $\mathcal{G}(L/2,L/2+j;t)$
as a function of time. The system is initialized in the density matrix
$\rho_0(2,0.5)$ and time evolved with the integrable (free) Hamiltonian
$H(0,0,0)$. The light cone's edge is spreading with velocity 
$v_{\rm max}=2\textrm{max}_{k}\epsilon_+'(k)\big|_{J_2=0;\delta=0}=2 J_1$, 
see Eq.~\fr{dispersion}. The signal outside of the light cone decays exponentially in time.} 
\label{fig:lightconeJ205FF}
\end{figure}

In order to set the stage we first consider quenches in the
non-interacting case $U=0$. We prepare the system in the
density matrix $\rho_0(\beta,\delta_i)$ and, to keep things as simple
as possible, time evolve with $H(0,\delta_f,0)$. We focus on the
single particle Green's function~\fr{Eq:GreenFun}. An example of the
light cone effect is shown in Fig.~\ref{fig:lightconeJ205FF}. The
real and imaginary parts of ${\cal G}(L/2+j,L/2,t)$
are seen to be very small outside a light cone that spreads with a
velocity that equals the maximal group velocity of elementary
excitations of the post-quench Hamiltonian. 
In the absence of interactions the structure of the light cone can be
straightforwardly analyzed. For ease of presentation we restrict
ourselves to odd separations $j$, choosing (even) $L/2$ as the
reference point. For odd separations the Green's function is real, see
Eq.~\fr{Eq:oddreal} (even separations can be treated in complete
analogy). In the thermodynamic limit our object of interest 
${\frak g}(j,t)={\cal G}(L/2,L/2+j;t)$ is thus given by
\bea
{\frak g}(j,t)&=&\int_0^\pi \frac{\text{d}k}{2\pi} e^{ikj
-i\varphi_k(\delta_i)}
\Big[ n_{++}(k)-n_{--}(k)\nn
&-&n_{+-}(k)e^{i\eps_{+-}(k)t}+n_{-+}(k)e^{i\eps_{-+}(k)t}\Big].\label{eq:gf}
\eea
The contribution due to the $n_{++}$ and $n_{--}$ terms is time independent
and plays no role in the light cone effect at large separations
$j$. We now simplify the problem further by taking $\delta_f=0$.
The relevant part of the integral is then given by
\bea
{\frak g}(j,t)&\simeq& \int_{0}^\pi \frac{\text{d}k}{\pi}e^{ikj}
\tilde n_{+-}(k) \cos\left({4t\cos k}\right).
\label{integral}
\eea
where we introduced 
\be
\tilde n_{+-}(k)= \frac{i}{2}\sin(\varphi_k(\delta_i))\tanh\left(\frac{\beta}{2}\varepsilon_+^{(0)}(k)\right)\,.
\ee
The structure of $\tilde n_{+-}(k)$ is such that there are no branch
points in the complex $k$-plane. The only singularities are simple
poles at positions determined by the argument of the $\text{tanh}$.
Depending on the ratio $j/t$ we have to distinguish between three
regimes. It is convenient to define a parameter $\gamma$ by
\be
\gamma=\sqrt{\frac{\frac{\pi ^2}{4 \beta ^2}+1}{1-\delta_i^2}}>1.
\ee
\subsubsection{Interior of the light cone: $j<2v_{\rm max}t$}
\label{interior}
The first regime is characterized by $j<2v_{\rm max} t$,
where $v_{\rm max}=2$ is the maximal group velocity of elementary
excitations. As long as ${2v_{\rm max}t}$ is  sufficiently larger
than $j\gg 1$, the $k$-integral can be evaluated by a straightforward
stationary phase approximation, which gives
\bea
{\frak g}(j,t)&\sim&\frac{1}{2\sqrt{2\pi t}}{\frak f}\left(\frac{j}{2v_{\rm max}t}\right)\nn
&&\qquad\times{\rm Im}\Big[ e^{i\left(\sqrt{(2v_{\rm max}t)^2-j^2}
+j\, \text{arcsin}\left(\frac{j}{2v_{\rm max}t}\right)-\frac{\pi }{4}\right)}\Big]\,,\nn
{\frak f}(z)&=&
 \frac{z\delta_i \tanh\big(\beta\sqrt{1-(1-\delta_i^2)z^2}\big)}
{\sqrt{1-(1-\delta_i^2)z^2}\ |1-z^2|^{1/4}}.
\label{Eq:inlc}
\eea
We see that inside the light cone the real part of the Green's
function displays an oscillating power-law decay with exponent $3/2$
(at late times).
\subsubsection{Short-time regime: $2v_{\rm max}t \gamma<j$}
In this regime the integrand features two saddle points 
in the complex $k$-plane at ${k_\pm=\frac{\pi}{2}\pm i
  \textrm{arccosh}\left(\frac{j}{2v_{\rm  max}t}\right)}$. When
deforming the integration contour to pass through $k_+$,
one encounters at least the simple pole at
$k=\frac{\pi}{2}+i\textrm{arccosh}(\gamma)$. The  
leading contribution to the integral stems from this pole. For large
$j$ we then obtain
\bea
{\frak g}(j,t)&\sim&A(j)
e^{-\left(j\ \text{arccosh}(\gamma)-2v_{\rm max} t \sqrt{\gamma^2-1}\right)}\ ,\nn
A(j)&=&\frac{\delta_i \cos\big((j-1)\pi/2\big)}{2\beta
  (1-\delta_i^2)\sqrt{\gamma^2-1}}.
\label{Eq:st}
\eea
This shows an exponential increase in time.

\subsubsection{Intermediate regime: ${j}<2v_{\rm max}t \gamma<j \gamma$.}
\label{intermediate}
Interestingly there exists an intermediate regime that separates the
interior of the light cone from the early time behaviour. 
For times ${j}<2v_{\rm max}t \gamma<\gamma j$ and large values of $j$,
the integral \fr{integral} can be evaluated by deforming the
integration contour into the complex plane until it passes through the
saddle point $k_+=\frac{\pi}{2}+i\text{arccosh}\left(\frac{j}{2v_{\rm
      max}t}\right)$. As long as ${\rm Im}(k_+)$ is
smaller than the imaginary parts of the simple poles of the integrand,
which defines the intermediate regime, the leading contribution to the
integral \fr{integral}  can be obtained by a saddle point
approximation. This gives 
\bea
{\frak g}(j,t)&\sim&
\frac{(-1)^{\frac{(j-1)}{2}}}{4\sqrt{2\pi t}} {\frak f}\left(\frac{j}{2v_{\rm max}t}\right) \nn
&&\quad\times e^{-\left(j\, \text{arccosh}\left(\frac{j}{2v_{\rm max}t}\right)-\sqrt{j^2-(2v_{\rm max}t)^2}\right)}\,.
\label{ginter}
\eea
Equation~\fr{ginter} provides a good approximation of the integral only ``far
enough'' from the light cone $j=2 v_{\rm max}t$ (as it exhibits a
singularity at $j=2 v_{\rm max}t$).

\begin{figure}[ht]
\includegraphics[width=0.5\textwidth]{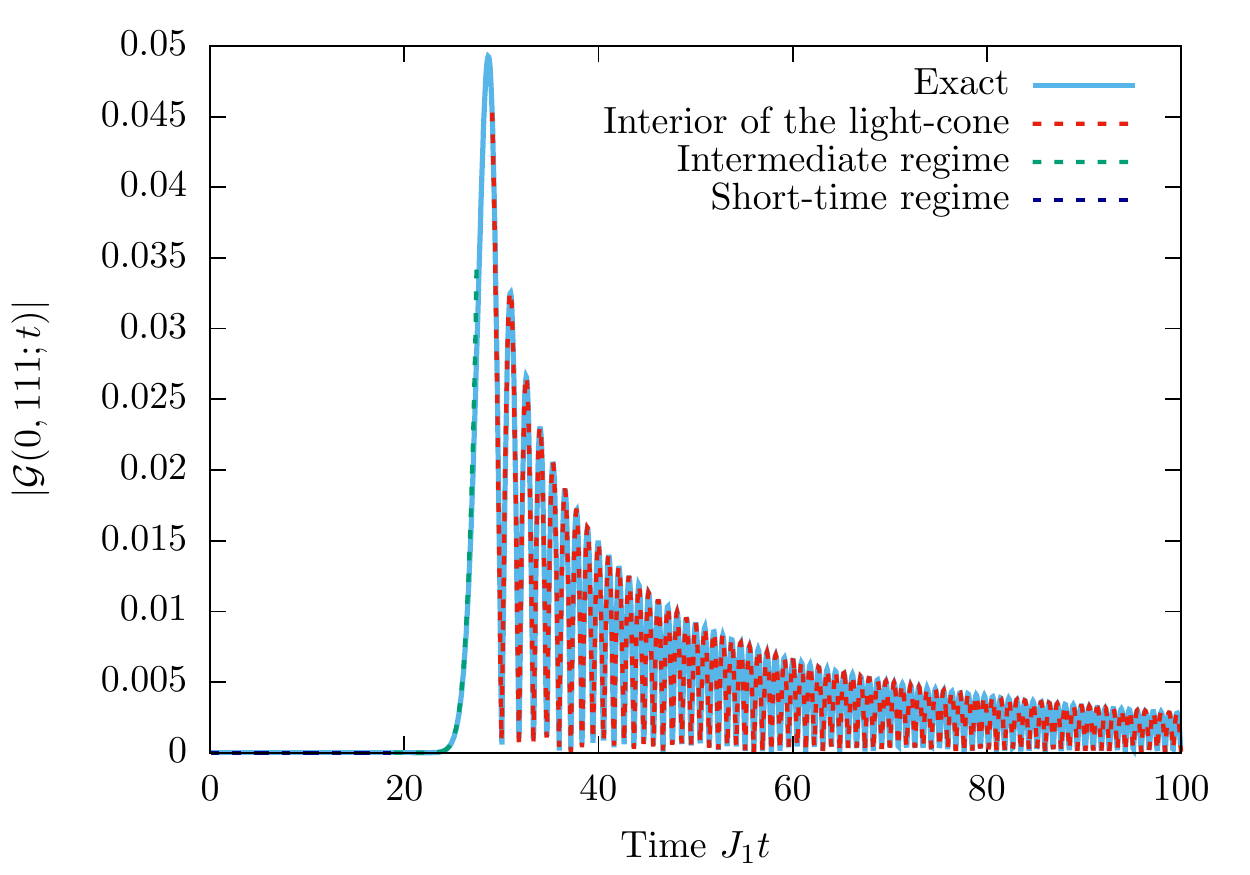}
\caption{Time evolution of ${\cal G}(0,111;t)$. The full line is obtained numerically integrating the exact expression \fr{integral}, while the dashed lines are obtained using the asymptotic expansions \fr{Eq:inlc}, \fr{Eq:st}, \fr{ginter}.} 
\label{fig:Cut}
\end{figure}

Fig.~\ref{fig:Cut} reports a representative fixed-separation cut of
${\cal G}(0,j;t)$ (here we have used invariance under translations by two
sites to shift the reference position to zero), comparing the
expansions \fr{Eq:inlc}, \fr{Eq:st}, and \fr{ginter} with the exact
expression obtained by numerical integration of \fr{integral}. The
agreement is clearly excellent. 

\subsubsection{Proximity of the light cone}
\begin{figure}[ht]
\includegraphics[width=0.45\textwidth]{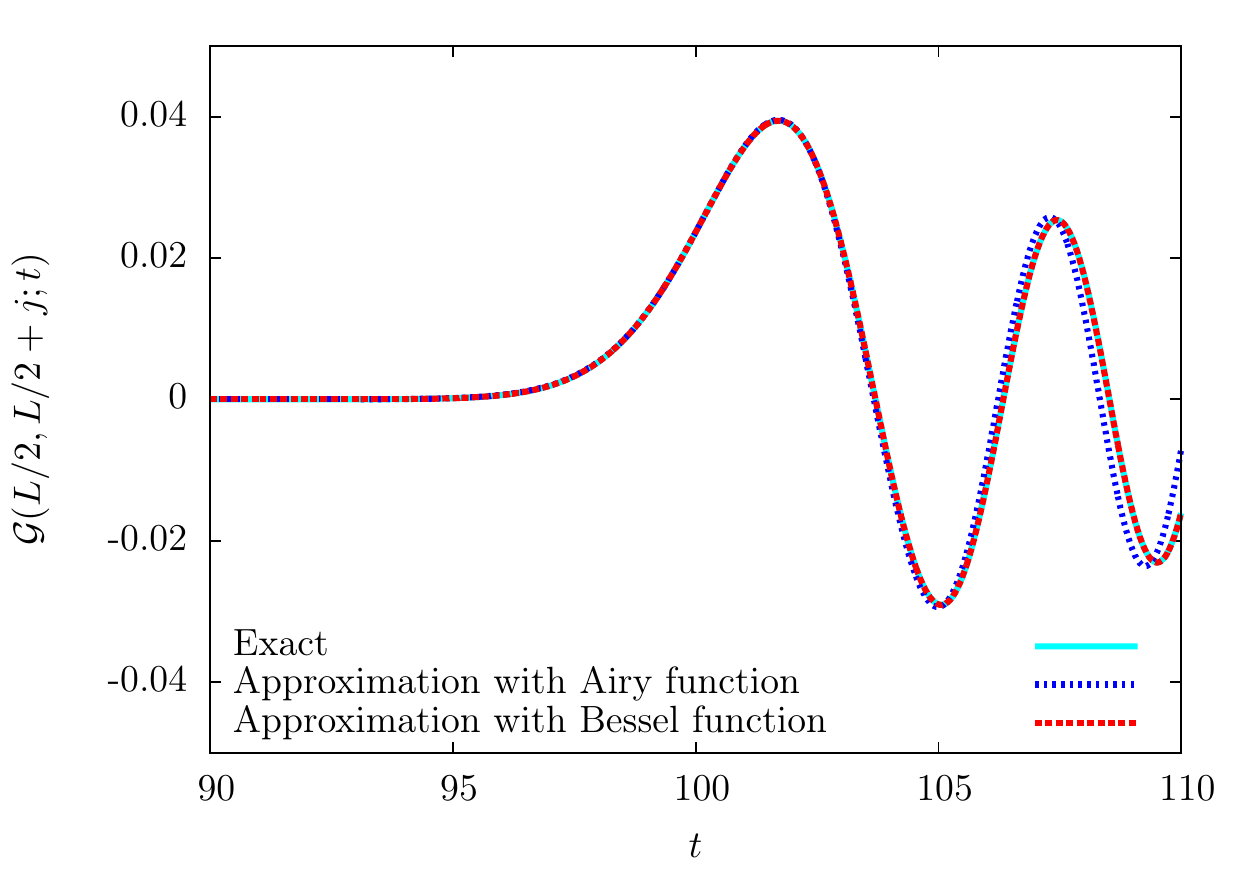}
\caption{Comparison of the Green's function $\mathcal{G}(0,j;t)$ for
an infinite system as a function of time at fixed distance $j=401$
with the approximations \eqref{eq:BesselSol} and
\eqref{eq:AirySol}. The system is initialized in the density matrix 
$\rho_0(2,0.5)$ and time evolved with the free Hamiltonian
$H(0.5,0,0)$. }
\label{fig:Bessel_Airy_approx}
\end{figure}

For $\delta_f=0$ we can furthermore describe the regime close to the
light cone $j=2v_{\text{max}}t$. Starting from equation
\eqref{integral}, we again deform the integration contour into the
complex plane so that it passes through the saddle point at $k_+$.
We then make the approximation that we can replace the factor
$\widetilde n_{+-}(k)$ in the integrand by its value at the saddle point
$\widetilde n_{+-}(k_+)$. Deforming the contour back to the real line
we obtain the following approximation to the integral
\be
\mathfrak{g}(j,t)\simeq \mathfrak{h}\left(\frac{j}{2v_\text{max}t}\right)\text{Im}\left[\int_0^\pi \frac{\text{d}k}{\pi}\, e^{ikj+4ti\cos{k}}\right],
\ee
where 
\be
\mathfrak{h}(z)=\frac{z \delta_i}{2} \frac{\tanh\left(\beta\sqrt{1-(1-\delta_i^2)z^2}\right)}{\sqrt{1-(1-\delta_i^2)z^2}}
\ee
Carrying out the integral then gives
\be
\mathfrak{g}(j,t)\simeq(-1)^{\frac{(j-1)}{2}}\, \mathfrak{h}\left(\frac{j}{2v_\text{max}t}\right) J_j(4t) \, . \label{eq:BesselSol}
\ee
As can be seen from Fig.~\ref{fig:Bessel_Airy_approx}, this is a very good
approximation. For large $j$ we can furthermore expand the Bessel
function for large orders.\cite{AbramowitzStegun} This
allows us to recover the results of the stationary phase/saddle point
approximations for the intermediate regions and the interior of the
light cone. In the vicinity of the light cone,
i.e. $2v_\text{max}\approx j$, we find 
\begin{align}
\mathfrak{g}(j,t)\simeq& \frac{(-1)^{\frac{(j-1)}{2}}}{{2}^{-{1}/{3}}j^{{1}/{3}}}\, \mathfrak{h}\left(\frac{j}{2v_\text{max}t}\right)  \text{Ai}\left(\frac{ j-2v_\text{max}t}{{2}^{-{1}/{3}}j^{{1}/{3}}}\right)\,. 
\label{eq:AirySol}
\end{align}
As shown in Fig.~\ref{fig:Bessel_Airy_approx} this provides a very
good approximation of $\frak{g}(j,t)$ close to the light cone.
Similar results involving the Airy function have been previously
obtained for the propagation of fronts in ``inhomogeneous quantum
quenches'' in tight-binding and XX models.\cite{{HRS},{ER},{VSDH}}
The setting in these works is different in that the initial state
features a step-like density profile, and the observable of interest
is the time evolution of the density. 

\subsubsection{``Width'' of the light cone}
We are now in a position to define the width of the light cone. 
This is most conveniently done by considering the approximation
\fr{eq:AirySol}. At large values of $j$ most of the variation of
${\frak g}(t,j)$ in the vicinity of the light cone is due to the Airy
Function. Irrespective of the details of how one defined the width
$\sigma(j)$, it then scales as 
\be
\sigma(j)\propto j^{1/3} \, .
\ee
This behaviour is the same as for the inhomogeneous quenches in
lattice models\cite{{HRS},{ER},{VSDH}}, and was argued to hold 
for inhomogeneous quenches in perturbed conformal field theories
as well \cite{BD:PertCFT}.

\subsubsection{$\delta_f\neq0$ case}
When the final dimerization is non-zero the analytic structure of the
integrand in \fr{eq:gf} becomes more complicated: $\epsilon_{+-}(k)$
has branch cuts in the complex $k$-plane. In addition, the number of
saddle points increases. This complicates the analysis of the
short-time regime.  Nonetheless, the behaviour of the Green's function
in the intermediate regime,  $C j <2 v_{\rm max} t < j$ ($C<1$), and
in the interior of the light cone, $2 v_{\rm max} t >j$, can be
determined as above. The structure remains very similar: for $2 v_{\rm
  max} t >j$  the Green's function decays as $t^{-\frac{3}{2}}$ and
for  $C j <2 v_{\rm max} t < j$ it decays exponentially, but  there
are oscillatory contributions multiplying the exponential
decay. Importantly, the ``width'' $\sigma(j)$ of the light cone scales
always as $j^{1/3}$.

\subsection{Interacting case}
\begin{figure}[ht]
\includegraphics[width=0.45\textwidth]{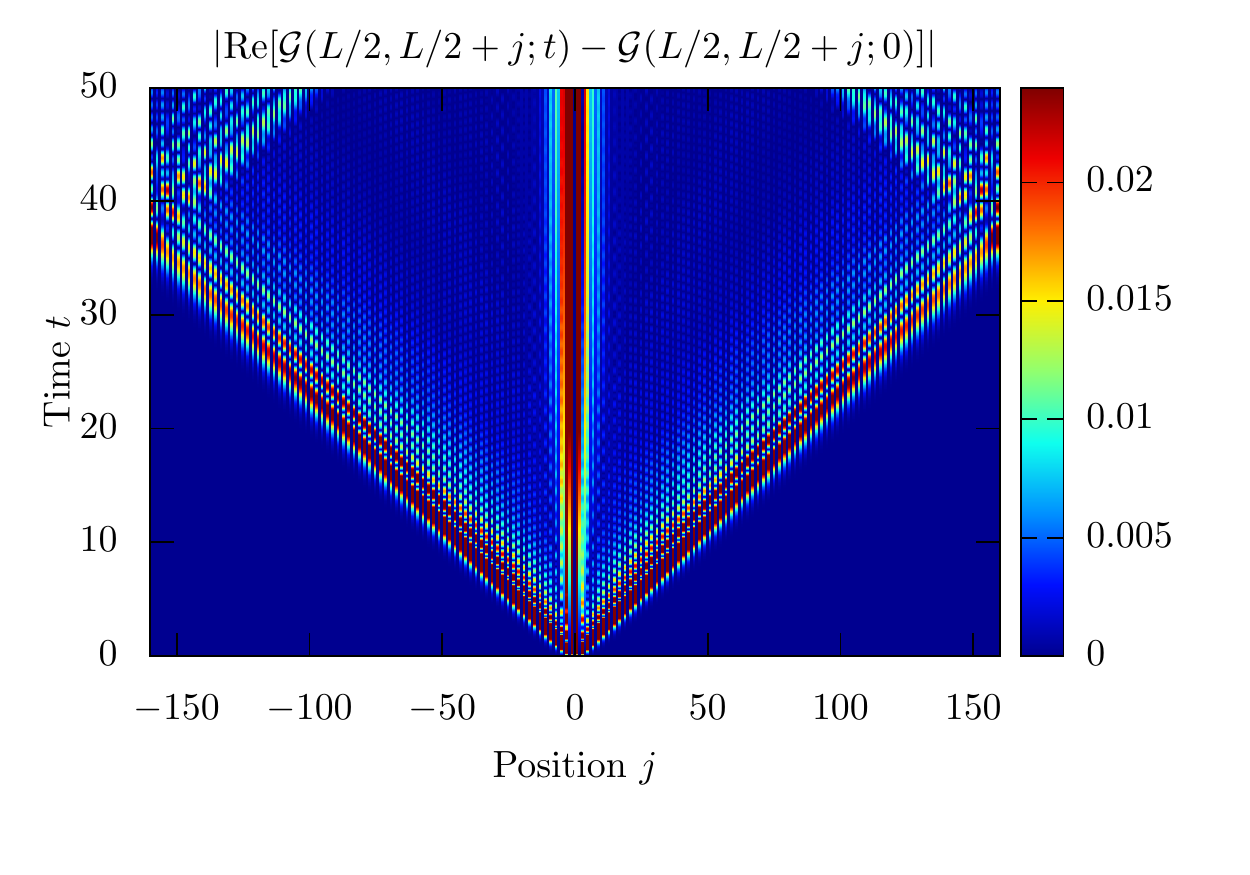} \quad
\includegraphics[width=0.45\textwidth]{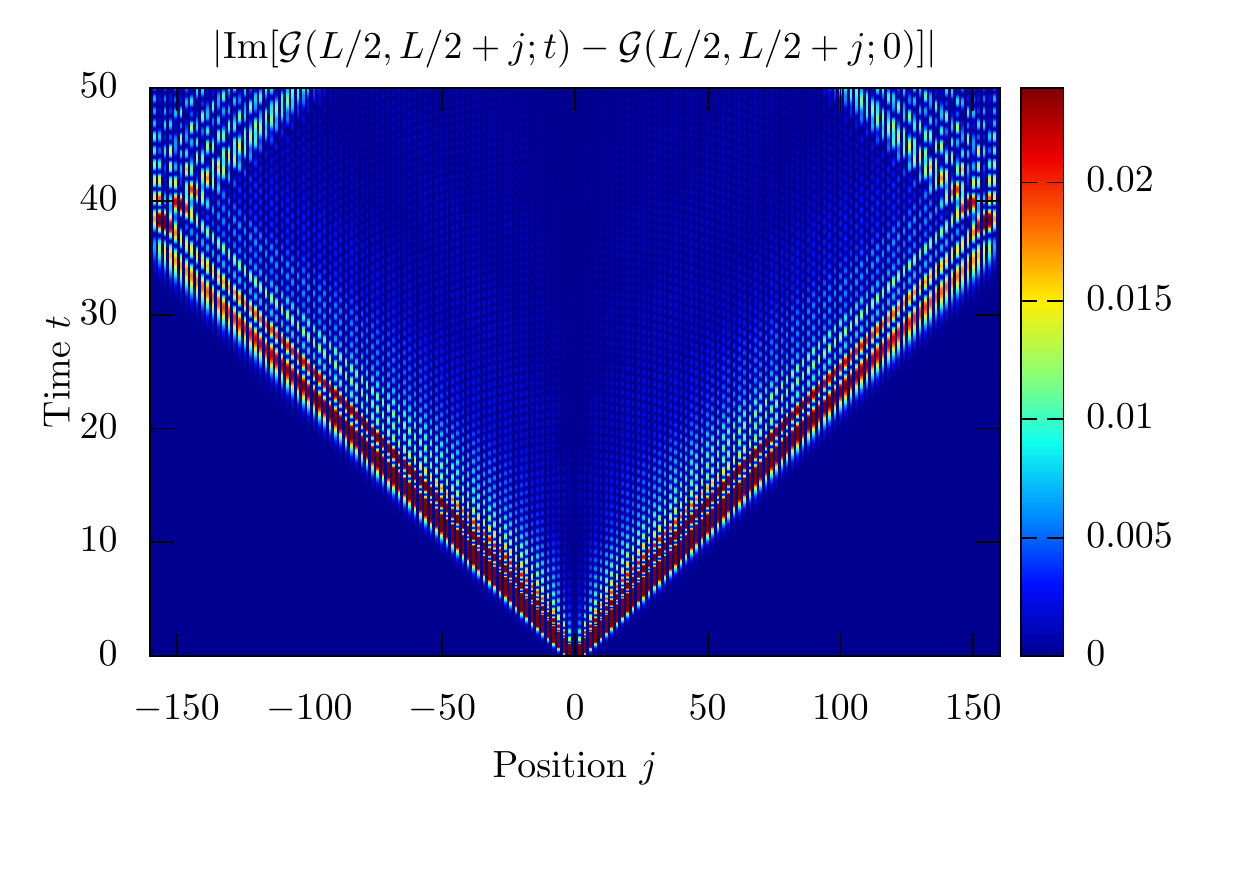}
\caption{Dynamics of the full Green's function $\mathcal{G}(L/2,L/2+j;t)$
as a function of time. The system is initialized in the density matrix
$\rho_0(2,0.5)$ and time evolved with the non-integrable Hamiltonian
$H(0.5,0,0.4)$. The signal outside of the light cone decays exponentially in time.} 
\label{fig:lightconeinteracting}
\end{figure}

\begin{figure}[ht]
\includegraphics[width=0.45\textwidth]{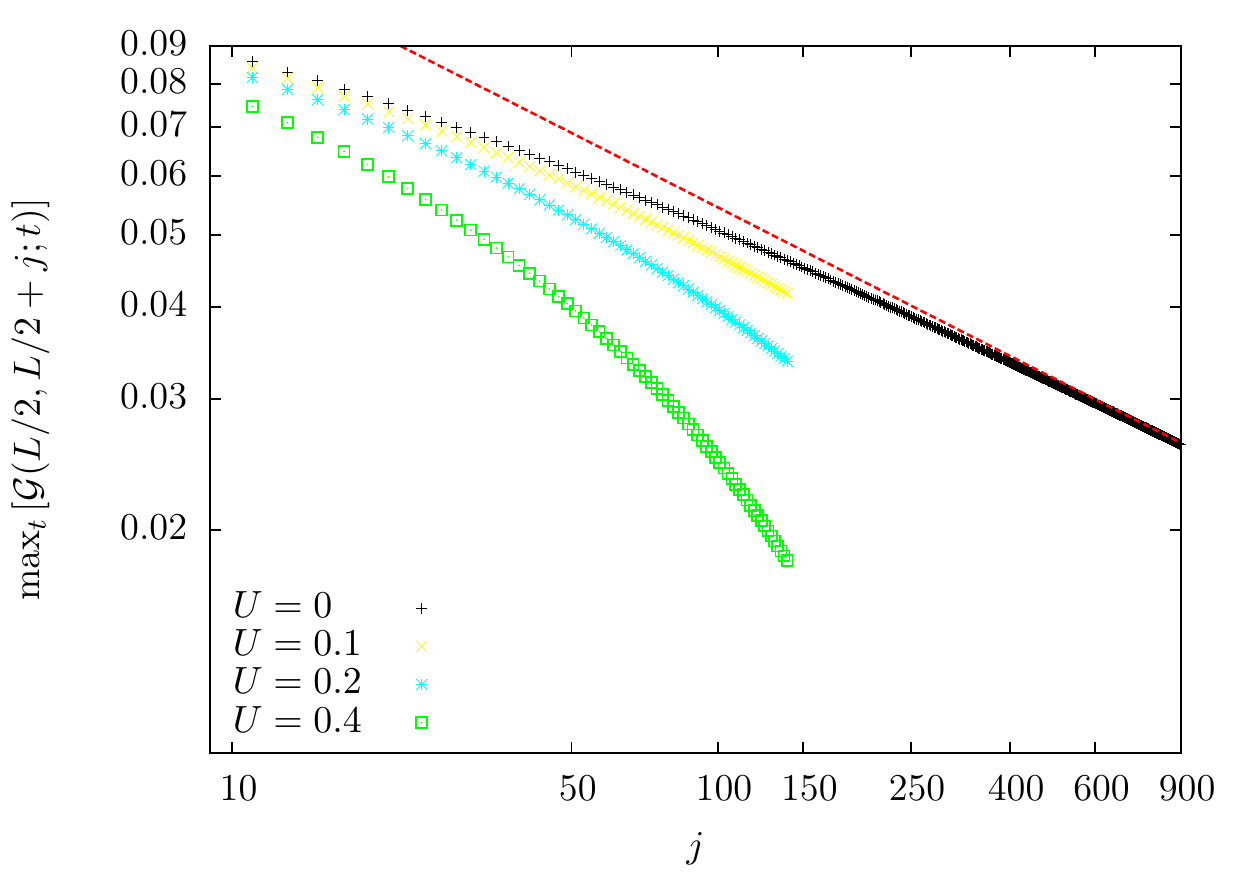}
\caption{Double logarithm plot of ${\rm max}_t{\cal G}(L/2,L/2+j;t)$ as a function of the odd separation $j$, for a system of length $L=1920$ ($U=0$) and $L=320$ ($U \neq 0$), initialized in the state $\rho_0(2,0.5)$ and evolved with $H(0.5,0,U)$ for $U=0, 0.1, 0.2, 0.4$. Different symbols correspond to different values of the interaction and the red dashed line is $\propto j^{-1/3}$.} 
\label{fig:peak}
\end{figure}

The structure of the light cone in the interacting case
remains qualitatively similar, see Fig.~\ref{fig:lightconeinteracting}. 
The most marked difference is that the height of the maximum at fixed separation now displays a much faster decay in $j$ as is shown in Fig.~\ref{fig:peak}. 

\begin{figure}[ht]
\includegraphics[width=0.45\textwidth]{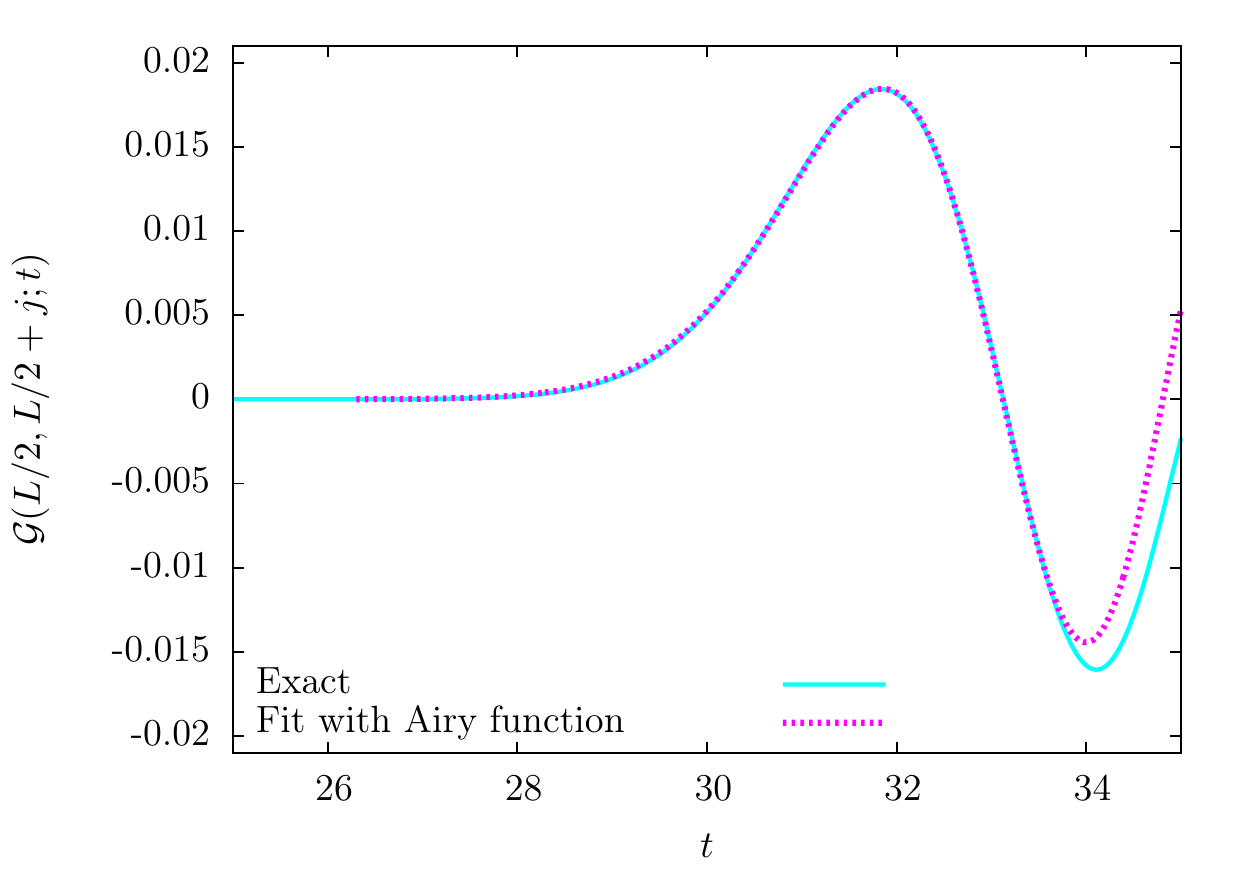}
\caption{Airy Function fit of the Green's function $\mathcal{G}(L/2,L/2+j;t)$
as a function of time at fixed distance $j=137$. The system is
initialized in the density matrix $\rho_0(2,0.5)$ and time evolved
with the non-integrable Hamiltonian $H(0.5,0,0.4)$.} 
\label{fig:airy_fit}
\end{figure}

\begin{figure}[ht]
\includegraphics[width=0.45\textwidth]{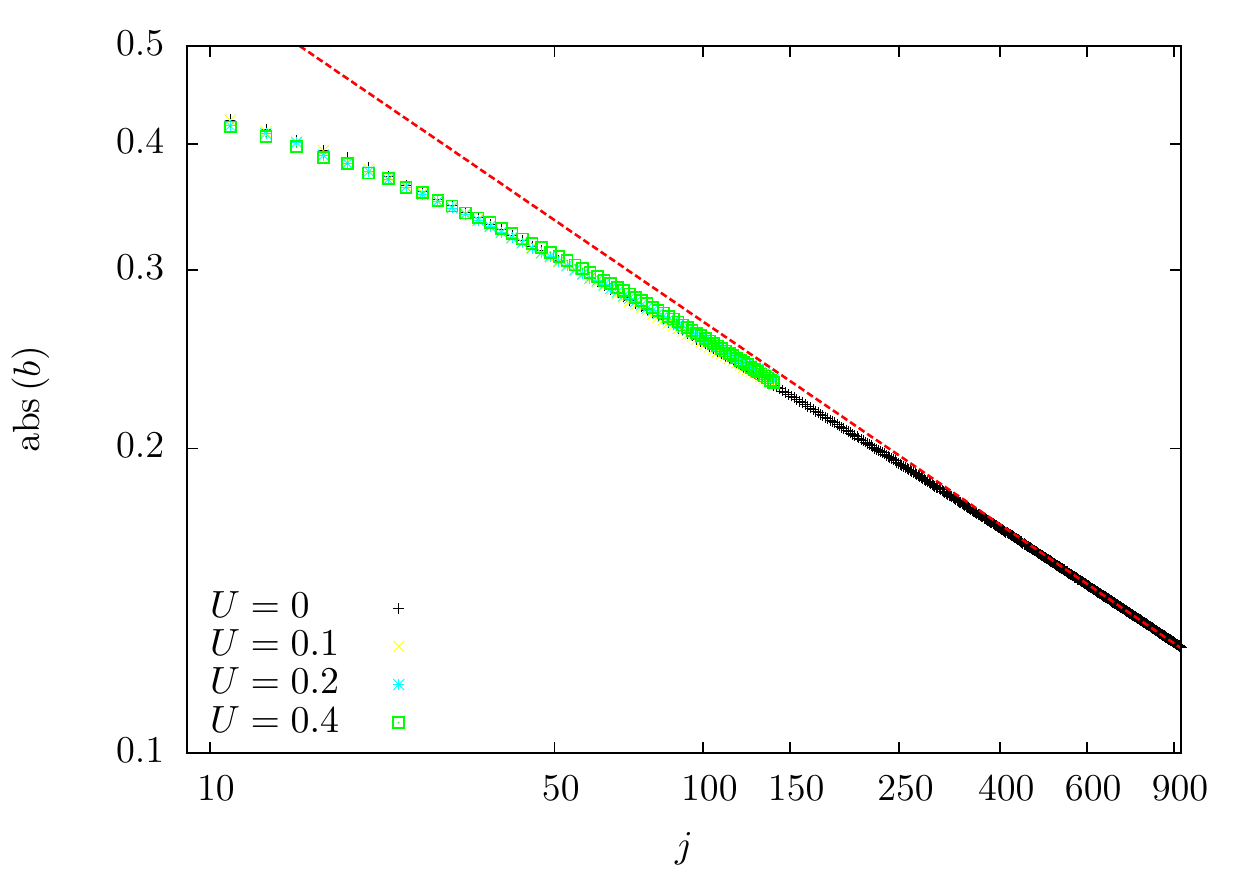}
\caption{Double logarithm plot of the fitting parameter $b$ (\emph{cf}. Eq.~\fr{Eq:fitlightcone}) as a function of $j$. The system, of length ${L=1920}$ for $U=0$ and ${L=320}$ for $U\neq0$, is
initialized in the density matrix $\rho_0(2,0.5)$ and time evolved
with the non-integrable Hamiltonian $H(0.5,0,U)$ with $U=0,0.1,0.2,0.4$. Different symbols correspond to different values of $U$ and the red dashed line is $\propto j^{-1/3}$.} 
\label{fig:scalingwidth}
\end{figure}

In order to determine the width of the light cone we fit the Green's
function in the vicinity of the light cone by an expression of the
form
\be
{\cal G}(L/2,L/2+j,t)= a\text{Ai}\left(b(2v_\text{max}t-j)\right)\,,
\label{Eq:fitlightcone}
\ee
where $a$, $b$ and $v_{\rm max}$ are fit parameters ($a$ and $b$ are
j-dependent). This provides an excellent description in the proximity
of the light cone for a wide range of $j$ as is shown for an example in
Fig.~\ref{fig:airy_fit}. 
The width $\sigma(j)$ of the light cone can be extracted from the
$j$-dependence of the parameter $b$, and we find that
\be
\sigma(j)\propto j^\alpha\qquad \alpha<1\,.
\ee 
The main difficulty we face is that we can only reach separations of
around $160$ sites, which imposes serious limitations to the precision
with which we can determine the scaling exponent $\alpha$. In the
non-interacting case we need to consider extremely large values of $j$
to observe the scaling $\sigma(j)\propto j^{1/3}$. Comparing the scaling
in the free and interacting case for different values of $U$ in the
regime accessible to us, we do not find a significant dependence of
the scaling exponent on the interaction strength $U$, and all of our
results are compatible with an exponent $\alpha=1/3$, see Fig.~\ref{fig:scalingwidth}. 
We have also
considered some quenches with $\delta_f>0$ and come to identical
conclusions. 

\section{Quantum Boltzmann equation}
\label{sec:qbe}

An important question is whether the set \fr{Eq:EOM} of coupled
integro-differential equations can be simplified for late times by
removing the time integration, in analogy with standard quantum
Boltzmann equations (QBE).\cite{ESY:QBE,LS:QBE} 
Given that the structure of \fr{Eq:EOM} is rather different to the
standard QBE case, \emph{cf.} Ref.~[\onlinecite{ESY:QBE}], it is
not \textit{a priori} clear that this is possible. More precisely, as 
\fr{Eq:EOM} includes an ${\cal O}(U)$ term intimately related to the
existence of a PT plateau, it is far from obvious that
the solutions of the EOM will only depend on $t$ through the rescaled
variables $\tau = U^2 t$ at late times, as is the case in the standard
QBE framework.\cite{ESY:QBE,LS:QBE}

\subsection{Simplifying the EOM}
\begin{figure}[ht]
\includegraphics[width=0.425\textwidth]{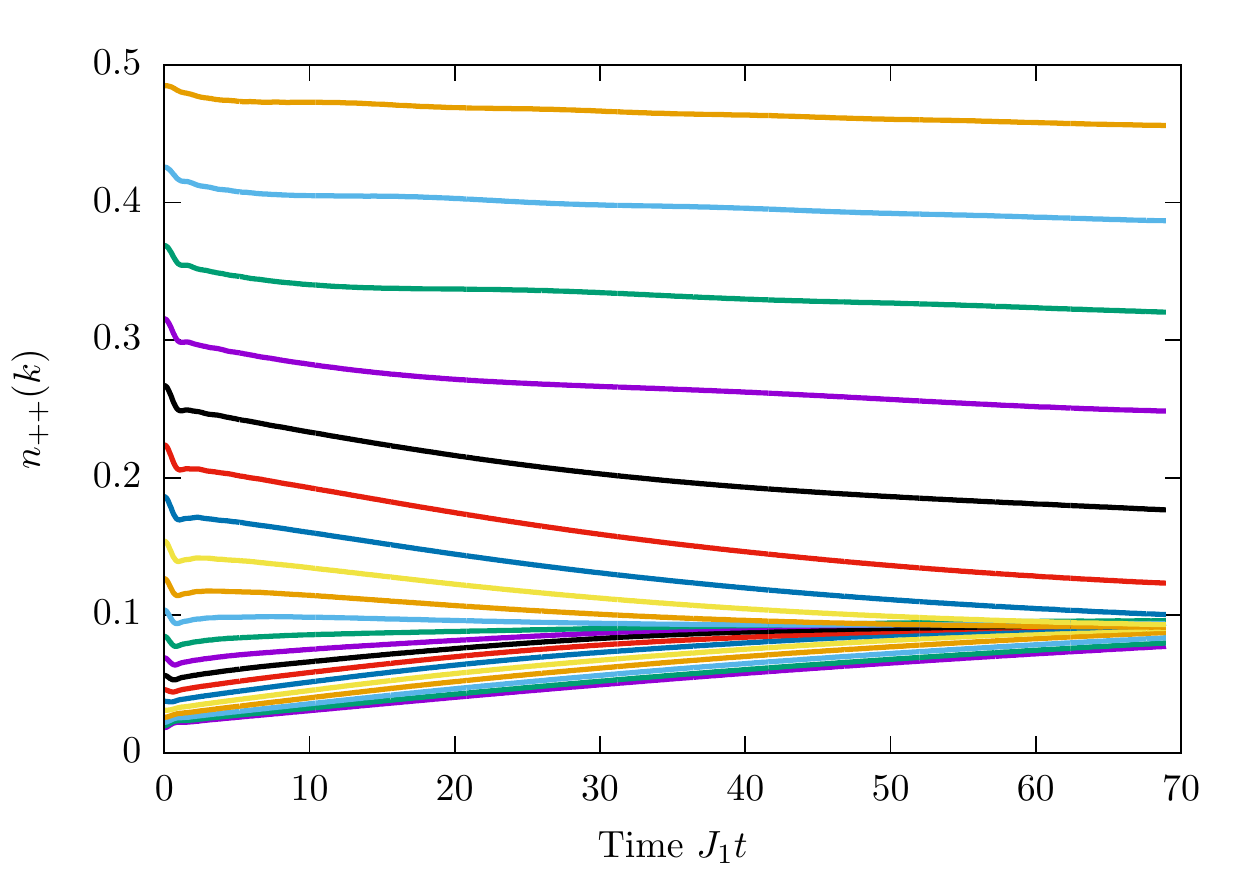}
\caption{The time dependence of the Bogoliubov mode occupation numbers
  $n_{++}(k)$ for a system of size $L=320$ initialized in the density matrix
$\rho(2,0.5)$ and time evolved with $H(0.375,0,0.4)$. The different
lines are different $k$-modes (we restrict our attention to $0\leq k \leq \pi/2$, as
$n_{\mu\nu}(k,t)=\mu\nu n_{\mu\nu}(\pi-k,t)$, and plot every fourth
$k$-mode). } 
\label{Fig:nktJ20375}
\end{figure}

\begin{figure}[ht]
\includegraphics[width=0.425\textwidth]{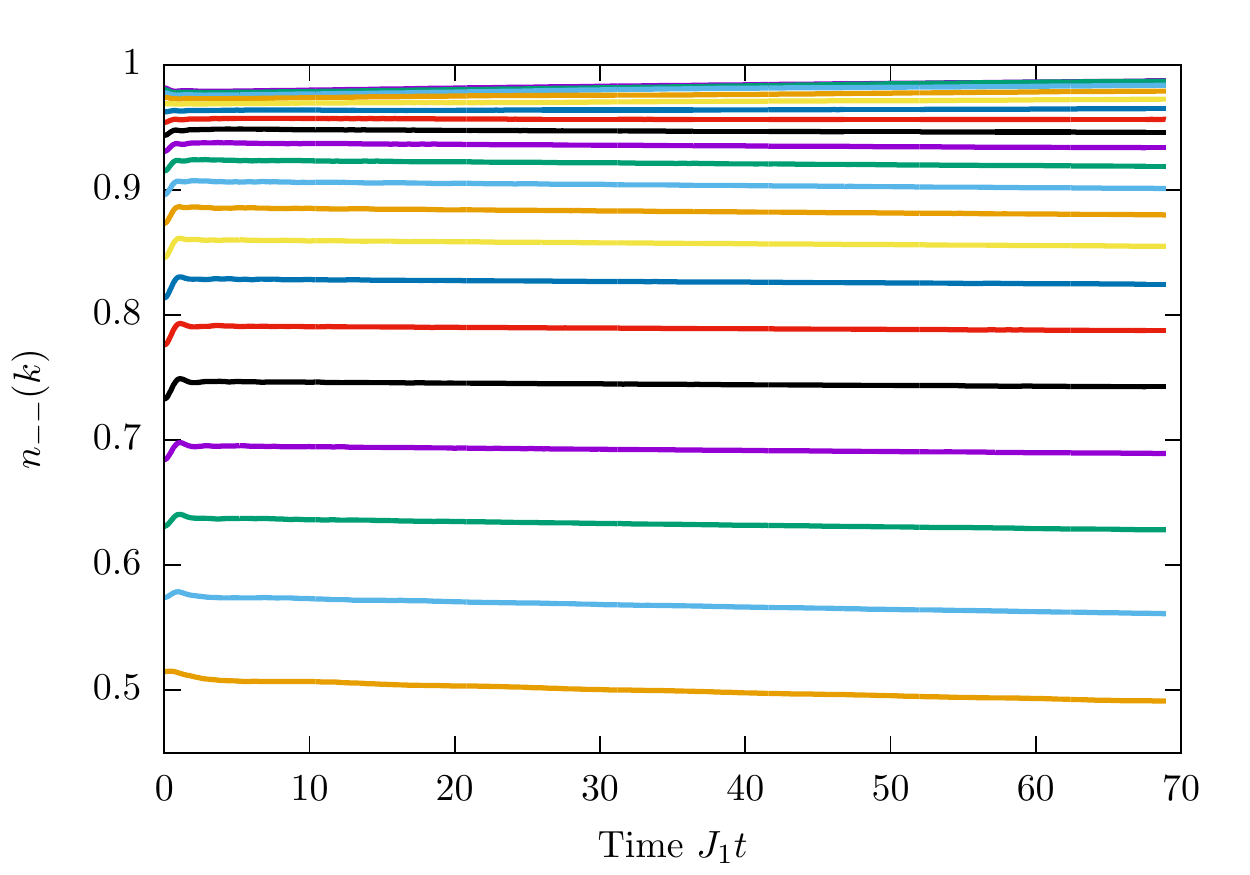} 
\caption{Same as Fig.~\ref{Fig:nktJ20375} for $n_{--}(k)$.}
\label{Fig:nktJ20375b}
\end{figure}

\begin{figure}[ht]
\includegraphics[width=0.425\textwidth]{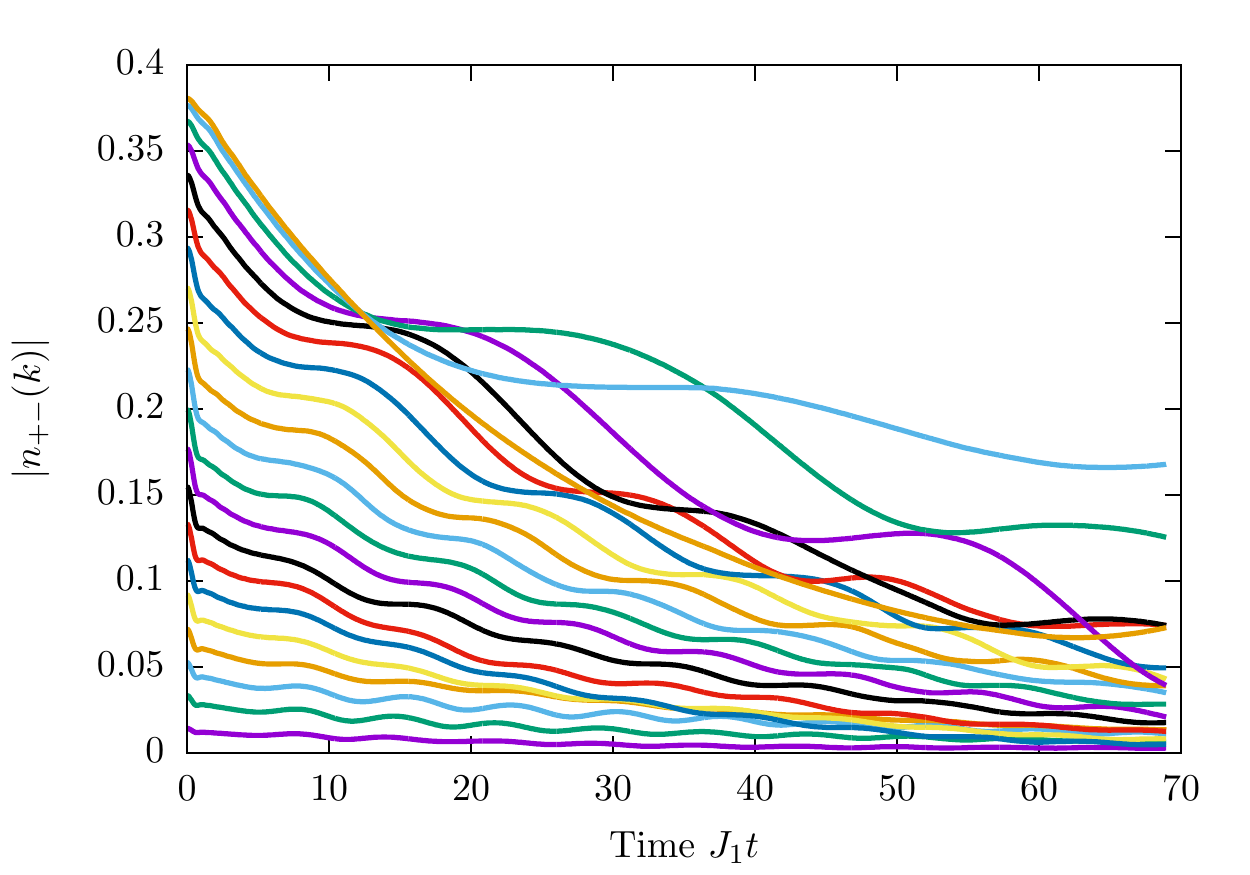} 
\caption{Same as Fig.~\ref{Fig:nktJ20375} for $|n_{+-}(k)|$.}
\label{Fig:nktJ20375c}
\end{figure}

In all the cases that we have analyzed, the ``off-diagonal'' two-point functions
$n_{\mu\bar\mu}(k;t)$ become negligible at sufficiently late times and
small values of $U$, see
Figs.~\ref{Fig:nktJ20375}--\ref{Fig:nktJ20375c} for representative
examples. This leads us to formulate the following approximation
\be
n_{\mu\bar{\mu}}(k,t)\approx 0 \qquad \text{for}\,\,t\gg {U}^{-1}\ ,
\label{Eq:Assumpionoffdiag}
\ee
where we have introduced the notations $\bar\mu=-\mu$. We note that
our approximation is consistent with relaxation of
$n_{\mu\nu}(k,t)$ towards their thermal values at late times as
this would suggest $n_{\mu\bar\mu}(k;\infty) \sim {\cal O}(U)$ and 
$n_{\mu\mu}(k;\infty) \sim {\cal O}(U^0)$.

We now use the approximation \fr{Eq:Assumpionoffdiag} to simplify the
EOM. We drop the equation for $n_{+-}(k,t)$ and retain only those for
$n_{++}(k,t)$ and $n_{--}(k,t)$. These do not contain ${\cal O}(U^0)$ 
contributions on the right-hand side of \fr{Eq:EOM}, but they do feature
${\cal O}(U^1)$ terms, which can be cast in the form
\be
8 U \textrm{Im}\left\{\left[A_{\bar\mu}(k)  +B_{\bar\mu}(k,t)\right]n_{\mu \bar\mu}(k)e^{i\epsilon_{\mu\bar\mu}(k)t}\right\}\,,
\label{Eq:drivingOU}
\ee
where we have introduced notations
\begin{align}
& A_{\mu}(k) \equiv  \sum_{\gamma}\sum_{q>0} V_{\bar\mu\gamma\gamma\mu}(k,q,q,k) n_{\gamma\gamma}(q)\,,\label{Eq:OUnotdec}\\
& B_{\mu}(k,t) \equiv \sum_{\gamma}\sum_{q>0} V_{\bar\mu\gamma\bar\gamma\mu}(k,q,q,k) n_{\gamma\bar\gamma}(q)e^{i\epsilon_{\gamma\bar\gamma}(q)t}\,.\label{Eq:OUdec}
\end{align}
The function $B_{\mu}(k,t)$ sums an oscillating phase multiplied by a
smooth function and consequently decays to zero at late times (see
below). $A_{\mu}(k)$, however, is independent of time and its presence
generally complicates our analysis of the long time limit. An
exception occurs for the special value $\delta_f = 0$, where
\be
V_{\eta\gamma\gamma\bar\eta}(k,q,q,k)|_{\delta_f=0}=0\quad\forall \eta,\gamma\,,\forall k,q\,,
\ee 
and concomitantly $A_{\mu}(k)$ vanishes. For the rest of the section
we will focus on this special case, and show that the remaining 
${\cal O}(U)$ terms do not contribute in the ``Boltzmann scaling limit'' 
\be
 U\to0\quad\text{and}\quad t\to\infty\quad \text{with}\quad \tau = t
 U^2 \quad\text{fixed}\, .
 \label{Eq:Boltzscallim}
\ee
This then implies that we may use a QBE description at late times. We return 
to the general case $\delta_f\neq 0$ in Sec.~\ref{sec:Boltzdeltafinite}.

\subsection{QBE for $\delta_f=0$}
\label{sec:Boltdelta=0}

In the scaling limit \fr{Eq:Boltzscallim}, the diagonal EOM~\fr{Eq:EOM} for 
$\delta_f=0$ become
\begin{align}
\partial_{\tau}{n}_{\mu\mu}(k,\tau) & = 
\limb\, 4iU^{-1} \left(B_{\mu}(k,t) n_{\bar \mu \mu}(k)e^{i\epsilon_{\bar\mu\mu}(k)t}+\textrm{c.c.}\right)\nn
& +\limb \sum_{\bm p{>0}, \bm \nu} \int_0^{t} \textrm{d} s~ e^{i E_{\bm \nu}(\bm p)(t-s)} F_{\bm \nu}^{\mu}(\bm p;k; s )\ .
\label{AC:EOM}
\end{align}
Here $\limb$ denotes the Boltzmann scaling limit \fr{Eq:Boltzscallim}
and we have collected the integrand of the $s$-integral into a single
function $F_{\bm \nu}^{\mu}(\bm p;k; s)$ to lighten notations. At late
times in the scaling limit \fr{Eq:Boltzscallim} the $B_\mu(k,t)$-term
can be evaluated by a stationary phase approximation
\begin{align}
&\limb \frac{4i}{U}B_{\mu}(k,t) n_{\bar \mu \mu}(k)e^{i\epsilon_{\bar\mu\mu}(k)t}\nn
&=\limb \frac{U}{\tau^{3/2}} \sin\left(\frac{\tau \epsilon_{+-}(0)}{U^2}-\frac{\pi}{4}\right) {\mathcal V}_{\mu}(k)e^{i\epsilon_{\bar\mu\mu}(k)\tau U^{-2}}\nn
&=0~.
\label{AC:decay}
\end{align}
Here ${\mathcal V}_{\mu}(k)$ is an amplitude depending on the initial
state and the vertex function
\be
{\mathcal V}_{\mu}(k)=\frac{2 \mu  \delta_i \sin(k) n_{\bar \mu\mu}(k)}{\sqrt{2 \pi} (\epsilon''_{+-}(0))^{3/2}}  \frac{|\cos(k)|}{\cos(k)}\tanh(\beta_i) \,.
\ee
The exponent $3/2$ in \fr{AC:decay} is a consequence of $n_{+-}(k,0)$
being zero at the saddle point $k=0$, \emph{cf.} Eq.~\fr{AB:Occval}. 
From Eq.~\fr{AC:decay}, we conclude that the ${\cal O}(U)$ terms do not
contribute in the scaling limit when $\delta_f = 0$.

This leaves us with the ${\cal O}(U^2)$ contribution on the right-hand
side of Eq.~\fr{AC:EOM}. According to our basic approximation
\fr{Eq:Assumpionoffdiag}, this can be simplified in the long time
limit because the off-diagonal two point functions $n_{+-}(k;s)$ can be
neglected for $s\gg U^{-1}$. We now make the further assumption that
the diagonal mode occupation numbers $n_{\mu\mu}(k;s)$ depend on $s$
only through $sU^2$, \emph{i.e.} they are very slowly varying. This
assumption is again motivated by numerical results obtained by
integrating the full EOM, see Figs.~\ref{Fig:nktJ20375} and
\ref{Fig:nktJ20375b} for a representative example. The ${\cal O}(U^2)$
terms can then be treated as follows. We introduce an intermediate time-scale  $U^{-1}\ll \bar t\ll t$ and split the time integral into two parts
\bea
&&\int_0^{t} \textrm{d}s~  e^{i E_{\boldsymbol{\nu}}(\bm p)(t-s)} F^\mu_{\boldsymbol{\nu}}(\bm p;k; s )\nn
&& \hspace{1cm} = \int_0^{\bar t} {\textrm{d} s}~ e^{i E_{\boldsymbol{\nu}}(\bm p)
  (t-s)} F^\mu_{\boldsymbol{\nu}}(\bm p;k; s)\nn
&& \hspace{1.5cm} + \int_{0}^{t-\bar t}  \textrm{d} s~  e^{i E_{\boldsymbol{\nu}}(\bm p) s}F^{\mu}_{\boldsymbol{\nu}}(\bm p;k; t-s) ~.
\label{AD:simpint}
\eea
We then make the assumption that the first term
on the right hand side of \fr{AD:simpint} does not contribute in our scaling
limit, while the remaining integral can be treated as follows. We
first replace $F^{\mu}_{\boldsymbol{\nu}}(\bm p;k; s)$ by its smooth
part $F^{\mu}_{\boldsymbol{\nu}}(\bm p;k; s)_{{\rm slow}}$  depending
on $s$ only through $U^2s$, this is motivated by numerical analysis as
discussed above. We then add
an infinitesimal convergence factor $i \eta$, to
$E_{\boldsymbol{\nu}}(\bm p)$ in the exponential factor of the
integrand in order to ensure the convergence of the integral. In
principle the parameter $\eta$ should be taken to zero after the
Boltzmann scaling limit has been performed. In practice we keep $\eta$
small but finite and ensure that our results for the Green's function
only depend very weakly on it in the time interval
considered. This procedure is equivalent to the regularization adopted
in Refs.~[\onlinecite{FMS:NIQBE}, \onlinecite{FMS:QBE}]. 
Next we expand the function $F^{\mu}_{\boldsymbol{\nu}}(\bm p;k;
t-s)_{{\rm slow}}$ around $t$, which gives
\begin{align}
&\int_{0}^{t-\bar t}  \textrm{d} s~  e^{i (E_{\boldsymbol{\nu}}(\bm p)+i\eta) s}
F^{\mu}_{\boldsymbol{\nu}}(\bm p;k; t-s)_{{\rm slow}} =\notag\\
&=\sum_{n=0}^{\infty}\! \frac{(-1)^{n}}{n!} \frac{{\rm d}^n}{{{\rm d}s}^n } 
F^{\mu}_{\boldsymbol{\nu}}(\bm p;k; t)_{{\rm slow}} \int_{0}^{t-\bar t} 
\!\!\! \!\!\!\textrm{d} s~  e^{i (E_{\boldsymbol{\nu}}(\bm p)+i\eta) s} s^n\,.
\end{align} 
As $F^{\mu}_{\boldsymbol{\nu}}(\bm p;k; s)_{{\rm slow}}$ is a function of $U^2 s$ 
its derivatives are suppressed by factors of $U^{2}$ in the Boltzmann scaling limit, which implies that we only need to retain the first contribution. Evaluating the first integral we have
\begin{align}
\!\!\!\limb \int_{0}^{t-\bar t}\!\!\!\!\!\!\!\! \textrm{d} s~ e^{i [E_{\boldsymbol{\nu}}(\bm p)+i\eta]
  (t-s)}=
\frac{i}{E_{\boldsymbol{\nu}}(\bm p)+i\eta} \equiv D( E_{\boldsymbol{\nu}}({\bm p}))~.
\label{AD:Reg}
\end{align}
Putting everything together
we arrive at a QBE for the mode occupation numbers
\bea
\partial_{\tau}{n}_{\mu\mu}(k,\tau) &=&  - \sum_{\gamma,\eta}\sum_{p,q{>0}} 
\widetilde K^{\gamma\eta}_{\mu}(p,q|k) n_{\gamma\gamma}(p,\tau) n_{\eta\eta}(q,\tau)\nn
&-&\sum_{\gamma,\eta,\epsilon}\sum_{p,q,r{>0}} \
\widetilde L^{\gamma\eta\epsilon}_{\mu}(p,q,r|k)n_{\gamma\gamma}(p,\tau)  \nn
&&\hspace{2cm} \times
n_{\eta\eta}(q,\tau)n_{\epsilon\epsilon}(r,\tau)\ .
\label{Eq:BB}
\eea
Here the kernels are given by 
\bea
&&\widetilde K^{\gamma_1\gamma_2}_{\alpha}(k_1,k_2|q) = 4 \sum_{k_3,k_4{>0}} \sum_{\nu,\nu'}  
\widetilde X^{\gamma_1\gamma_2\nu\nu'|\nu\nu'\gamma_2\gamma_1}_{{\bm k}|{\bm k}'} (\alpha|q),\nn
&&\widetilde L^{\gamma_1\gamma_2\gamma_3}_{\alpha}(k_1,k_2,k_3|q) = 8
\sum_{\nu}\sum_{k_4{>0}} \widetilde X^{\gamma_1\gamma_2\gamma_3\nu|\nu\gamma_3\gamma_2\gamma_1}_{{\bm k}|{\bm k}'}(\alpha|q)\nn
&&\hspace{3.2cm} - 16\sum_{\nu} \widetilde X^{\gamma_1\gamma_2\nu\gamma_2|\gamma_3\nu\gamma_3\gamma_1}_{k_1k_2k_1k_2|k_3k_1k_3k_1}(\alpha|q),\nn
&&\widetilde X^{{\bm\gamma}|{\bm\alpha}}_{{\bm k}|{\bm q}}(\alpha|q) =
Y^{\bm\gamma}_{\alpha\alpha}({\bm k},q)V_{\bm\alpha}({\bm q}) D(
E_{\bm\gamma}({\bm k}))\nn
&&\hskip 2cm
 - ({\bm \gamma},{\bm k})\leftrightarrow({\bm \alpha},{\bm q})~.\label{Eq:Boltkernels}
\eea

The Boltzmann equation has been derived in the scaling limit
\fr{Eq:Boltzscallim}. In practice, this limit cannot be accessed in
numerical computations starting from $t=0$. Instead, we keep $U$ small
but \emph{finite} and initialize the QBE at a \emph{finite} time $t_0 \gg U^{-1}$
using the occupation numbers computed up to $t=t_0$ with the 
full EOM~\fr{Eq:EOM}. 

\begin{figure}[ht]
\begin{tabular}{l}
\includegraphics[width=0.425\textwidth]{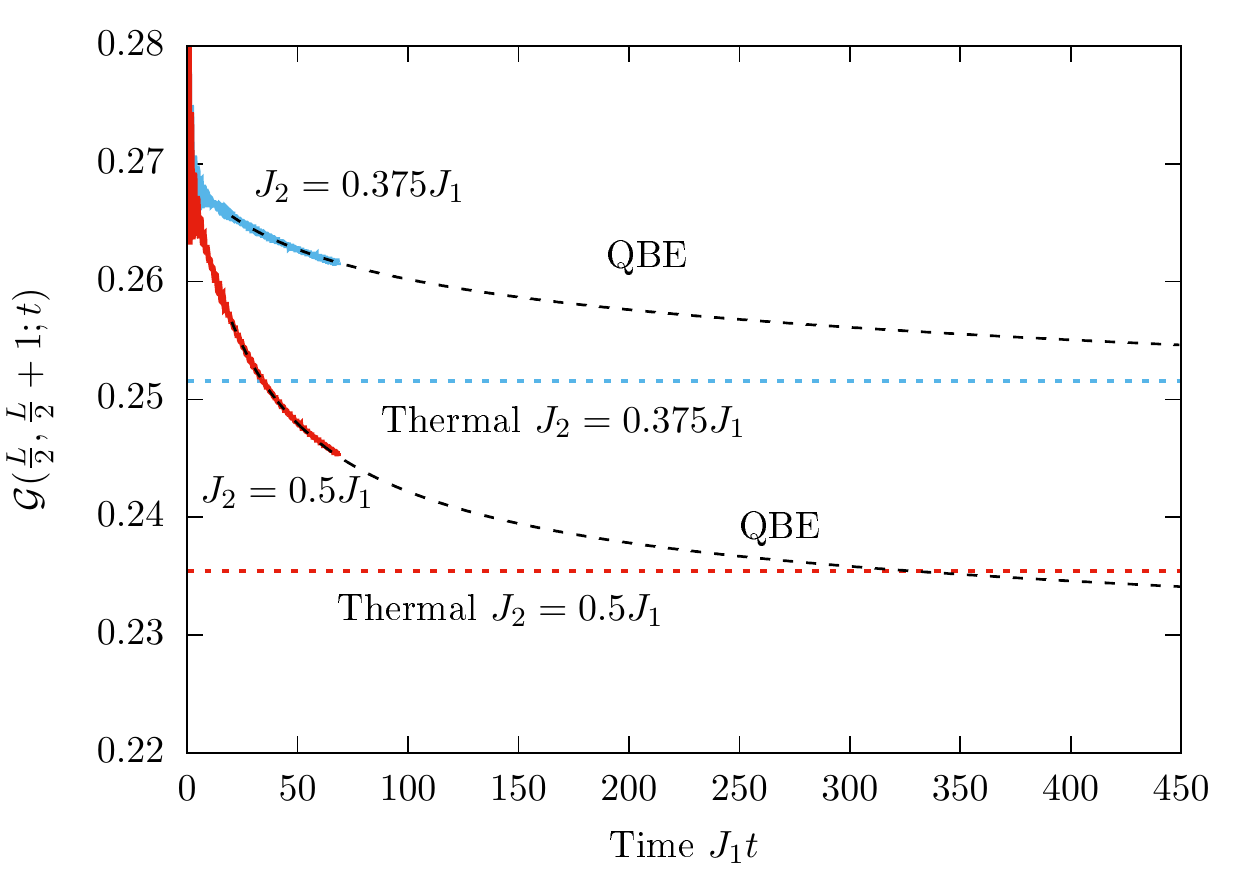}\\
\includegraphics[width=0.425\textwidth]{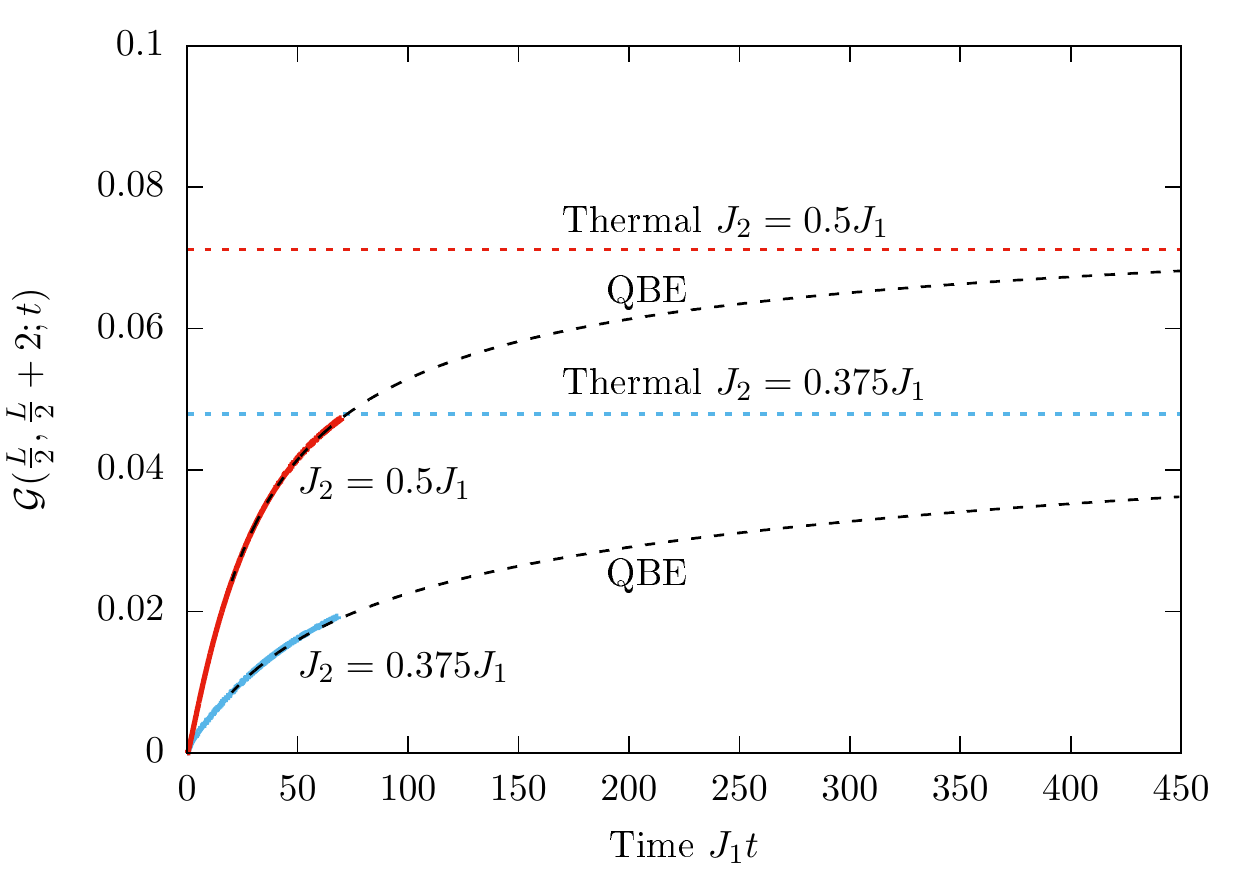}
\end{tabular}
\caption{
${\cal G}({L}/{2},{L}/{2}+j;t)$ with (upper) $j=1$ and (lower) $j=2$
for a system of size $L=320$ initially prepared in a state with
density matrix $\rho(2,0.5)$, and time evolved with $H(J_2,0,0.4)$.
Full lines show results obtained by integrating the EOM \fr{Eq:EOM},
and black dashed lines indicate the QBE results.}
\label{fig:QBEG12}
\end{figure}

In Figs.~\ref{fig:QBEG12} we present results obtained using the QBE
for cases where the system is initialized in the density matrix
$\rho_0(2,0.5)$ and time evolved with $H(J_2,0,0.4)$, where
$J_2 = 0.375,0.5$. The QBE is initialized at time $t_0 = 20$. Even for 
the relatively large value of $U=0.4$, we see that the results of the QBE 
are in good quantitative agreement with the full EOM. The agreement 
worsens for larger separations, which is not surprising as the Green's 
function itself becomes smaller.  At late times the mode occupation 
numbers also approach their thermal values.

This approach should not be taken too literally as we now explain.
The non-interacting Fermi-Dirac distribution (for arbitrary $\beta$
and $\mu$) is always a stationary solution of the QBEs \fr{Eq:BB}.
It is believed that for non-integrable models it is the only stationary 
solution.\cite{FMS:NIQBE} If this holds true, the value to which a 
mode occupation number relaxes is determined by the number 
density and the kinetic energy at the time the QBE is initialized.~\cite{LS:QBE,FMS:QBE} 
This means that in the late time limit we expect the QBE to converge to values that agree 
with the ``correct'' thermal values only up to corrections of order ${\cal O}(U)$.

\subsection{Scaling form of the Green's function}
\label{subsec:exponent}
\begin{figure}[t!]
\begin{tabular}{l}
\includegraphics[width=0.425\textwidth]{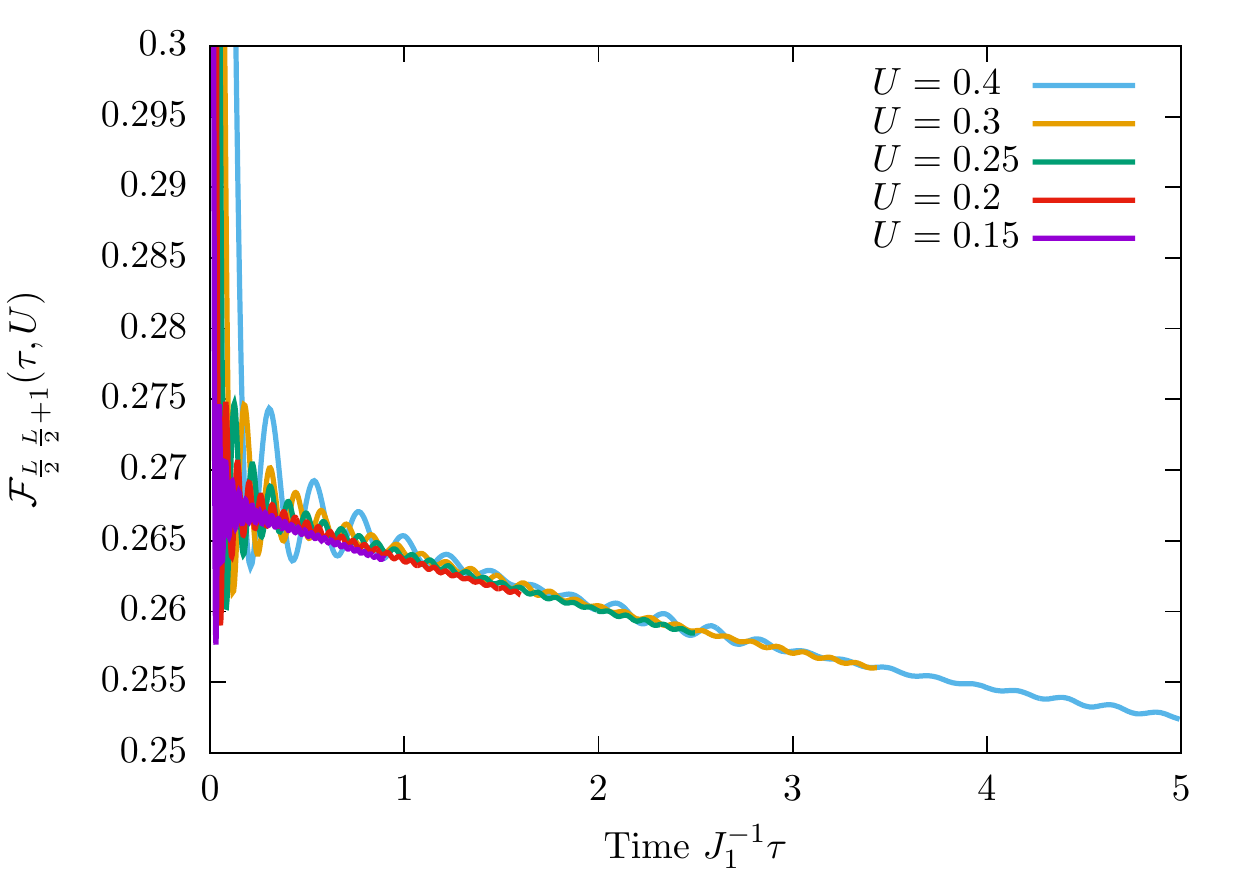} \\
\includegraphics[width=0.425\textwidth]{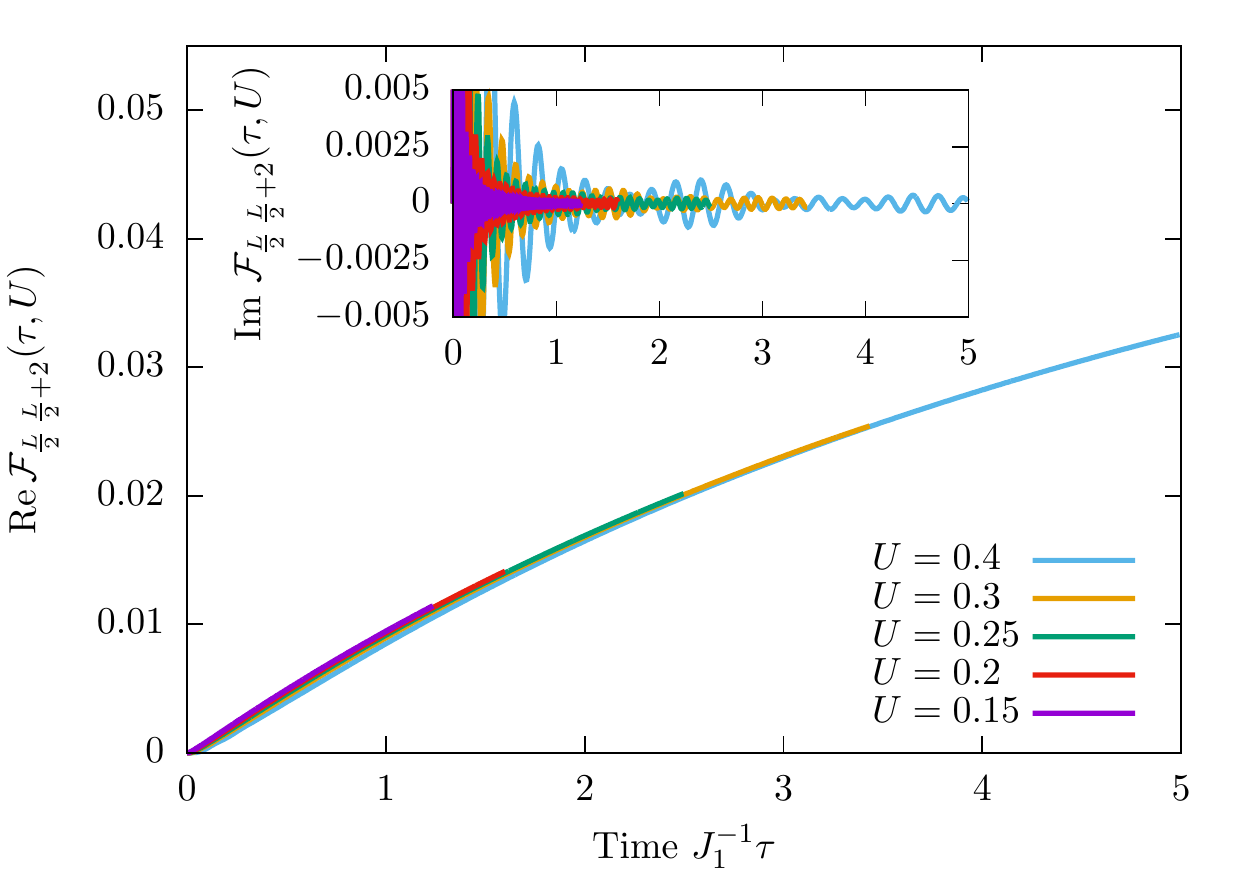}
\end{tabular}
\caption{Green's functions ${{\cal F}_{L/2\, L/2+1}(\tau,U)}$ (top) and ${{\cal F}_{L/2\, L/2+2}(\tau,U)}$ (bottom) 
obtained by numerical solution of the full EOM \fr{Eq:EOM} for a system
  prepared in the state with density matrix $\rho_0(2,0.5)$ and time
evolved with $H(0.5,0,U)$. Results for several values of $U$
are plotted against the rescaled variable $\tau=U^2 t$.
} 
\label{Fig:SF12}
\end{figure}

In the QBE framework the Green's function depends on $U$ both via
the rescaled time $\tau$, and through the initial conditions imposed
at time $t_0$. We express this as
\be
{\cal G}(i,j;t,U)= {\cal F}_{ij}(\tau,U)~,\qquad t > t_0\gg {U}^{-1}~.
\label{scalingform}
\ee 
Expanding ${\cal F}_{ij}(\tau,U)$ to leading order in $U$ (at fixed $\tau$) then gives 
\be
{\cal F}_{ij}(\tau,U)\sim{\cal F}_{ij}(\tau,0)+{\cal O}(U)~.
\label{Eq:scalingform}
\ee
We expect that the full EOM will give rise to the scaling form
\fr{scalingform}, \fr{Eq:scalingform} at sufficiently late times. 
This is indeed the case as shown in Fig.~\ref{Fig:SF12} for a
representative example.

\begin{figure}[t!]
\begin{tabular}{l}
\includegraphics[width=0.425\textwidth]{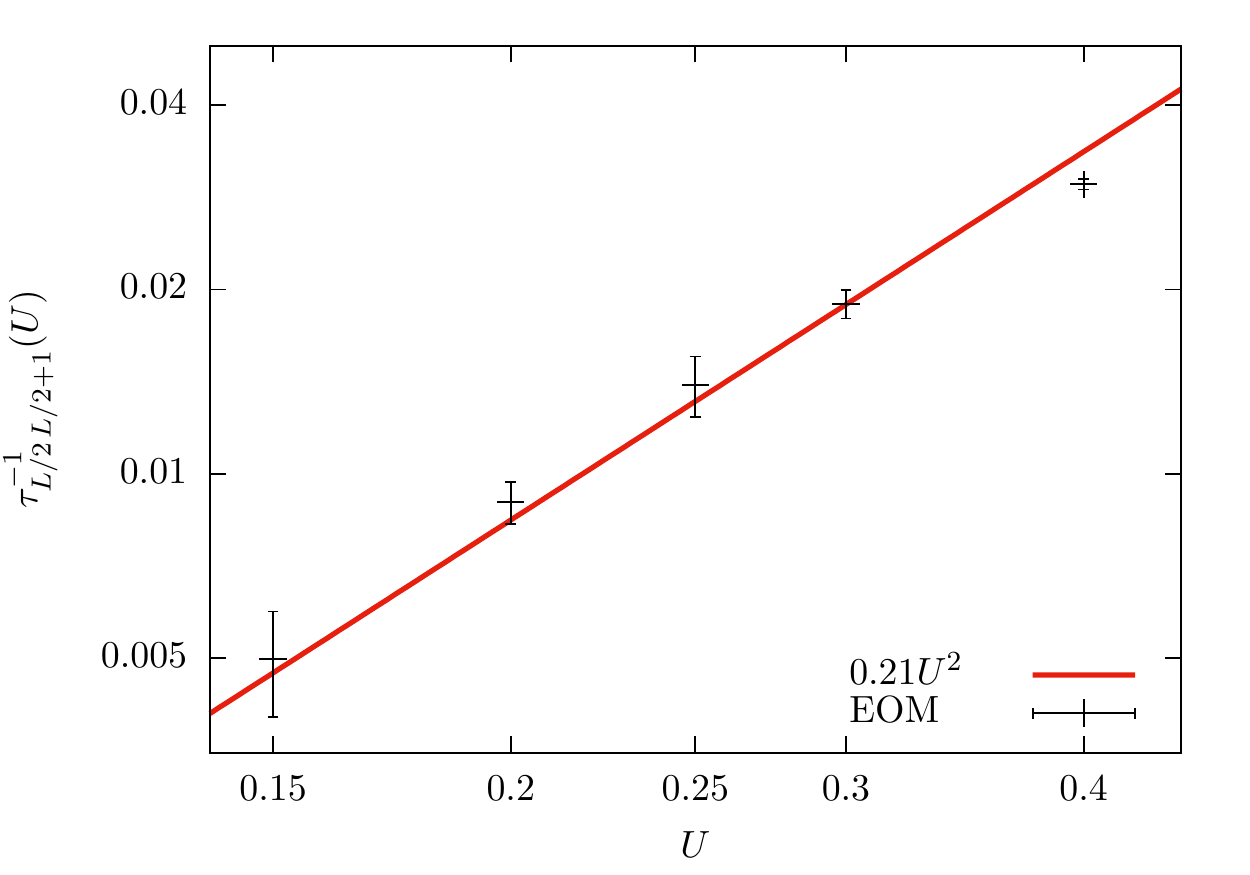} 
\end{tabular}
\caption{Double logarithmic plot of the $U$-dependence of the inverse
relaxation times $\tau^{-1}_{L/2\,L/2+1}(U)$ for a system initialized
in the density matrix $\rho_0(2,0.5)$ and time evolved with
$H(0.5,0,U)$. Relaxation times
are obtained by fitting the EOM results to the form \fr{Eq:ExpFit}.
Errors are estimated by varying the initial time at which the
exponential fit is applied.}  
\label{Fig:exponent}
\end{figure}

By virtue of its simpler structure, the QBE allows us to determine how
the exponent~\fr{Eq:ExpFit} scales with the interaction strength
$U$. To that end we consider the exponential fit \fr{Eq:ExpFit} that
we have found to give a good account of the intermediate time behaviour of the
Green's function. Expanding the inverse relaxation time in powers of $U$ 
\be
\tau^{-1}_{ij}(U)=\sum_{i=0}^{\infty} a_{i} U^{i}\, ,
\ee
we have
\bea
{\cal G}(i,j;t;U) &\approx& {\cal G}(i,j)_{\text{th}}\notag\\
&+& A_{ij}(J_2,0,U) e^{-t a_0-t U a_1-\tau  a_2}+\ldots
\label{Eq:ExpFitexpanded}
\eea
In order for this to be compatible with \fr{scalingform} we must have
$a_0=a_1=0$, $a_2\neq 0$, which gives
\be
\tau^{-1}_{ij}(J_2,\delta_f=0,U)= U^2 a_2 + O(U^3)\,.
\label{Eq:scalingexponent}
\ee  
As shown in Fig.~\ref{Fig:exponent} for a particular example, the
$U^2$-scaling in \fr{Eq:scalingexponent} is in good agreement with
inverse relaxation times extracted from the numerical solution of the EOM. 

We note that the scaling of inverse relaxation times with $U$ found
here differs from that obtained in Ref.~[\onlinecite{SK:EOM}]. In
contrast to our quench protocol, Ref.~[\onlinecite{SK:EOM}] considers
situations where the energy density in the initial state is ${\cal O}(U)$, 
which results in a $U^{4}$ scaling. 

\subsection{Mode occupation numbers}
\label{subsec:numbop}
In order to obtain further indicators that integrability breaking
perturbations lead to thermalization, we turn our attention to the
(Bogoliubov) mode occupation numbers $n_{\mu\mu}(q,t)$ themselves. The first
question to consider is whether we expect these quantities to relax at
all? The number operators are local in momentum space, and hence are
non-local in real space. It is then \textit{a priori} unclear whether they 
will relax at late times (see, however, Ref.~[\onlinecite{wright14}]).
Here we take a practical point of view: we simply follow the evolution
of $n_{\mu\mu}(k,t)$ on the time scales accessible to us, and compare
them to the appropriate thermal values $n_{\mu\mu}(k,\beta_{\rm
  eff},\mu_{\rm eff})$ of the putative stationary behaviour. The latter
are calculated by standard second order perturbation theory in
$U$; details are given in Appendix~\ref{app:pert}. 

In Fig.~\ref{Fig:nks0375} we show results for the mode occupation
numbers $n_{\mu\mu}(k,t)$ at several different times for a system of
size $L=320$ that has been prepared in the density matrix
$\rho(2,0,0.5,0)$ and evolved with $H(0.375,0,0.4)$ (see also the
example reported in Ref.~{[\onlinecite{BEGR:PRL}]}). For short and 
intermediate times $J_1t\lesssim70$ we use the full EOM, while 
later times are analyzed in the framework of the QBE. The QBE is
initialized at time $t_0 = 20$, and is in good agreement with the full
EOM until the latest times accessible by the latter. 
\begin{figure}[ht]
\begin{tabular}{l}
\includegraphics[width=0.425\textwidth]{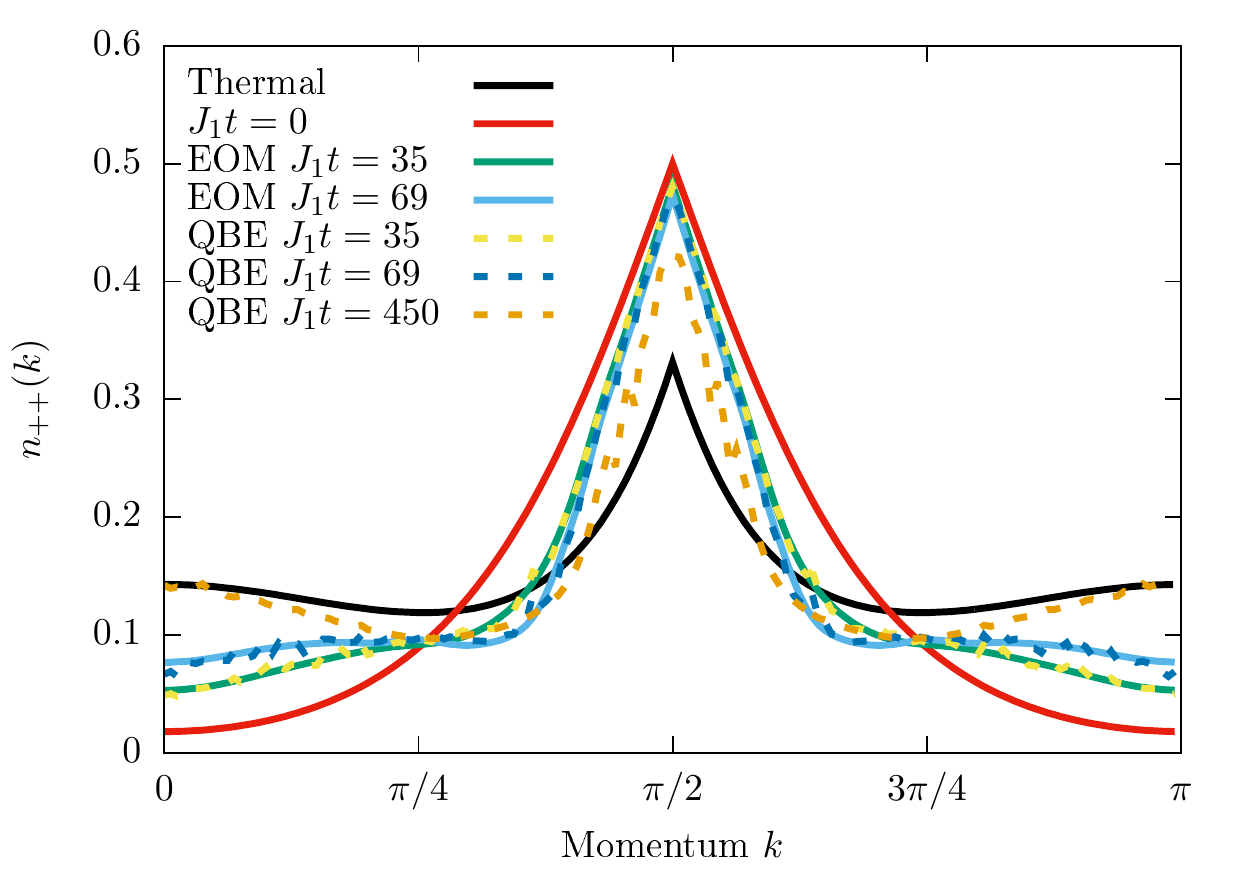} \\
\includegraphics[width=0.425\textwidth]{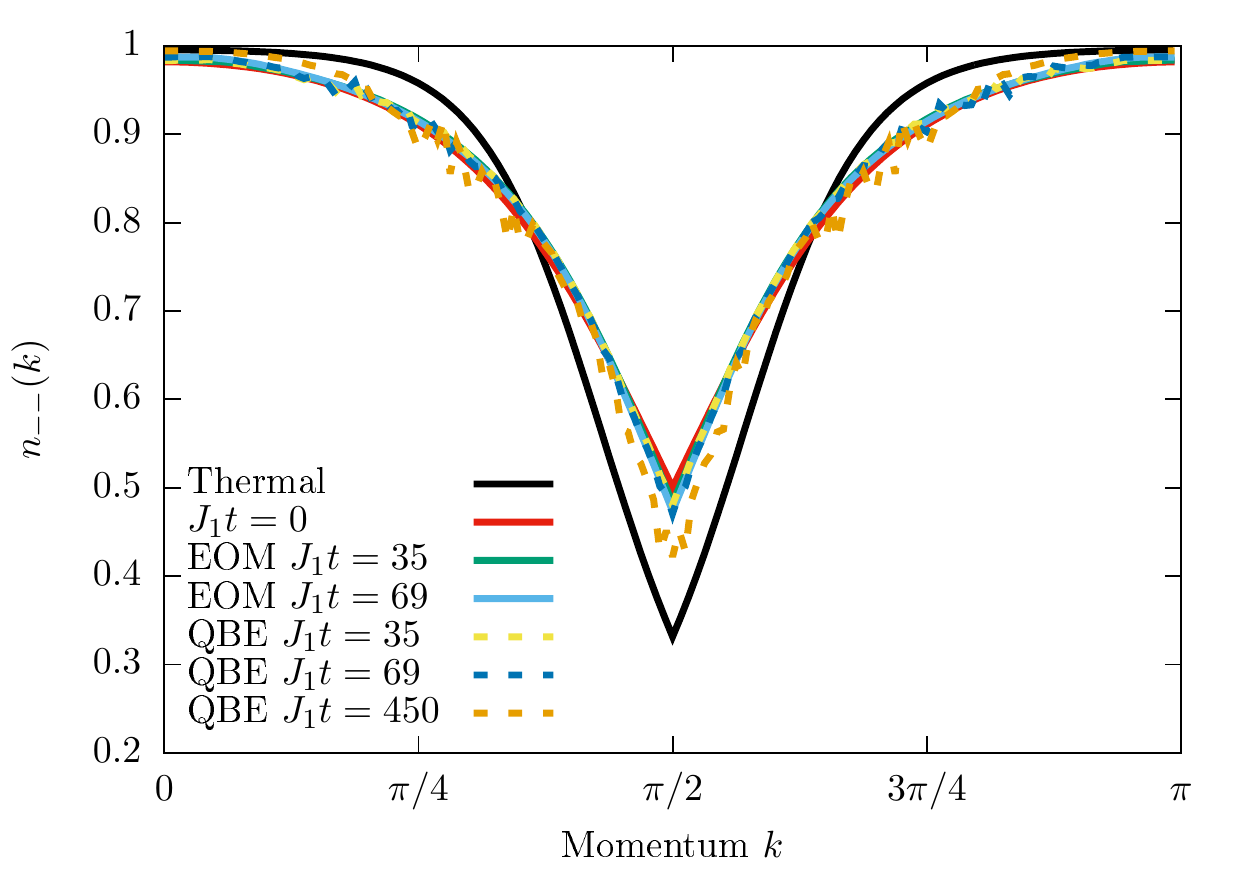}
\end{tabular}
\caption{Mode occupation numbers $n_{++}(k,t)$ (top) and
$n_{--}(k,t)$ (bottom) for a system of size $L=320$ that is
initialized in the thermal state $\rho_0(2,0.5)$ and time evolved with
$H(0.375,0,0.4)$. Solid (dotted) lines are obtained by integrating the EOM
(QBE). The solid black line is the thermal value found by means of
second order perturbation theory in $U$.}  
\label{Fig:nks0375}
\end{figure}
We observe that the mode occupation numbers slowly evolve towards
the values for a system at thermal equilibrium with the correct particle 
and energy densities, see Appendix~\ref{app:pert}.
In particular, at the latest time reached ($t=450$), the occupation
numbers $n_{++}(k,t)$ are close to the appropriate thermal
distribution. We recall that integration of the full EOM~\fr{Eq:EOM}
indicates that the ``off-diagonal'' occupation numbers $n_{+-}(k,t)$
approach their thermal value (zero) in an oscillatory fashion, see
Fig.~\ref{Fig:nktJ20375c}. These results suggest that \emph{the weak 
integrability breaking term induces thermalization of the system}. 

\subsection{Quantum Boltzmann equation for $\delta_f\neq0$}
\label{sec:Boltzdeltafinite} 
In the case $\delta_f\neq 0$ we again observe that the off-diagonal
two-point functions $n_{+-}(k,t)$ become ${\cal O}(U)$ at sufficiently late
times; for $Jt\sim100$ they are oscillating around the thermal value, found
by means of second order perturbation theory. However, this fact can no 
longer be exploited in a straightforward manner to obtain a closed system 
of equations for the mode occupation numbers $n_{\mu\mu}(k,t)$. 
This is because for $\delta_f \neq 0$ the right hand side of the 
EOM~\fr{Eq:EOM} for the diagonal components contains a term which 
does not decay in time
\be
8U ~\textrm{Im}\left[A_{\bar\mu}(k)
  n_{\mu\bar\mu}(k,0)e^{i\epsilon_{\mu\bar\mu}(k)t}\right]~. 
\label{Eq:driving}
\ee
Here $A_{\mu}(k)$ has been introduced in~\fr{Eq:OUnotdec}.

\begin{figure}
\begin{tabular}{l}
\includegraphics[width=0.425\textwidth]{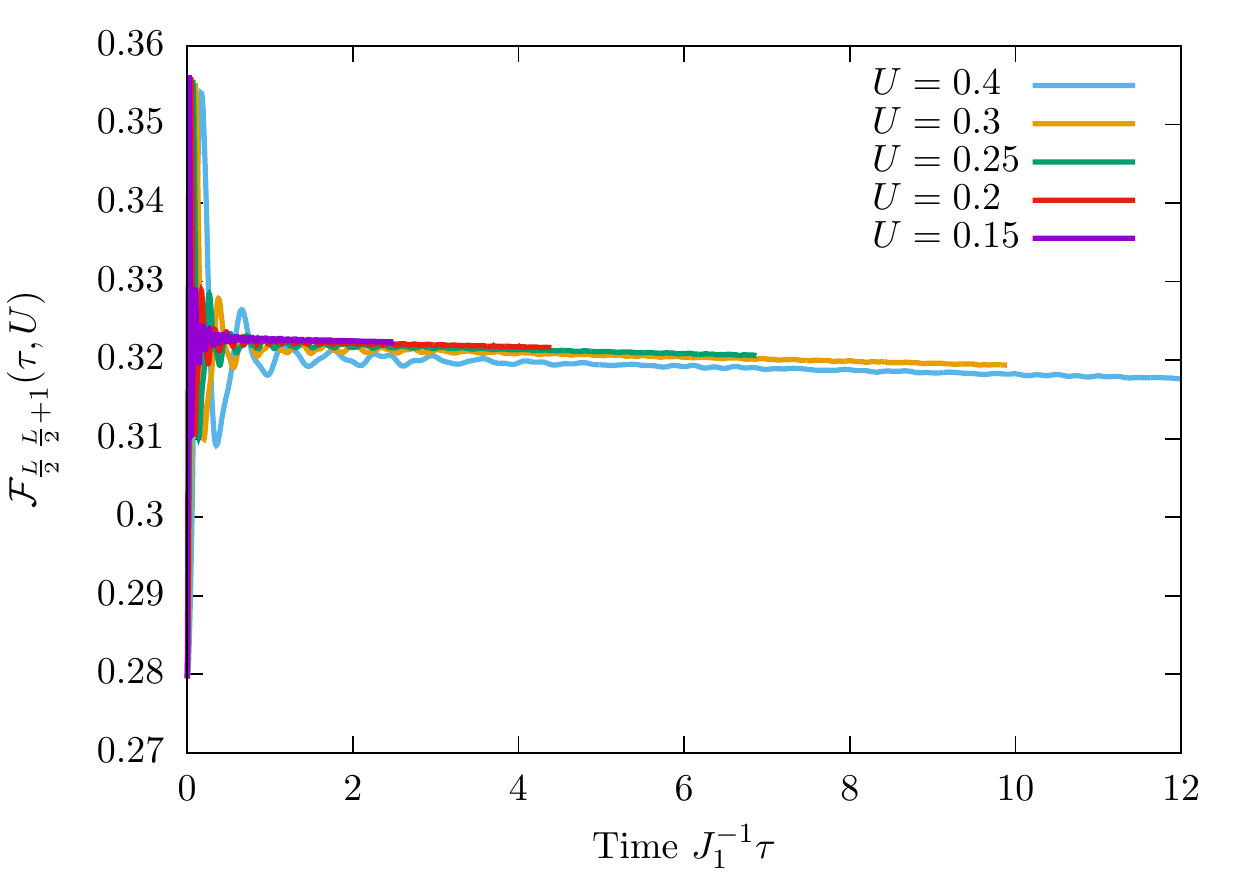} \\
\includegraphics[width=0.425\textwidth]{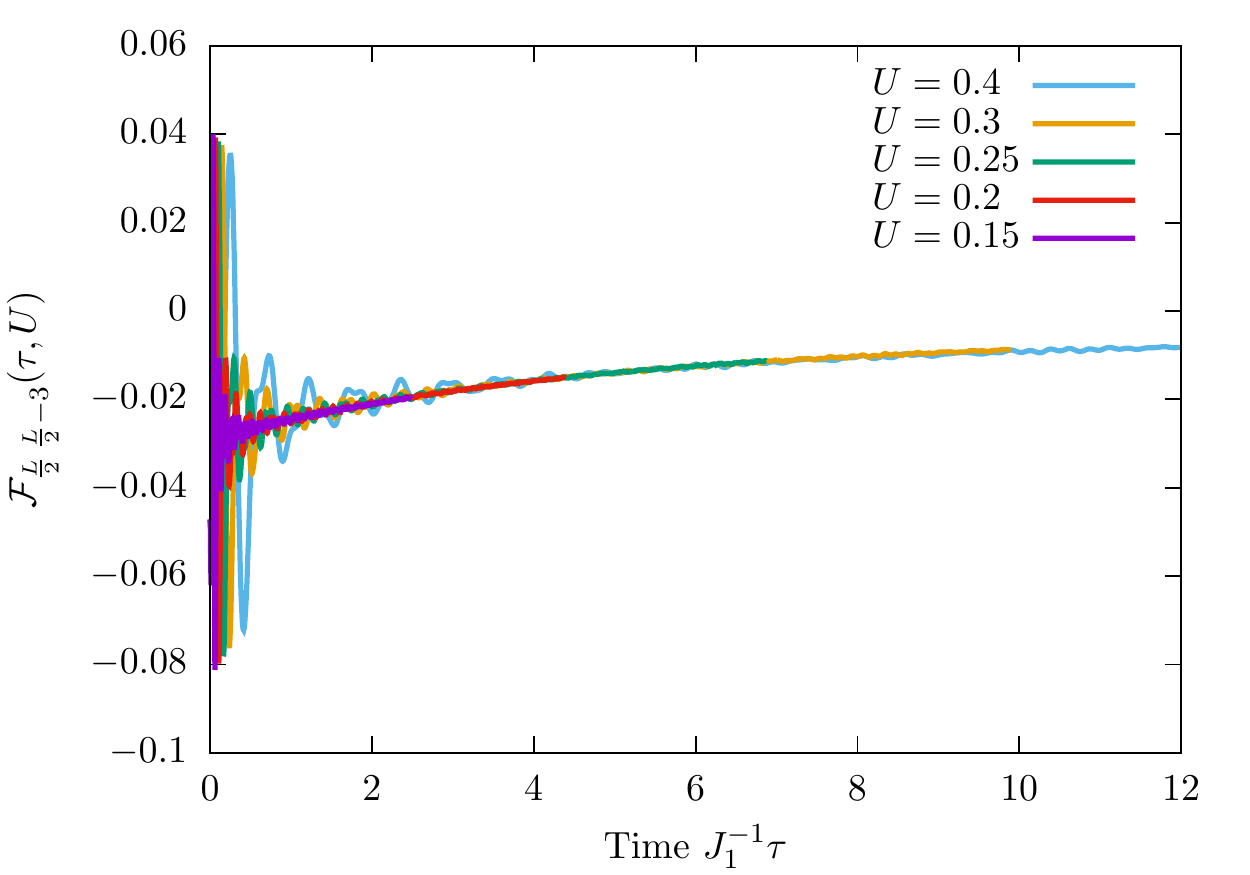}
\end{tabular}
\caption{(upper) ${{\cal F}_{L/2\, L/2+1}(\tau,U)}={{\cal G}(L/2,L/2+1;t,U)}$ and (lower)  ${{\cal F}_{L/2\, L/2-3}(\tau,U)}={{\cal G}(L/2,L/2-3;t,U)}$, the system is initially prepared in the state $\rho_0(2,0)$ and 
evolved with the Hamiltonian $H(0.5,0.4,U)$, for different values of $U$. The time evolution is obtained by numerical solution of the full 
EOM \fr{Eq:EOM} and plotted as a function of the rescaled variable $\tau=U^2 t$.}
\label{Fig:SFBC1m1}
\end{figure}

\begin{figure}
\begin{tabular}{l}
\includegraphics[width=0.425\textwidth]{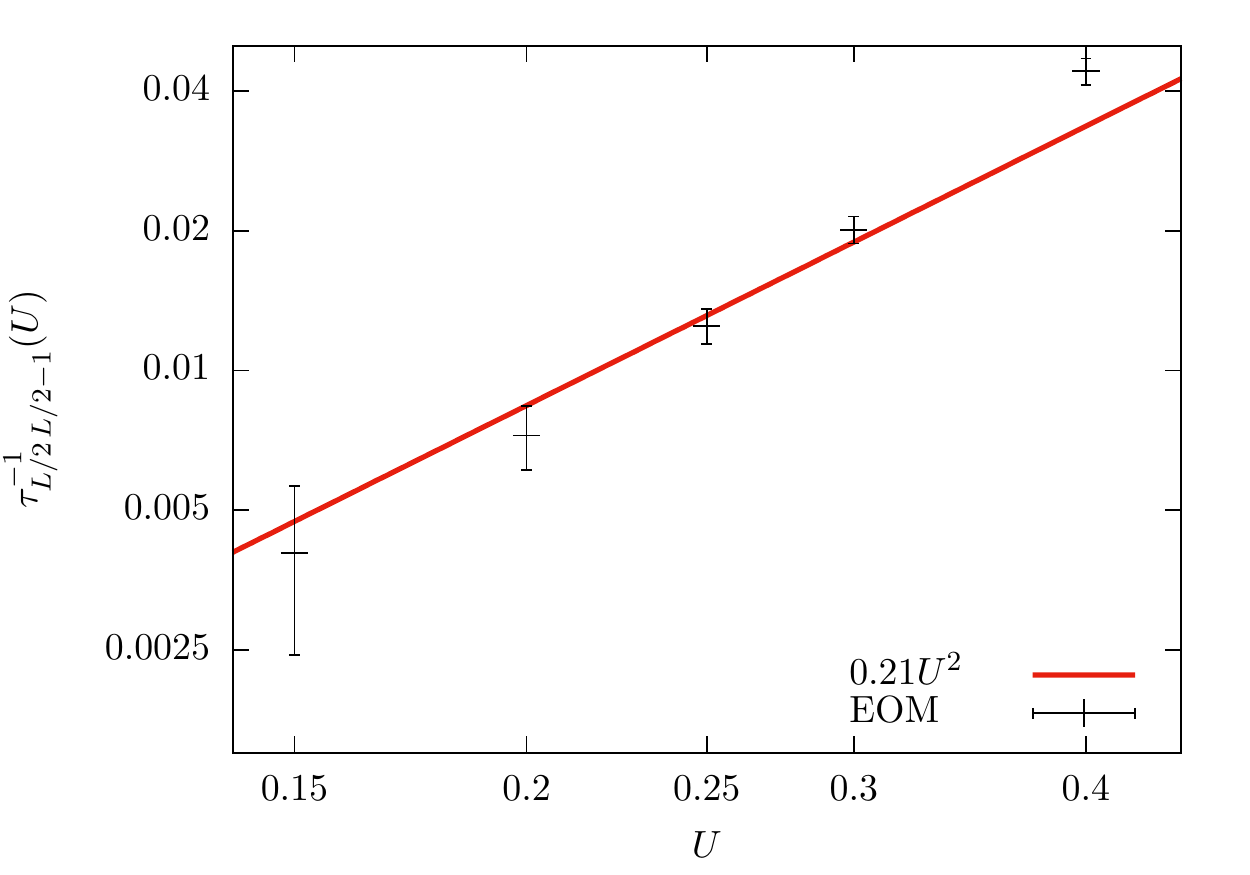} \\
\includegraphics[width=0.425\textwidth]{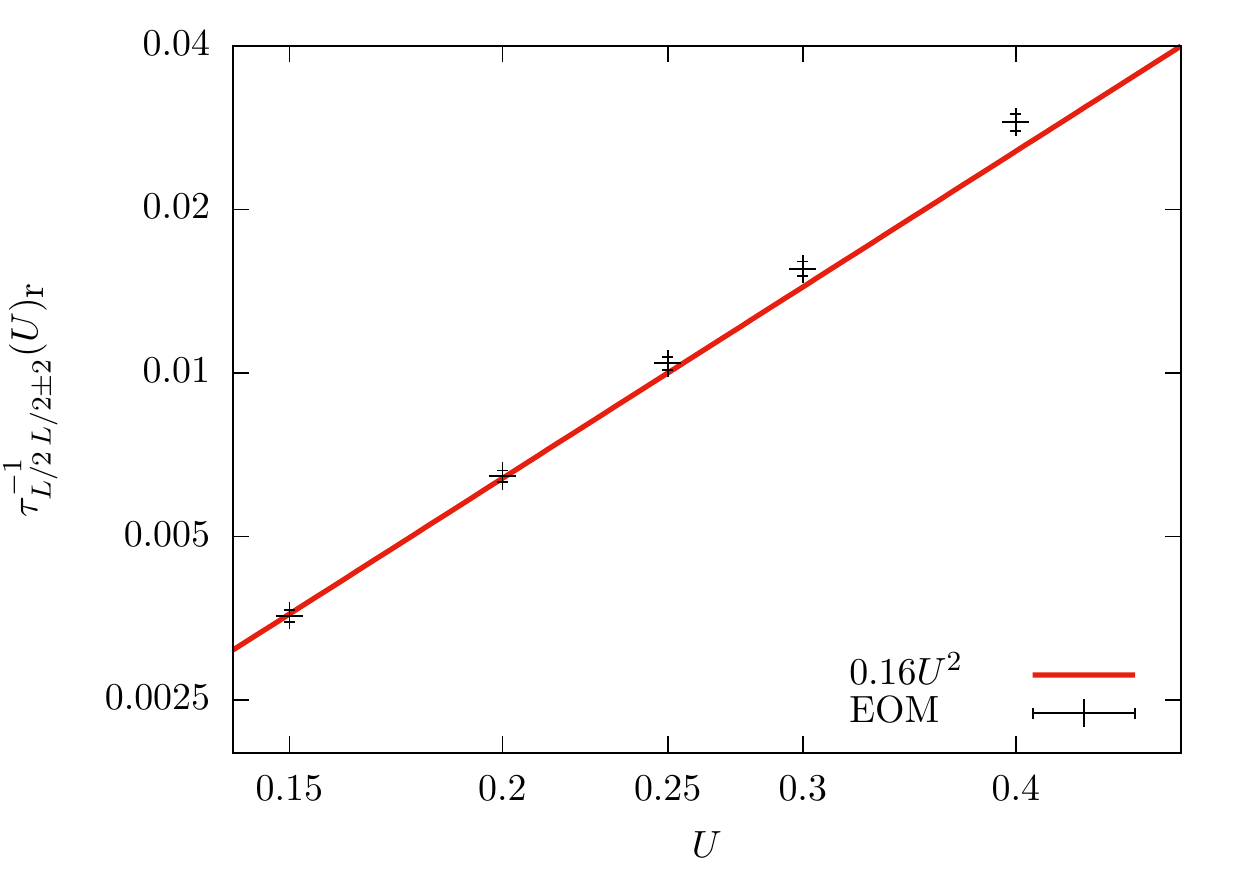}
\end{tabular}
\caption{The $U$ dependence of the exponents (upper)
  $\tau^{-1}_{L/2\,L/2-1}(U)$ and (lower)
  $\tau^{-1}_{L/2\,L/2+2}(U)_{\textrm{r}}$ obtained from the
  exponential fit \fr{Eq:ExpFit}, plotted in logarithmic
  scale.\cite{notefig} In the lower panel the error bars are contained
  within the symbols. Data is presented for the time-evolution from
  the initial state $\rho_0(2,0)$ with Hamiltonian $H(0.5,0.1,U)$ for
  $U=0.15,0.2,0.25,0.3,0.4$ (see also Figs.~\ref{Fig:SFBC1m1}). 
The errors are estimated by varying the initial time at which the exponential fit is applied.}
\label{Fig:genericexponent}
\end{figure}

In spite of this, local observables computed from our numerical solutions of the EOM exhibit an approximate scaling collapse at sufficiently late times as is shown in Fig.~\ref{Fig:SFBC1m1}
for two representative examples. This in turn allows us to repeat the arguments of the previous subsection for generic $\delta_f$, and suggests that the inverse decay times scale as
\be
\tau^{-1}_{ij}(J_2, \delta_f, U)\propto U^2.
\label{Eq:scalingexponentgeneric}
\ee
In other words, {\em the time scale for thermalization is proportional to $U^{-2}$ 
for generic $\delta_f$}. This is consistent with our solution of
the full EOM on the accessible time scales, \emph{cf.} Fig.~\ref{Fig:genericexponent}. 

In order to obtain a quantum Boltzmann like equation for $\delta_f\neq
0$ we proceed as follows. The numerical solutions of the full EOM indicate
that the mode occupation numbers exhibit small amplitude,
high-frequency oscillations on top of a smoothly varying
part. Separating these two components using a low pass filter ${\cal
  L}$ we have  
\bea
n_{\mu\mu}(k,t) &=& {\cal L}[n_{\mu\mu}(k,t)] + (1-{\cal L})[n_{\mu\mu}(k,t)],\nn
&=& s_{\mu}(k,U^2 t) + \Delta_\mu(k,t). 
\label{lowpass}
\eea
The low-pass filter ${\cal L}$ separates the slowly varying contributions
$s_\mu(k,U^2t)$ from the rapidly oscillating (small amplitude) parts
$\Delta_\mu(k,t)$. We expect that the late time behaviour of local
observables will not depend on the oscillatory parts. This expectation
is based on the observation that a stationary phase approximation
applied to the momentum sum would show that these contributions are
suppressed, \emph{cf}. Eq~\fr{Eq:GreenFun}. Applying the low-pass
filter ${\cal L}$ to the equation of motion \fr{Eq:EOM}, where the
off-diagonal two-point functions $n_{+-}$ have been neglected, and
considering the Boltzmann limit \fr{Eq:Boltzscallim}, we find that
$s_\mu(k,\tau)$ \emph{satisfy the Boltzmann equation~\fr{Eq:BB}  
with $n_{\mu\mu}(k,\tau) \to s_\mu(k,\tau)$}. Further details are
presented in Appendix~\ref{app:BrunoC}.
In Fig.~\ref{Fig:G12BC} we show comparisons between results obtained
from the quantum Boltzmann equation~\fr{Eq:BB} for $s_\mu(k,t)$ to the
full EOM~\fr{Eq:EOM}. We see that the agreement is quite satisfactory,
which gives us some confidence in the above line of argument.
\begin{figure}
\begin{tabular}{l}
\includegraphics[width=0.425\textwidth]{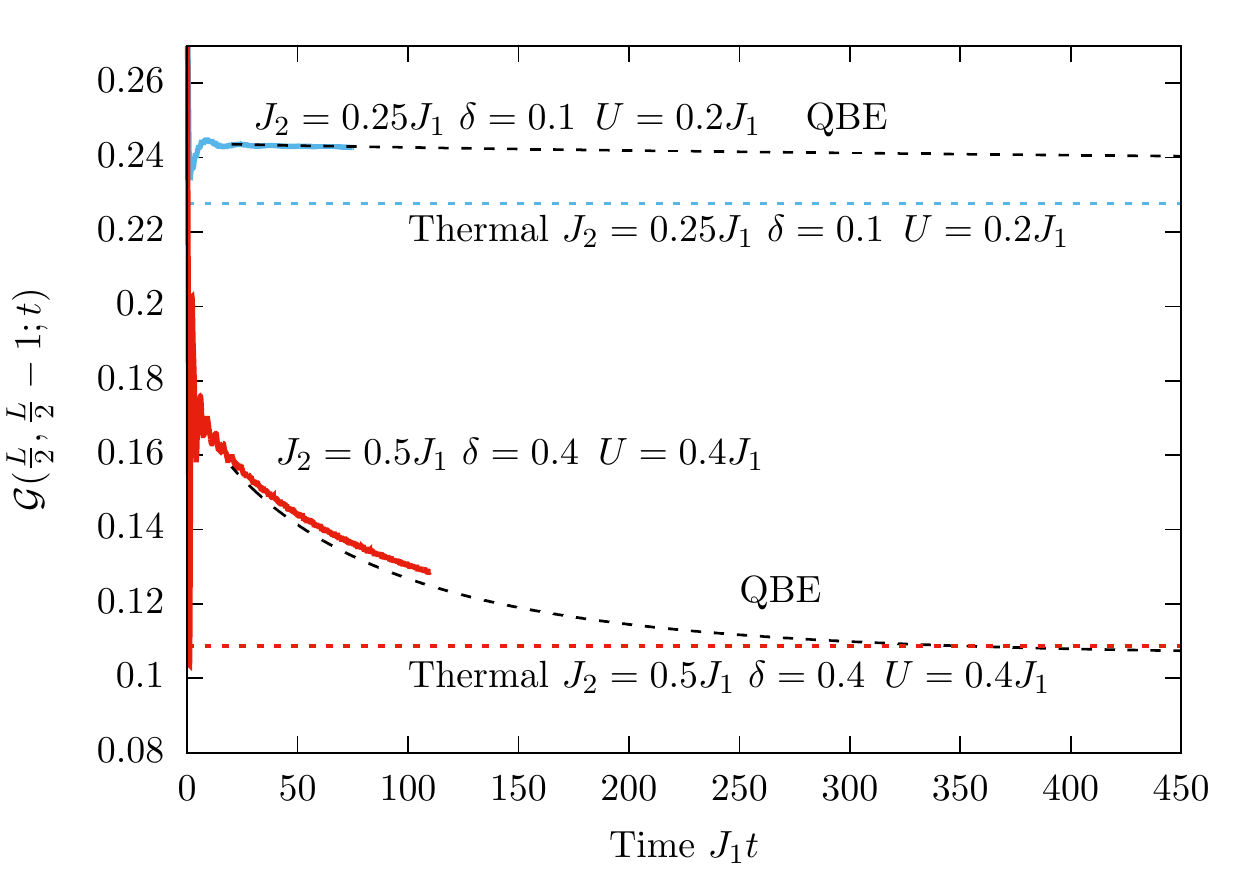} \\
\includegraphics[width=0.425\textwidth]{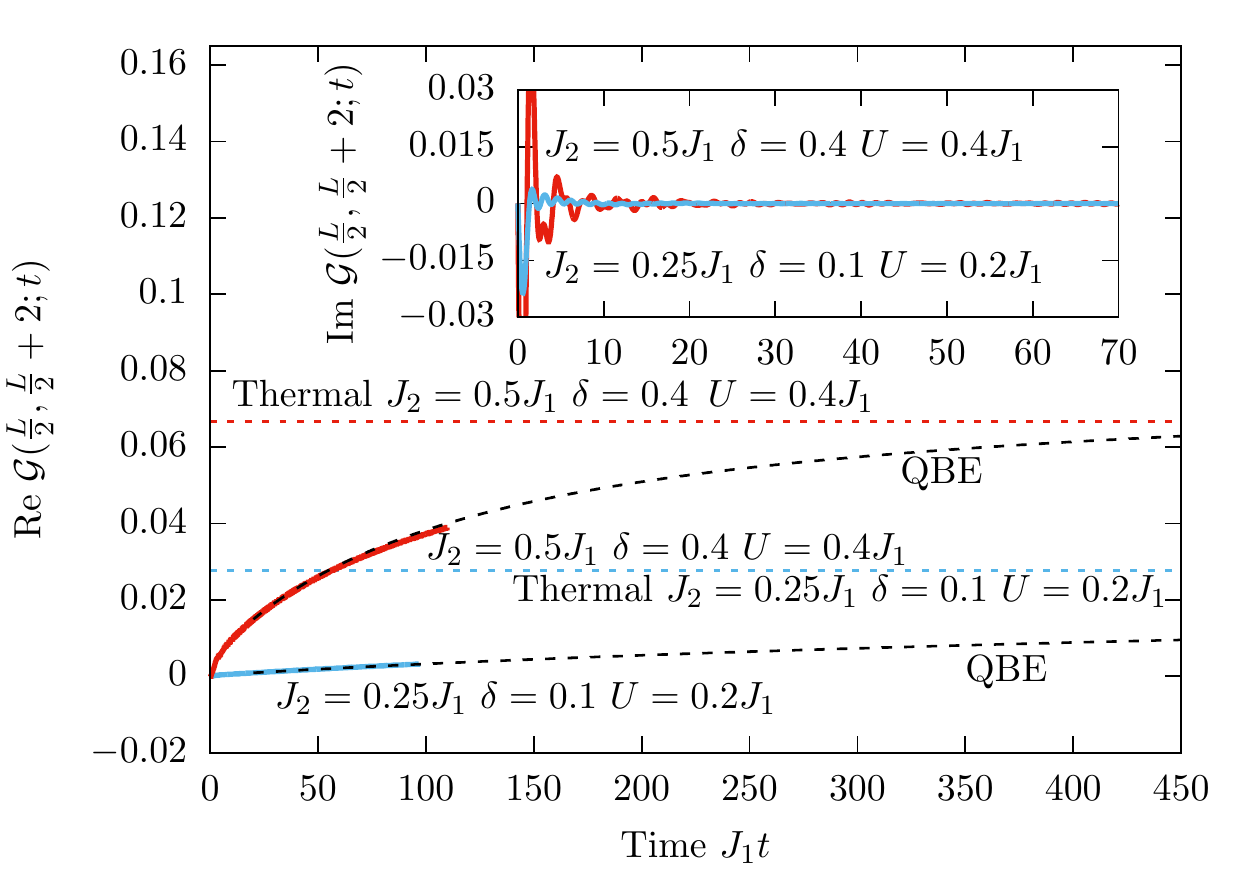}
\end{tabular}
\caption{${\cal G}({L}/{2},{L}/{2}-1;t)$ (upper panel) and ${\cal
    G}({L}/{2},{L}/{2}+2;t)$ (lower panel) for two systems with
  Hamiltonians  $H(0.25,0.1,0.2)$,  
$H(0.5,0.4,0.4)$ and size $L=320$ initially prepared in a thermal state \fr{thermal} 
with density matrix $\rho_0(2,0)$. The full lines are obtained by
integrating the EOM \fr{Eq:EOM} and the black dashed lines are found
by means of the QBE. } 
\label{Fig:G12BC}
\end{figure}

To conclude our discussion of the $\delta_f \neq 0$ case, we note that
the occupation numbers $s_\mu(k,\tau)$ appear to approach their
thermal values (computed by second order perturbation theory) in the
long time limit.

\section{Breaking the $\textrm{U(1)}$ symmetry}
\label{sec:u1b}
One of the key features of the class of models~\fr{Eq:Ham} is that they possess
a global $\textrm{U}(1)$ symmetry associated with particle number
conservation, see Sec.~\ref{sec:model}. We expect
PT to be robust with respect to breaking this symmetry.
To check whether this is indeed the case we have investigated quantum
quenches to the class of models
\begin{align}
\label{Eq:pertXY}
{\cal{H}}(\gamma,h,U)&=\frac{J}{2}\sum_{i}\left[ c^\dag_ic^{\phantom{\dag}}_{i+1}+ \gamma c^\dag_i c^\dag_{i+1}+\textrm{h.c.}\right]\nn
&+J h\sum_{i}c^\dag_ic^{\phantom{\dag}}_{i}+U\sum_{i}c^\dag_ic^{\phantom{\dag}}_{i}c^\dag_{i+1}c^{\phantom{\dag}}_{i+1}~,
\end{align}      
which are related to the Heisenberg XYZ chain in a magnetic field by a
Jordan-Wigner transformation. The model \fr{Eq:pertXY} becomes
integrable in several limits
\begin{enumerate}
\item{}
For $U=0$ the model is non-interacting;
\item{}For ${\gamma = 0}$ it is equivalent to the spin-1/2 Heisenberg XXZ 
chain in an external magnetic field; 
\item{}For $h = -U$, where it is equivalent to the {spin-1/2}
  Heisenberg XYZ chain. 
\end{enumerate}
In order to apply the EOM formalism we prepare the system in an
initial density matrix with respect to which Wick's theorem holds. Our
choice is
\be
\sigma_0=\sigma(\beta,\gamma_i,h_i)=\frac{e^{-\beta
    {\cal H}(\gamma_i,h_i,0)}}{\textrm{Tr}\left[e^{-\beta
      {\cal H}(\gamma_i,h_i,0)}\right]}.
\ee
We then time evolve with ${\cal H}(\gamma_f,h_f,U)$ and are interested in
the following Green's functions
\bea
\mathcal{G}_{+-}(i,j;t)&=&\textrm{Tr}\left[c^{{\dag}}_i(t)
  c^{\phantom{\dag}}_j(t)\sigma_0\right]\ ,\nn
{\cal G}_{++}(i,j;t)&=&\textrm{Tr}\left[c^{{\dag}}_i(t)
  c^{{\dag}}_j(t)\sigma_0\right] .
\label{Gab}
\eea
Details regarding the implementation of the EOM formalism are
presented in Appendix~\ref{app:EOMnoU1}. Figure~\ref{Fig:noU1G} shows
results for the Green's functions \fr{Gab} for a system prepared in
the density matrix $\sigma(\infty,0.2,0)$ and time evolved with
${\cal H}(0.5,0.1,U)$.
\begin{figure}[ht]
\includegraphics[width=0.425\textwidth]{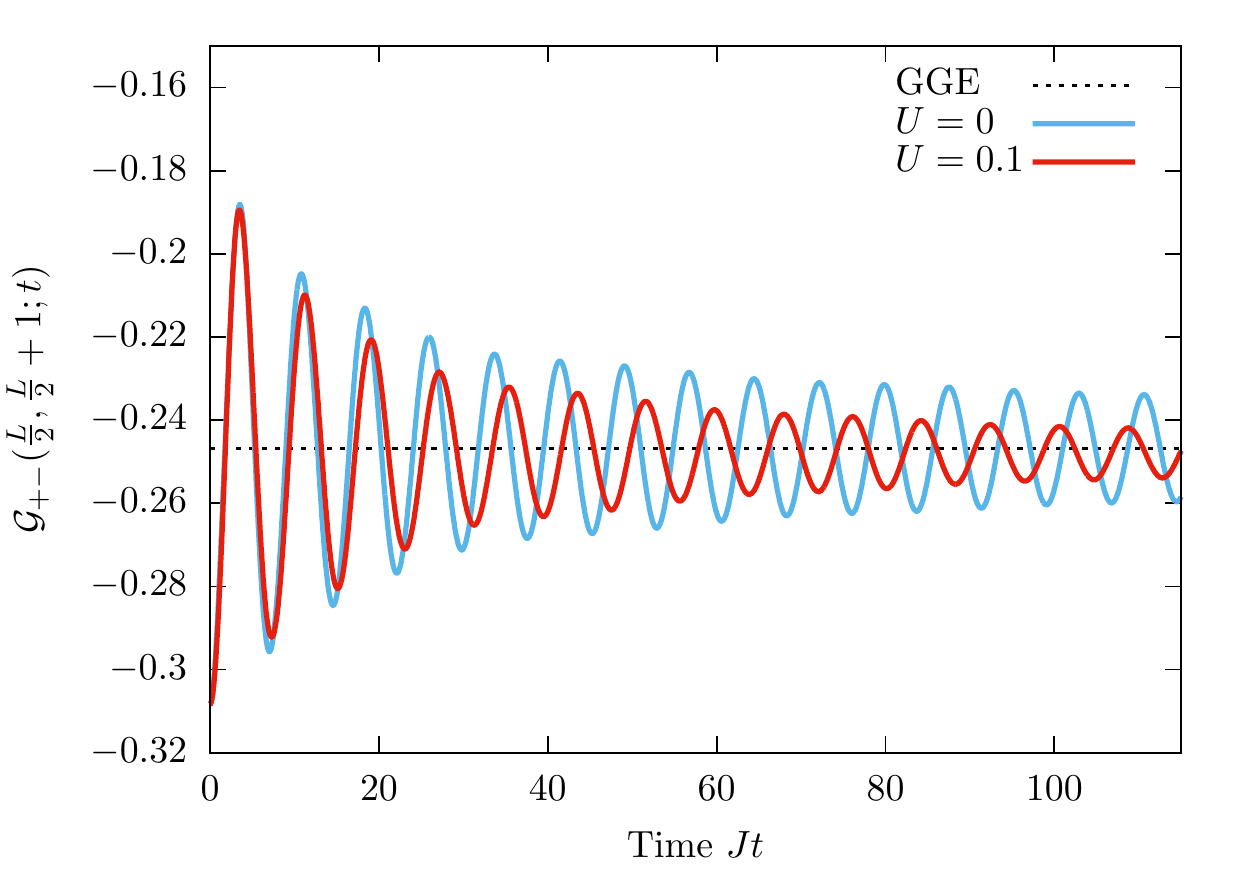}\\
\includegraphics[width=0.425\textwidth]{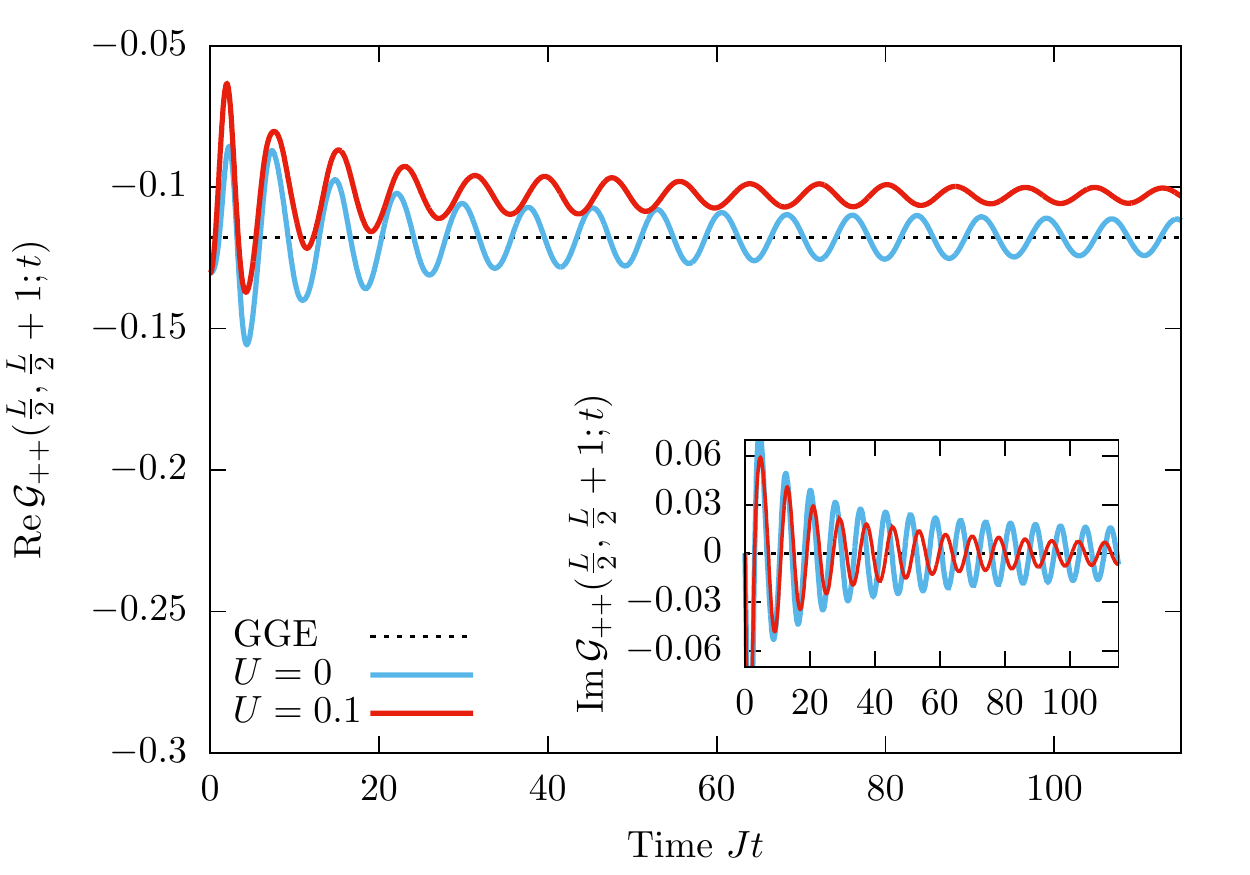}
\caption{${\cal G}_{+-}(L/2,L/2+1;t)$ (top) and ${\cal
 G}_{++}(L/2,L/2+1;t)$ (bottom) for a system prepared in the state
$\sigma(\infty,0.2,0)$ and time evolved with ${\cal H}(0.5,0.5,0.1)$. Dotted lines denote the non-interacting GGE.}
\label{Fig:noU1G}
\end{figure}
In the integrable case $U=0$ we observe relaxation towards the appropriate
GGE. On the other hand, in the non-integrable case $U=0.1$ we observe a
PT plateau. This analysis establishes that
PT is robust under $U(1)$ symmetry breaking.

\subsection{Pre-relaxation}
\label{subsec:MF}
An interesting limit of the model \fr{Eq:pertXY} is the case when $h=O(U)$,
\emph{i.e.} the chemical potential is included in the
integrability-breaking perturbation. Then the ``unperturbed'' integrable
model is related to the XY chain in zero magnetic field by a
Jordan-Wigner transformation. The XY chain is known to possess 
infinitely many local conservation laws $\{\mathcal Q_n\}$ that
satisfy a non-abelian commutation algebra,\cite{Fagotti14} as well as
a infinite set of mutually commuting local conserved charges
$\{\mathcal I_n\}$.

Reference~[\onlinecite{BF15}] investigated how the additional conservation
laws influence the time evolution in the presence of an interacting
perturbation which breaks this structure. The problem was studied
using a novel mean-field-like technique which was conjectured to be
accurate for times $t\sim U^{-1}$. It was found that observables 
show highly non-trivial behaviour if one starts from an initial state
in which some of the additional charges have non-zero expectation
values. In this situation, observables rapidly relax towards values
close to the unperturbed GGE prediction (which describes the
stationary state for $U=0$) and then, at times $t \sim U^{-1}$,
drift away. Two scenarios are possible at this point: either the
observables relax \textit{towards a second nonthermal plateau} which
is ${\cal O}(U^0)$ different from the first, or they \emph{show persistent
  oscillations} within the entire accessible time-window. This
phenomenon was termed \emph{pre-relaxation} because it describes the
crossover between two \emph{non-thermal} behaviours. In the case where
the introduced perturbation breaks integrability, the system is
believed to eventually thermalize, although the thermalization
time-scales are beyond the time regime that is accessible to the
method of Ref.~[\onlinecite{BF15}].

\begin{figure}[ht]
\centering
\includegraphics[width=0.425\textwidth]{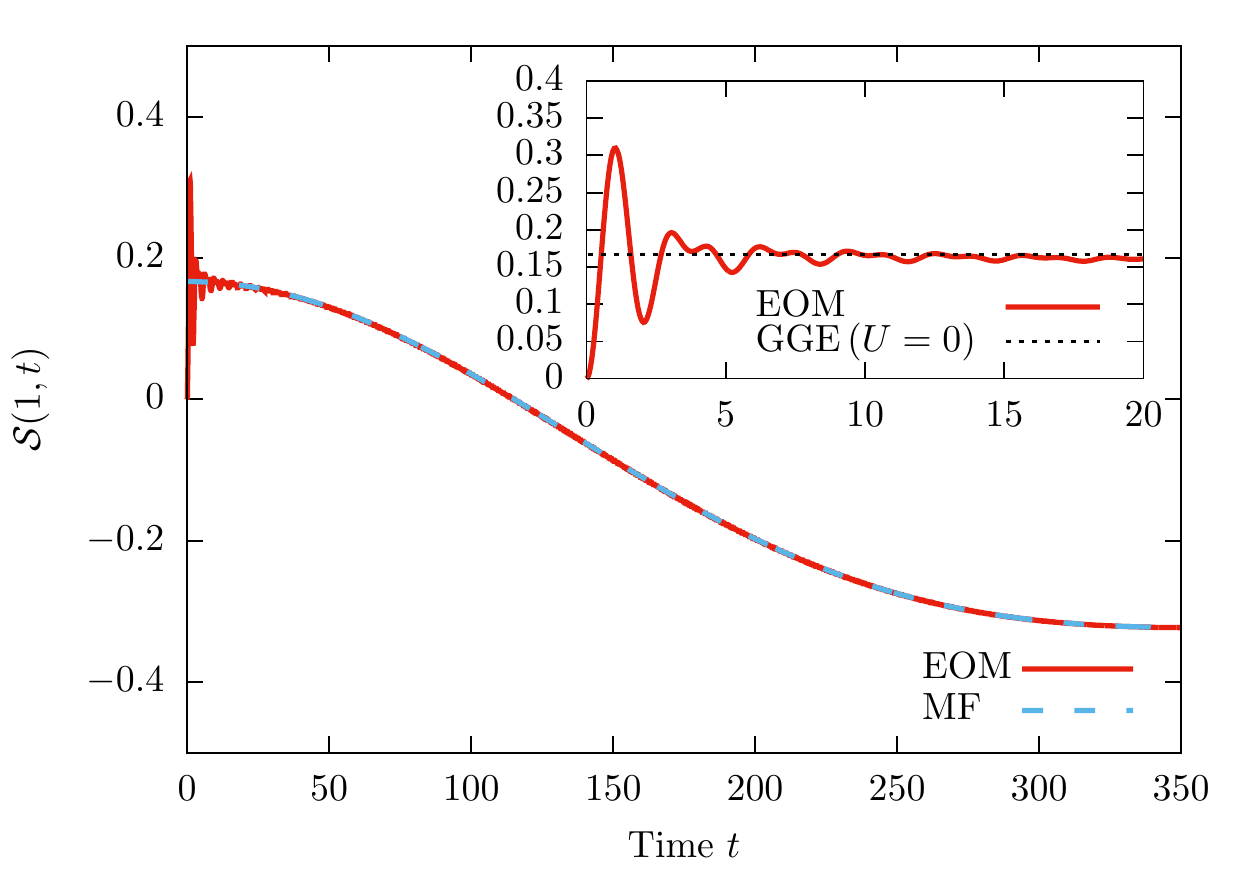}
\caption{$\mathcal{S}(1;t)$ (\emph{cf}. Eq.~\fr{Eq:S}) evolved by
$\mathcal{H}(2,0.005,0.005)$, starting from the ground state of the
Majumdar-Ghosh Hamiltonian.\cite{MG:1969} The full line is obtained
by integration of the first order EOM (reported in Appendix
\ref{app:EOMnoU1}), while the dashed lines are the prediction of the
mean-field approach developed in [\onlinecite{BF15}]. The inset
shows a shorter time window where the observable lies on the first quasi-stationary plateau (black dotted line).} 
\label{Fig:EOMvsMF}
\end{figure}

The Hamiltonian~\fr{Eq:pertXY} for $h={\cal O}(U)$ is precisely one of
the cases considered in Ref.~[\onlinecite{BF15}], once spins are
mapped to fermions. Moreover, the time scale $t \sim U^{-1}$ on
which the mean-field-like approach was conjectured to be accurate is
amenable to analysis by our first-order EOM and it is interesting to
compare the two results. To perform the comparison, we consider the
correlation function   
\begin{align}
\mathcal{S}(i;t)\equiv\mathrm{Re}\left[\mathcal{G}_{++}(i+1,i;t)-\mathcal{G}_{+-}(i+1,i;t)\right]\,,
\label{Eq:S}
\end{align}
which corresponds, through a Jordan-Wigner transformation, to
$\braket{\sigma^x_i\sigma^x_{i+1}}(t)$ in the spin model. As the
additional ``non-abelian'' local conservation laws ${\cal Q}_n$ of the
XY chain change sign under translation by one site, we require an initial state that is not invariant under translations by one site. In order
for the ${\cal Q}_n$ to have non-vanishing expectation values. To
compare to the results of Ref.~[\onlinecite{BF15}], we take the
initial state to be the ground state of the Majumdar-Ghosh (MG)
Hamiltonian,\cite{MG:1969} which is invariant only under translations
by two sites. Expectation values in the initial state can be
calculated using Wick's theorem, as required for the EOM to be
applicable, and the initial Green's functions read  
\bea
\mathcal{G}_{+-}(2i-r,2j-s;0)\big|_{\rm MG}&=&\delta_{i-r,j-s}\,,\,\,\, r,s=0,1\nn
{\cal G}_{++}(i,j;0)\big|_{\rm MG}&=& 0\,.
\eea
In Fig.~\ref{Fig:EOMvsMF} we report a comparison between the
mean-field approach of Ref.~[\onlinecite{BF15}] and the first order
EOM. The mean-field solution starts from the prediction of the
non-interacting GGE and it is in excellent agreement with the EOM. The
agreement is almost perfect because the interaction is very small in
the case considered, but in general we do expect ${\cal O}(U)$  
differences between the two results.

\section{Conclusions}
\label{sec:conc}
In this work we have used equation of motion techniques to investigate
prethermalization in a class of one dimensional fermion models with
weak integrability breaking perturbations. Our integrable model is a
non-interacting theory, and the role of the integrability breaking
perturbation is played by density-density interactions. We focus on
the time evolution of the single-particle Green's function after
initializing the system in a density matrix that is not an eigenstate
of the time evolution operator. 

The non-equilibrium evolution in the
non-interacting theory is non-trivial in our setup and provides an
important point of reference. As expected expectation values of local
operators relax towards GGEs in this case. When
we turn on a weak integrability breaking perturbation of strength $U$,
we observe long lived PT plateaux: the Green's function
(for finite separations) relaxes towards constant values that differ
from the ones in the GGE at order ${\cal O}(U)$, and are compatible
with the deformed GGE description proposed in
Ref.~[\onlinecite{EsslerPRB14}]. We have verified that PT
occurs irrespective of whether particle number is conserved. This is
in accord with expectations based on the CUT approach to
non-equilibrium evolution\cite{MK:prethermalization,EsslerPRB14}: as
long as the integrability-breaking perturbation merely \emph{dresses} the
elementary excitations of the non-interacting theory we expect PT to
occur. For very weak perturbations $U$ the PT plateau is stable
throughout the time window accessible to us. By increasing $U$ we are
able to observe a crossover between PT and evolution towards a thermal
steady state. The corresponding crossover time scales as $U^{-2}$.

We have used our EOM methods to analyze the structure of light-cones
in the single-particle Green's function. Light cone effects provide a
direct probe of quasi-particle properties, and it is clearly an
interesting question how these are affected by integrability breaking
terms. We observed that in all cases the exterior and interior regions
of the light-cone are separated by an ``intermediate'' regime, the width
of which appears to scale with a universal exponent
$t^{1/3}$ irrespective of whether or not the post-quench Hamiltonian is
integrable. In contrast, the maximum values of the real and imaginary
parts of the single-particle Green's function, \emph{i.e.} our ``signals'', exhibit a markedly faster decay in time $t$ in the non-integrable case as compared to the integrable one.

Our work raises a number of issues deserving of further investigation.
First, our work suggests that PT is rather robust provided the
integrability breaking perturbation does not dramatically alter the
nature of quasi-particle excitations. It would be interesting to
analyze examples where we know confinement to occur, an example being
the transverse field Ising model in a weak longitudinal magnetic
field.\cite{confinement,confinementnoneq, RMCKT:confinementnoneq} Unfortunately such
situations cannot be accommodated in the EOM approach, because the
perturbation is non-local in terms of the elementary fermions. Second,
our approach is by construction uncontrolled. We have checked that our
results are in excellent agreement with existing t-DMRG
results,\cite{EsslerPRB14} but further checks are highly desirable. We
also expect our truncation of the infinite hierarchy of EOMs to become
inaccurate at late times. It would be interesting to try to implement
a truncation scheme that incorporates effects of the four particle
cumulant. This is numerically very demanding, but would open the
possibility of exploring the late time regime.\cite{Luxetal}

Finally, it would be very interesting to investigate the analogous
set of questions for a weak perturbation to a strongly interacting
integrable model. 
\acknowledgments

We thank John Cardy, Maurizio Fagotti, Stefan Kehrein, Robert Konik
and Wei Ku for useful discussions surrounding this work. FE and NR thank Salvatore Manmana for the previous collaboration [\onlinecite{EsslerPRB14}] on this problem. This work was
supported by the EPSRC under grants EP/I032487/1 (BB and FHLE) and
EP/J014885/1 (FHLE), the Isaac Newton Institute for Mathematical
Sciences under grant EP/K032208/1, the ERC under Starting Grant 279391
EDEQS (BB), the U.S. Department of Energy under Contract
No. DE-SC0012704 (NJR), and by the Clarendon Scholarship fund (SG). 

\appendix

\section{Further details on the time evolution of the Green's function}\label{app:details}

In this appendix we collect some results on how the time evolution of
${\cal G}(j,l;t)$ is influenced by (i) particle-hole symmetry, (ii)
final dimerization, (iii) sign of the interaction.

\subsection{The role of particle-hole symmetry}\label{sec:phsymmetry}

\begin{figure}[t!]
\begin{tabular}{l}
\includegraphics[width=0.425\textwidth]{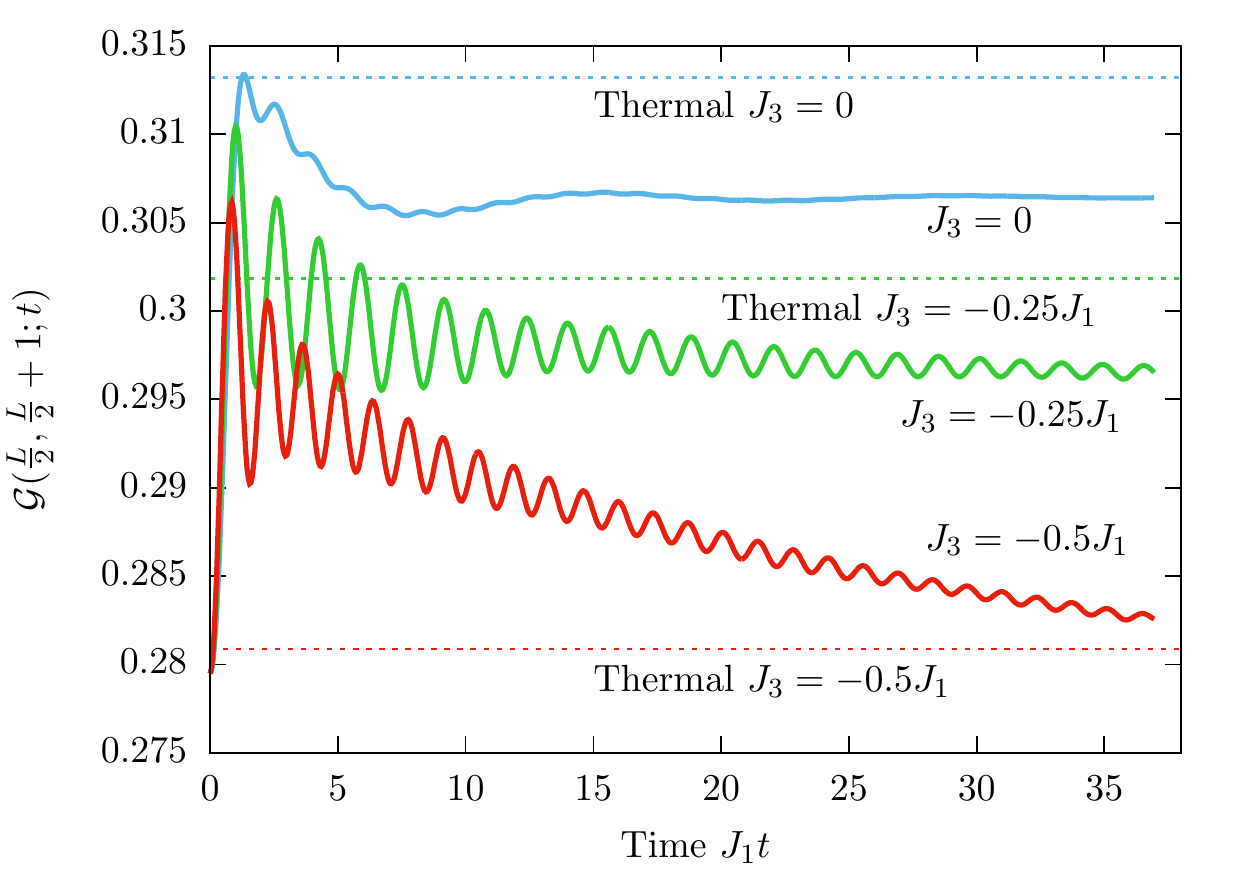}
\end{tabular}
\caption{Green's function ${\cal G}({L}/{2},{L}/{2}+1;t)$ for a
systems of sizes $L=320, 384$ that is prepared in the density matrix
$\rho_0(2,0)$ and time-evolved with $H_{3}(J_3,0.1,0.4)$
[\emph{cf}. \fr{Eq:H3n}]. The expected steady state thermal values
computed by ED for $L=16$ are shown by dotted lines. We note that the
cases with $J_3 \neq0$ exhibit pronounced finite-size effects in the
ED data.}
\label{fig:J3dep}
\end{figure}

\begin{figure}[h]
\includegraphics[width=0.5\textwidth]{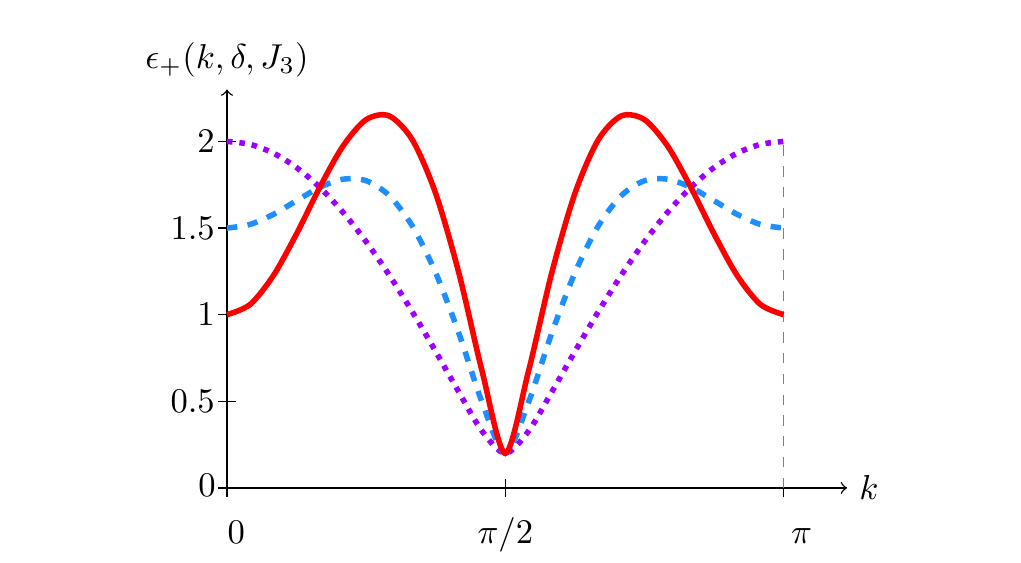}
\caption{Dispersion relation $\epsilon_+(k,\delta,J_3)=-\epsilon_-(k,\delta,J_3)$ [see Eq.~\fr{Eq:dispJ3}] for the two bands of Bogoliubov fermions in the non-interacting model with $J_1=1$, $\delta=0.1$ and $J_3=0$ (dotted), $J_3=-0.25$ (dashed), $J_3=-0.5$ (solid). Increasing $J_3$ leads to additional crossings at a fixed energy.} 
\label{fig:dispJ3}
\end{figure}

For a given interaction strength $U$, we have seen that the addition of 
next-neighbour hopping ($J_2 > 0$) to the Hamiltonian has a significant effect
on the time evolution of local observables -- they show a marked drift towards
their thermal values. To investigate whether this effect is related to the 
breaking of particle-hole symmetry by the $J_2$ term, we study the
time evolution with a modified Hamiltonian where we replace
the next-neighbour hopping with a next-next-neighbour term, $J_3$. Such
a modification preserves particle-hole symmetry, whilst modifying the
single-particle dispersion to introduce more crossings at a fixed
energy. Specifically, we consider
\be
H_{3}(J_3, \delta, U)=H(0, \delta, U)-J_3\sum_{i=1}^{L} \left(c^{\dag}_i c^{\phantom{\dag}}_{i+3}+\rm{h.c.}\right).
\label{Eq:H3n}
\ee
The EOM analysis proceeds as before, provided one uses the appropriately modified single-particle dispersion and Bogoliubov angle
\begin{align}
& \!\!\!\epsilon_\eta(k,\delta_f,J_3) = 2\eta \sqrt{(\cos(k)+J_3\cos(3k))^2+\delta_f^2\sin^2(k)}~,\notag\\
& \!\!\!e^{-i \varphi_k(\delta_f,J_3)}=\frac{-(\cos k+J_3\cos(3k))+ i \delta_f\sin k}{\sqrt{(\cos(k)+J_3\cos(3k))^2+ \delta_f^2\sin^2k}}~.\label{Eq:dispJ3}
\end{align} 
In Fig.~\ref{fig:J3dep}, we show that \textit{drifting towards the thermal 
values also occurs when particle-hole symmetry is preserved}. 
Instead, it appears that the presence of multiple crossings at fixed energy 
(``scattering channels'') in the single particle dispersion (\emph{cf.}
Fig.~\ref{fig:dispJ3}) is the key ingredient for observing the drift towards 
thermalization in achievable time scales. As for the case with $J_2$,
the higher the degeneracy at fixed energy in the single particle
dispersion, the stronger the drifting becomes. We stress that there is no
enhancement of the effective interaction with increasing $|J_3|$
($cf.$ Sec.~\ref{subsec:nnnhopping}), as the bandwidths of both bands
are (slightly) \textit{increased} by the addition of $|J_3|$.  

As an aside, we note that the initial relaxation towards the PT
plateau is much slower in the $J_3 \neq 0$ case. For large positive
$J_3$, relaxation takes places at times much larger than those
reachable with the EOM. The slow decay of oscillation towards PT can
be understood from the leading order EOM~\fr{Eq:firstorderEOM1}, see also
Ref.~[\onlinecite{EsslerPRB14}]. Inserting the solution of
Eq.~\fr{Eq:firstorderEOM1} into Eq.~\fr{Eq:GreenFun}, one obtains the
prethermal behaviour of the Green's function. By means of a stationary
phase analysis, it can be seen that the relaxation of
the Green's function towards the PT plateau is
generically $t^{-1/2}$  for $J_3 \neq 0$, compared to $t^{-3/2}$ when $J_3=0$. 
For the cases reported in Fig.~\ref{fig:J3dep}, the
leading $t^{-1/2}$ term has a small pre-factor, and one effectively
sees oscillations whose amplitude decays as $1/t$. 

\subsection{Dependence on dimerization parameter $\delta_f$}
\label{subsec:deltadep}
We now turn to the dependence of the post-quench dynamics on the
nearest-neighbour dimerization $\delta_f$, which without loss of
generality can be taken in the range $0 \leq \delta_f \leq 1$. 
\begin{figure}[t!]
\begin{tabular}{l}
\includegraphics[width=0.425\textwidth]{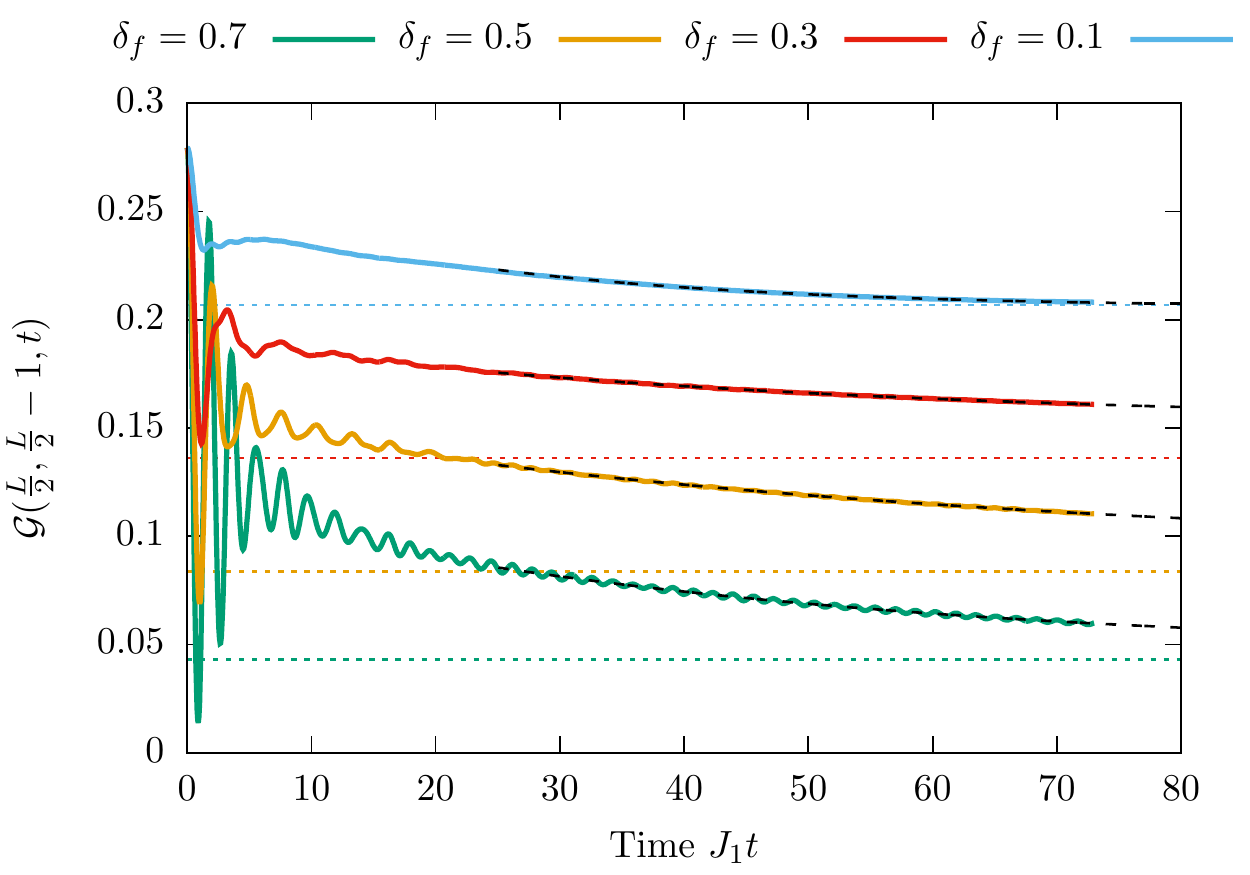} 
\end{tabular}
\caption{The Green's function ${\cal G}({L}/{2},{L}/{2}- 1;t)$ for 
a system of size $L=320$ that is prepared in the density matrix $\rho_0(2,0)$
and time-evolved with $H(0.5,\delta_f,0.4)$ [\emph{cf}. \fr{Eq:Ham}]
with $\delta_f=0.1,0.3,0.5,0.7$ (top to bottom). The expected steady state thermal
values are shown by dotted lines.} 
\label{Fig:dfdep}
\end{figure}
In Fig.~\ref{Fig:dfdep} we show the time-evolution of the Green's function
when the system is initialized in the density matrix $\rho_0(2,0)$ and time evolved
with $H(0.5,\delta_f,0.4)$ for a range of values of $\delta_f$. As we increase $\delta_f$ we initially observe a decrease in the
inverse relaxation times. 

\subsubsection{Restoration of translational symmetry: $\delta_f = 0$}

\begin{figure}
\begin{tabular}{l}
\includegraphics[width=0.425\textwidth]{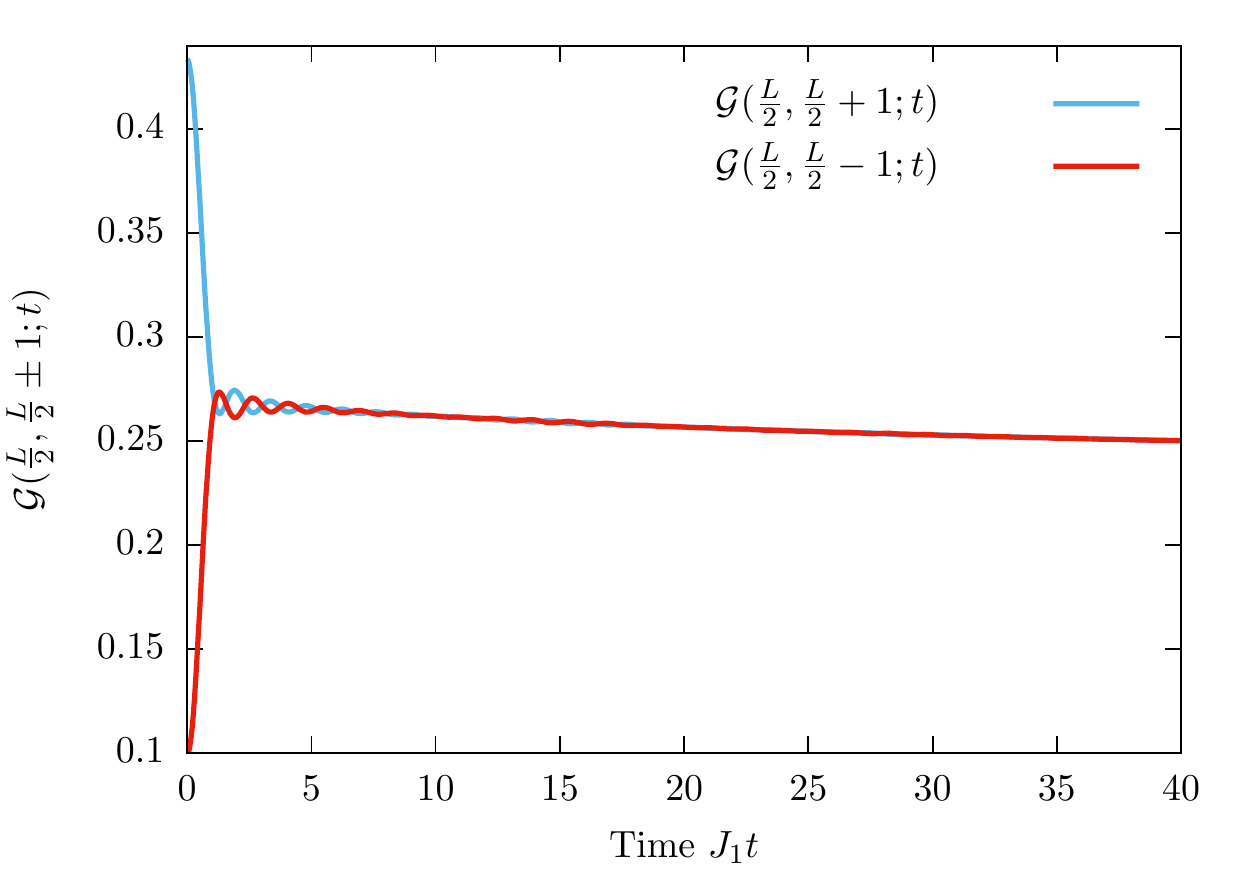}
\end{tabular}
\caption{Time evolution of $\mathcal{G}(L/2,L/2\pm 1;t)$
for a system initialized in the density matrix $\rho_0(2,0,5)$
~\fr{thermal} and time evolved with the translationally invariant
Hamiltonian $H(0.5,0,0.4)$.}
\label{Fig:SymmetryRestoration}
\end{figure}

In the special case $\delta_f=0$ our post-quench Hamiltonian is
translationally invariant by one site (rather than two). This allows
us to address the issue of translational symmetry restoration: if we
start in an initial state that is invariant only under translations by
two sites, is one-site translational symmetry restored at long times
after the quench? To address this question, we consider the
time-evolution of the initial density matrix~\fr{thermal} with 
$\beta_i = 2$, $\delta_i = 0.5$ with the Hamiltonian~\fr{Eq:Ham} with
$J_2 = 0.5$, $\delta_f = 0$ and $U =
0.4$. Figure~\ref{Fig:SymmetryRestoration} show results for the
Green's function with separation $\pm1$. In both cases we see
that there is a rapid restoration of translational symmetry -- by time
$J_1 t \sim 10$ the results are already extremely close. The situation
for larger separations is completely analogous, but the time scale
after which symmetry restoration is seen is pushed back, as expected.

\subsection{Attractive interactions}
\begin{figure}
\begin{tabular}{l}
\includegraphics[width=0.425\textwidth]{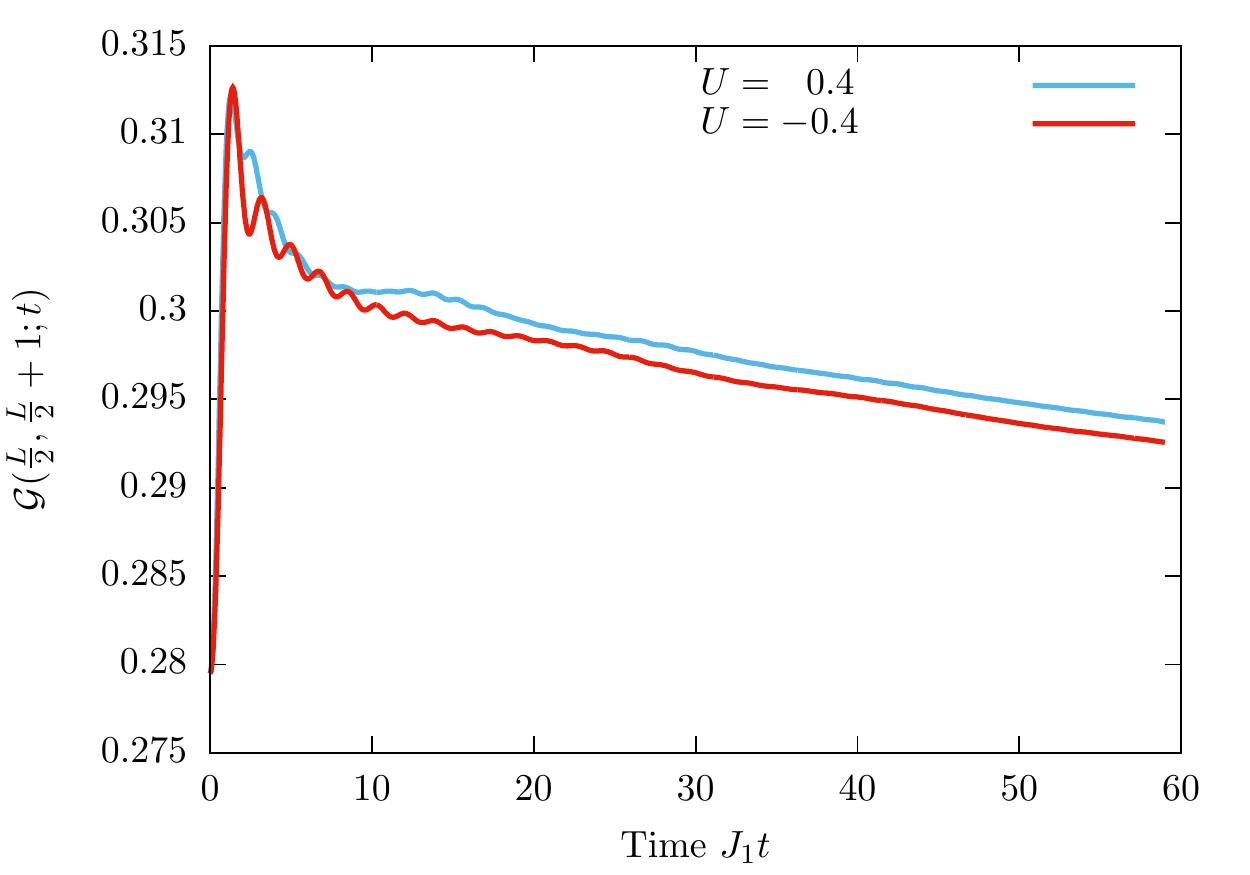}
\end{tabular}
\caption{The Green's function $\mathcal{G}(L/2,L/2 +1;t)$ with time-evolution generated 
by (blue) $H(0.5,0.1,0.4)$ and (red) $H(0.5,0.1,-0.4)$. The system starts from the thermal 
state~\fr{thermal} with $\beta_i = 2$, $J_2 = 0$, $\delta_i = 0$ and $U=0$.}
\label{Fig:negativeU}
\end{figure}

In the main text we have focused on repulsive interactions $U>0$ in the
Hamiltonian~\fr{Eq:Ham}. Here we briefly consider the case 
of attractive interactions. In Fig.~\ref{Fig:negativeU} we compare the
results for the time evolution of the Green's function at separation
$1$ for interactions strengths $U=\pm 0.4$. We see that the results
look broadly similar. The most marked difference is observed for
intermediate times ($5\lesssim t\lesssim 30$ in the figure). The same
holds true for larger separations. This is accordance with our
expectation: in the intermediate (PT) time-window the Green's
functions in the two cases are generically order $U$ different, but at
late enough times their evolution is described by the same quantum
Boltzmann equation.

\section{Perturbative calculation of the thermal values}
\label{app:pert}
In this appendix we compute the thermal expectation values
\bea
n_{\mu\nu}(k) &\equiv& \braket{\alpha^\dag_\mu(k) \alpha_\nu(k)}\\
&\equiv&\frac{1}{Z}\textrm{Tr}\left[\alpha^\dag_\mu(k) \alpha_\nu(k) e^{-\beta_\text{eff}(H-\mu N)}\right]~,
\label{AEeq:expval}
\eea
where $Z\equiv\textrm{Tr}\left[e^{-\beta_\text{eff}(H-\mu N)}\right]$ and we use the 
shorthand notation $H\equiv H(J_2,\delta_f,U)$. The inverse temperature $\beta_\text{eff}$ 
and the chemical potential $\mu_\text{eff}$ are fixed by requiring
\bea
\braket{ H}_{0}&=&\frac{1}{Z}\textrm{Tr}\left[ H e^{-\beta_\text{eff}( H-\mu_\text{eff} N)} \right]~,\\
\braket{ N}_{0}&=&\frac{1}{Z}\textrm{Tr}\left[ N e^{-\beta_\text{eff}( H-\mu_\text{eff} N)} \right]~, 
\eea
where $\braket{\cdot}_{0}$ is the expectation value in the initial state. In order to compute the 
finite-temperature mode occupation numbers~\fr{AEeq:expval}, we compute the thermal propagator
\bea
G_{\mu\nu}(\tau,k) &=& \braket{T_\tau\left[ \alpha^\dag_\mu(\tau,k) \alpha_\nu(0,k)\right]}~,\\
\alpha^\dag_\mu(s,k) &\equiv& e^{s H}\alpha^\dag_\mu(k)e^{-s H}~,
\eea
using finite-temperature perturbation theory to the order $U^2$. The thermal mode occupation numbers 
can be recovered from the thermal propagator using
\be
\label{eq:on}
n_{\mu\nu}(k)=\lim_{\tau\rightarrow 0^+} G_{\mu\nu}(\tau,k)=\frac{1}{\beta_\text{eff}}\sum_{\omega_n} G_{\mu\nu}(\omega_n,k) e^{i \omega 0^+}~.
\ee 

Writing the Green's function in matrix notation, the single particle self energy $\mathbf\Sigma(\omega_n,k)$ 
is defined by the Dyson equation
\be
\mathbf{G}(\omega_n,k)^{-1}=\mathbf{G_0}(\omega_n,k)^{-1}-\mathbf\Sigma(\omega_n,k) \label{Eq:Dyson}
\ee
where 
\be
(G_0)_{\mu\nu}(\omega_n,k)=\frac{\delta_{\mu\nu}}{i\omega_n-\bar{\eps}_\mu(k)}
\ee
and $\bar\epsilon_{\eta}(k)\equiv \epsilon_{\eta}(k)-\mu_\text{eff}$. The Feynman rules read:
\begin{fmffile}{rules}
\begin{align}
\parbox{20mm}{\begin{fmfgraph*}(80,50)
   \fmfleft{i}
   \fmfright{o}
   \fmf{fermion,label=$\omega_n,,k,,\eta$}{i,o}
\end{fmfgraph*}}\quad\qquad
&=\quad \frac{1}{i \omega_n -\bar\epsilon_{\eta}(k)}~,\nn
\parbox{20mm}{\begin{fmfgraph*}(120,75)
\fmfleft{i1,i2}
\fmfright{o1,o2}
\fmf{fermion,label=${\omega_{n_4}},,{k_4},,{\eta_4}$}{i1,v1}
\fmf{fermion,label=${\omega_{n_1}},,{k_1},,{\eta_1}$}{v1,o1}
\fmf{fermion,label=${\omega_{n_3}},,{k_3},,{\eta_3}$}{i2,v1}
\fmf{fermion,label=${\omega_{n_2}},,{k_2},,{\eta_2}$}{v1,o2}
\end{fmfgraph*}}
\qquad\qquad
&= 4 U V_{\eta_1\eta_2\eta_3\eta_4}(k_1,k_2,k_3,k_4)~. \nonumber
\end{align}
\end{fmffile}
with conservation of $\omega$ at each vertex. Therefore the diagrams contributing to the self 
energy to second order are
\begin{fmffile}{1storder}
\begin{align}
\hspace{-3cm}\Sigma^{(1)}_{\mu\nu}(k):\hspace{1.1cm}\parbox{20mm}{\begin{fmfgraph*}(120,80)
\fmfleft{i1}
\fmfright{o1}
\fmf{scalar,label=${\omega_{n}},,{k},,{\mu}$}{i1,v1}
\fmf{scalar,label=${\omega_{n}},,{k},,{\nu}$}{v1,o1}
\fmf{fermion,label=${\omega_{m}},,{q},,{\gamma}$}{v1,v1}
\end{fmfgraph*}} \notag
\end{align}
\end{fmffile}
\begin{fmffile}{2ndorder1}
\begin{align}
\Sigma^{(2)}_{\mu\nu}(k)_{[1]}:& &\parbox{20mm}{\begin{fmfgraph*}(150,85)
\fmfleft{i1}
\fmfright{o1}
\fmf{scalar,label=${\omega_{n}},,{k},,{\mu}$}{i1,v1}
\fmf{fermion,label=${\omega_{m_2}},,{q_2},,{\gamma_2}$,right,tension=0.3}{v1,v2}
\fmf{fermion,label=${\omega_{m_1}},,{q_1},,{\gamma_1}$,tension=0.3}{v2,v1}
\fmf{fermion,label=${\omega_{m_3}},,{q_3},,{\gamma_3}$,left,tension=0.3}{v1,v2}
\fmf{scalar,label=${\omega_{n}},,{k},,{\nu}$}{v2,o1}
\end{fmfgraph*}}\qquad\qquad\qquad\qquad\quad~
\notag 
\end{align}
\end{fmffile}
\begin{fmffile}{2ndorder2}
\begin{align}
\hspace{-3cm}\Sigma^{(2)}_{\mu\nu}(k)_{[2]}:\hspace{1cm}\parbox{20mm}{
\begin{fmfgraph*}(130,85)
\fmfleft{i1,i2,i3}
\fmfright{o1,o2,o3}
\fmf{scalar,label=${\omega_{n}},,{k},,{\mu}$}{i1,v1}
\fmf{scalar,label=${\omega_{n}},,{k},,{\nu}$}{v1,o1}
\fmf{phantom}{i2,v2,o2}
\fmf{phantom}{i3,v3,o3}
\fmffreeze
\fmf{fermion,label=${\omega_{m_1}},,{q_1},,{\gamma_1}$,tension=1.8,right}{v1,v2}
\fmf{fermion,tension=1.8,right}{v2,v3,v2}
\fmf{fermion,label=${\omega_{m_3}},,{q_3},,{\gamma_3}$,tension=1.8,right}{v2,v1}
\fmfv{label=${\omega_{m_2}},,{q_2},,{\gamma_2}$}{v3}
\end{fmfgraph*}}
\nonumber
\end{align}
\vspace{0.7cm}\\
\end{fmffile}
where the incoming and outgoing legs are amputated. Evaluating these, we find:
\bw
\begin{align}
\Sigma^{(1)}_{\mu\nu}(k) &\equiv 4 U\sum_{q,\gamma} V_{\nu \gamma \gamma \mu}(k,q,q,k) n(\bar\epsilon_{\gamma}(q))~, \\
\Sigma^{(2)}_{\mu\nu}(k)_{[1]} &\equiv 
-4 U\sum_{\gamma_1,\gamma_3,q_1}\Sigma^{(1)}_{\gamma_1\gamma_3}(q_1)V_{\nu \gamma_1\gamma_3 \mu}(k,q_1,q_1,k)
n(\bar\epsilon_{\gamma_1}(q_1))\tilde n(\bar \epsilon_{\gamma_3}(q_1))f(\epsilon_{\gamma_1\gamma_3}(q_1))~,\\
\Sigma^{(2)}_{\mu\nu}(\omega,k)_{[2]} &\equiv 
8U^2\sum_{\{\gamma_i\}}\sum_{\{q_i\}}\Bigl\{V_{\nu\gamma_1\gamma_2\gamma_3}(k,q_1,q_2,q_3)
V_{\gamma_3\gamma_2\gamma_1\mu}(q_3,q_2,q_1,k)n(\bar\epsilon_{\gamma_1}(q_1))\notag\nn
&\qquad\qquad\qquad\qquad \times \tilde n(\bar\epsilon_{\gamma_2}(q_2))\tilde n(\bar\epsilon_{\gamma_3}(q_3))
f(i\omega+\bar\epsilon_{\gamma_1}(q_1)-\bar\epsilon_{\gamma_2}(q_2)-\bar\epsilon_{\gamma_3}(q_3))\Bigr\}~.
\end{align}
\ew
with the functions $\tilde n(x)\equiv 1-n(x)$, $n(x)$ being the Fermi-Dirac distribution 
and $f(x)\equiv \frac{e^{\beta_\text{eff} x}-1}{x}$. Using the Dyson equation for the propagator~(\ref{Eq:Dyson}), 
expanding to second order in the self energy and inserting into Eq.~\fr{eq:on}, we find the thermal mode occupation 
numbers up to ${O}(U^2)$ 
\bw
\begin{align}
n_{\mu\nu}(k,\beta_\text{eff},\mu) =&
\delta_{\mu,\nu}~n(\bar\epsilon_{\mu}(k))-(\Sigma^{(1)}_{\mu\nu}(k)+\Sigma^{(2)}_{\mu\nu}(k)_{[1]})n(\bar\epsilon_{\mu}(k))
\tilde n(\bar\epsilon_{\nu}(k)) f(\epsilon_{\mu\nu}(k))\nn
&+\sum_{\delta}\frac{\Sigma^{(1)}_{\mu\delta}(k)\Sigma^{(1)}_{\delta\nu}(k)}{\epsilon_{\mu\delta}(k)\epsilon_{\delta\nu}(k)\epsilon_{\mu\nu}(k)}
\Bigl(n(\bar\epsilon_{\mu}(k))\epsilon_{\delta\nu}(k)-n(\bar\epsilon_{\delta}(k))\epsilon_{\mu\nu}(k)+n(\bar\epsilon_{\nu}(k))\epsilon_{\mu\delta}(k)\Bigr)\nn
&+8U^2\sum_{\{\gamma_i\}}\sum_{\{q_i\}}\Bigl\{V_{\mu\gamma_1\gamma_2\gamma_3}(k,q_1,q_2,q_3
)V_{\gamma_3\gamma_2\gamma_1\nu}(q_3,q_2,q_1,k)\tilde n(\bar\epsilon_{\gamma_3}(q_3))\tilde n(\bar\epsilon_{\gamma_2}(q_2)) \nn
&\qquad\qquad\qquad\qquad\times n(\bar\epsilon_{\gamma_1}(q_1))G_{\mu\nu\gamma_1\gamma_2\gamma_3}(k,q_1,q_2,q_3)\Bigr\}~.
\end{align}
where we set
\begin{align}
G_{\mu\nu\gamma_1\gamma_2\gamma_3}(k,q_1,q_2,q_3)=&
\frac{1}{\epsilon_{\mu\nu}(k)}\sum_{\eta=\pm}\{(\delta_{\eta,\mu}-\delta_{\eta,\nu})n(\bar{\epsilon}_\eta(k))
f(E_{\eta \gamma_1\gamma_2\gamma_3}(k,q_1,q_2,q_3))\}~.
\end{align}
\ew
\section{Boltzmann Equation for $\delta_f\neq0$}
\label{app:BrunoC}

The numerical solution of the EOM \fr{Eq:EOM} for $\delta_f\neq 0$ suggests that at sufficiently late times the
occupation numbers assume the form
\be
n_{\mu\mu}(k,t)=s_{\mu}(k,U^2 t) + \Delta_\mu(k,t)\ ,
\label{Eq:Latetimes}
\ee
where $s_{\mu}(k,U^2 t)$ is a smooth, slowly varying function of time
and $\Delta_\mu(k,t)$ is a small highly oscillatory contribution. As
we are interested in the single particle Green's function in position
space, the contributions arising from $\Delta_\mu(k,t)$ will be
negligible at late times. As we will now argue, the smooth component
$s_{\mu}(k,U^2 t)$ fulfils a QBE in the Boltzmann limit
\fr{Eq:Boltzscallim}. 

Our starting point are the EOM for the occupation numbers, where we
neglect all contributions involving the off-diagonal two-point
functions $\{n_{+-}(k,t)\}$ as at late times they are ${\cal O}(U)$
and rapidly oscillating. The ``reduced'' EOMs can be written in the
following compact form    
\be
\dot n_{\mu\mu}(k,t) = U D_\mu(k,t) + U^2 I_\mu[\{n_{\nu\nu}\}](k,t)\ ,
\label{Eq:diagEOM}
\ee
where we introduced 
\be
\!\!\!D_\mu(k,t)= 8 \textrm{Im}\!\left[\left(A_{\bar\mu}(k)  +B_{\bar\mu}(k,t)\right)n_{\mu \bar\mu}(k)e^{i\epsilon_{\mu\bar\mu}(k)t}\right].
\ee
Here $A_{\bar\mu}(k)$ and $B_{\bar\mu}(k,t)$ are defined in
Eqs.~\fr{Eq:OUnotdec} and \fr{Eq:OUdec} respectively and
\bw
\begin{align}
I_\mu[\{n_{\nu\nu}\}](k,t) &\equiv - \int_0^t \!\textrm{d}s  \sum_{\bm \gamma}\sum_{k_1,k_2{>0}} \!\!\!\!K^{\bm\gamma}_{\mu\mu}(k_1,k_2; k; t-s) 
n_{\gamma_1\gamma_2}(k_1,s) n_{\gamma_3\gamma_4}(k_2,s)\delta_{\gamma_1,\gamma_2}\delta_{\gamma_3,\gamma_4}\notag\\
& -\int_0^t \!\textrm{d} s  \sum_{\vec{\gamma}}\sum_{k_1,k_2,k_3{>0}} \!\!\!\!\!L^{\vec{\gamma}}_{\mu\mu}(k_1,k_2,k_3; k; t-s) 
n_{\gamma_1\gamma_2}(k_1,s) n_{\gamma_3\gamma_4}(k_2,s)n_{\gamma_5\gamma_6}(k_3,s)\delta_{\gamma_1,\gamma_2}\delta_{\gamma_3,\gamma_4}\delta_{\gamma_5,\gamma_6}\,,
\label{Eq:FF}
\end{align}
where the kernels are defined in Eq.~\fr{Eq:kernelsEOM}. Substituting
\fr{Eq:Latetimes} into \fr{Eq:diagEOM} we obtain
\begin{align}
U^2\partial_\tau{s}_\mu(k,\tau)\big|_{\tau=U^2 t}+\partial_t\Delta_\mu(k,t)=U D_\mu(k,t)+ U^2 I_\mu[\{s_{\nu}+ \Delta_\nu\}](k,t)\,.
\label{Eq:differenttimescales}
\end{align}
\ew
We now remove the rapidly oscillating part of
\fr{Eq:differenttimescales} by acting with a low-pass filter and then
take the Boltzmann scaling limit \fr{Eq:Boltzscallim}. We employ
a filter of the form 
\bea
\mathcal{L}[f(t)] &\equiv& \int_{-\infty}^\infty 
{\rm d}s\,\, \ell(t-s,\omega_{\textsc{cut}})\ f(s)\ ,\nn
\ell(t,\omega)&\equiv&\int_{-\omega}^{\omega} \frac{{\rm
    d}\sigma}{2\pi} \,\,e^{i \sigma t} =\frac{\sin(\omega t)}{\pi
  t}\,. 
\eea
The cutoff frequency $\omega_{\textsc{cut}}$ is chosen such that
\bea
\mathcal{L}[\partial_t^n s_\mu(k,U^2t)] &=& \partial_t^n
s_\mu(k,U^2t)\,,\quad n=0,1,\ldots\ ,\nn
\mathcal{L}\left[{e^{i \omega t}}\right]&=&0\,,\qquad
\text{for}\qquad\omega={\cal O}\big(1\big)\,. 
\eea
These two requirements can be met by choosing
${\omega_{\textsc{cut}}\sim U}^\alpha$ with $0<\alpha<2$. Applying the
filter to the EOM~\fr{Eq:diagEOM} and taking the Boltzmann scaling
limit \fr{Eq:Boltzscallim} we find  
\begin{align}
\partial_\tau{s}_\mu(k,\tau)=\limb
\mathcal{L}\left[ I_\mu[\{s_{\nu}+ \Delta_\nu\}](k,t)\right]\,.
\end{align}
Here we have used that $|\epsilon_{\mu\bar{\mu}}(k)|$ is bounded from
below by a constant of order one, which implies that $D_{\mu}(k,t)$ is
rapidly oscillating at late times. Let us now study the effect of the
low-pass filter combined with the Boltzmann scaling limit on the functional
$I_\mu$. Because of the linearity of the filter we have  
\bea
&&\mathcal{L}[ I_\mu[\{s_{\nu}+ \Delta_\nu\}](k,t)]= 
\mathcal{L}\left[ I_\mu[\{s_{\nu}\}](k,t)\right]\nn
&&\qquad+\mathcal{L}\left[ I_\mu[\{s_{\nu}+ \Delta_\nu\}](k,t)- I_\mu[\{s_{\nu}\}](k,t)\right] \,.
\label{Eq:quadraticfiltered}
\eea
The term $I_\mu[\{s_{\nu}\}](k,t)$ is of the same form as the one
considered in Eq.~\fr{AD:simpint}. It can be cast in the form
\be
I_\mu[\{s_{\nu}\}]{(k,t)}=\sum_{\boldsymbol{q}>0,\boldsymbol{\lambda}}
\int_0^t {\rm d}s\, f^\mu_{\boldsymbol{\lambda}}(\mathbf{q},k;U^2 s)e^{\ri
  {E}_{\boldsymbol{\lambda}}(\mathbf{q})(t-s)}.
\label{Eq:oscillatoryterm}
\ee 
Importantly $f^\mu_{\boldsymbol{\lambda}}(\mathbf{q},k;U^2 s)$ depends
on the variable $s$ only through the combination $U^2 s$. In writing
\fr{Eq:oscillatoryterm} we have assumed that in the scaling limit we
can neglect the analogue of the first term in the right hand side of
Eq.~\fr{AD:simpint}, and we have replaced the integration boundary
$t-\bar{t}$ by $t$ in the remaining contribution (which is justified
by referring to the scaling limit). In order to ease notations we now
focus on a single term to the sums in \fr{Eq:oscillatoryterm} and
suppress all unnecessary indices. It is convenient to define a
function
\be
{\frak F}(s_1,E)=\int_0^{s_1} {\rm d}s_2\, f(U^2 (s_1-s_2))e^{\ri  {E}s_2}.
\ee
We now define the action of the low-pass filter by
\bea
{\cal L}[{\frak F}(t,E)]&=&\int_{-\infty}^0{\rm d}s_1
\frac{\sin(\omega_{\textsc{cut}} (t-s_1))}{\pi (t-s_1)} {\frak
  F}(s_1,E-i\eta)\nn
&+&\int_{0}^\infty{\rm d}s_1 \frac{\sin(\omega_{\textsc{cut}}
  (t-s_1))}{\pi (t-s_1)} {\frak F}(s_1,E+i\eta)\nn
&\equiv&\Sigma_1+\Sigma_2.
\label{Eq:firstBC}
\eea
where we have appropriately regularised the two integrals using an
infinitesimal parameter $\eta$
(\emph{cf}. \ref{sec:Boltdelta=0}). Importantly, we will exchange
limits and keep $\eta$ fixed when taking the Boltzmann scaling
limit. Using that the derivatives of $f$ are suppressed by powers
of~$U^2$, in the scaling limit we have
\bea
\Sigma_1&\rightarrow&\frac{i}{(E-i\eta)}\limb\int_{t}^\infty{\rm d}s_1
\frac{\sin(\omega_{\textsc{cut}} s_1)}{\pi s_1} f(U^2(t-s_1))\nn
&=&\frac{i f(\tau)}{E-i\eta} \limb \int_{-\infty}^\infty\!\!{\rm d}s_1
\frac{\sin(\omega_{\textsc{cut}} s_1)}{\pi
  s_1}\,\theta_{\textsc{h}}(s_1-t)\nn
&=&0\ .
\eea
Here $\tau=U^2 t$ and in the second step we have used that
$\sin(\omega_{\textsc{cut}}s_1)/s_1$ is oscillating and peaked around
$s_1=0$ with a width that scales as $\omega_{\textsc{cut}}^{-1}$,
while $f(U^2(t-s_1))$ is essentially constant in that window.
Going through the analogous steps for the second term in
\fr{Eq:firstBC} gives
\be
\Sigma_2\rightarrow\frac{i f(\tau)}{E+i\eta} \int^{\infty}_{-\infty}\!\!\!{\rm d}s_1  \frac{\sin(s_1)}{\pi s_1}=D(E) f(\tau)\,,
\ee
where $D(E)$ is defined in Eq.~\fr{AD:Reg}. Putting everything
together we conclude that
\begin{align}
\label{Eq:simpleterm}
\limb \mathcal{L}\left[I_\mu[\{s_{\mu}\}](k,t)\right]&= \tilde I_\mu[\{s_{\mu}\}](k,\tau)\,,
\end{align} 
where 
\bw
\begin{align}
\tilde I_\mu&[\{s_{\mu}\}](k,\tau) \equiv -\! \sum_{\gamma,\eta}\!\sum_{p,q{>0}} 
\!\widetilde K^{\gamma\eta}_{\mu}(p,q|k) s_{\gamma}(p,\tau) s_{\eta}(q,\tau)-\sum_{\gamma,\eta,\epsilon}\sum_{p,q,r{>0}} \
\!\widetilde L^{\gamma\eta\epsilon}_{\mu}(p,q,r|k)s_{\gamma}(p,\tau)  s_{\eta}(q,\tau)s_{\epsilon}(r,\tau)\,.
\label{firstterm}
\end{align}
The various kernels appearing in \fr{firstterm} are defined in
Eq.~\fr{Eq:Boltkernels}.
\ew
We now turn to the second term in \fr{Eq:quadraticfiltered}. This is
more difficult to treat, because of the oscillating contributions to
the occupation numbers. The difference $I_\mu[\{s_{\nu}+
  \Delta_\nu\}](k,t)- I_\mu[\{s_{\nu}\}](k,t)$ can be cast in the form
\be
\sum_{\boldsymbol{q}>0,\boldsymbol{\lambda}}\sum_{i}\int_0^t {\rm d}s\
h^{\mu,\,i}_{\boldsymbol{\lambda}}(\mathbf{q},k;U^2 s) 
e^{i \varepsilon_{\boldsymbol{\lambda},\,i}(\mathbf{q}) s}
e^{\ri  {E}_{\boldsymbol{\lambda}}(\mathbf{q})(t-s)}.
\label{Eq:oscillatoryterm2}
\ee 
In order to proceed we now make the assumption that, as a
function of $\mathbf{q}$,
$\varepsilon_{\boldsymbol{\mu},\,i}(\mathbf{q})$ is generically of
order one and vanishes only on a set of measure zero. When applying
the low-pass filter to \fr{Eq:oscillatoryterm2} it useful to
distinguish between two cases: (i)
$\varepsilon_{\boldsymbol{\mu},\,i}(\mathbf{q})\neq
{E}_{\boldsymbol{\mu}}(\mathbf{q})$ except on a set of measure zero;
(ii) $\varepsilon_{\boldsymbol{\mu},\,i}(\mathbf{q})=
{E}_{\boldsymbol{\mu}}(\mathbf{q})$. We now again ease notations by
focussing on a single term and suppressing all indices. It is
convenient to define a function
\be
{\frak H}(s_1,E)=e^{\ri\varepsilon s_1}
\int_0^{s_1} {\rm d}s_2\, h(U^2 (s_1-s_2))e^{\ri  {E}s_2}.
\ee
Proceeding as before, we define the action of the low-pass filter by
\bea
{\cal L}[{\frak H}(t,E)]&=&\int_{-\infty}^0{\rm d}s_1
\frac{\sin(\omega_{\textsc{cut}} (t-s_1))}{\pi (t-s_1)} {\cal
  H}(s_1,E-i\eta)\nn
&+&\int_{0}^\infty{\rm d}s_1 \frac{\sin(\omega_{\textsc{cut}}
  (t-s_1))}{\pi (t-s_1)} {\frak H}(s_1,E+i\eta)\nn
&\equiv&\Sigma_3+\Sigma_4.
\label{Eq:secondBC}
\eea
Considering the first term, we have in the scaling limit
\bea
\Sigma_3&\rightarrow&\limb
\int_{-\infty}^0{\rm d}s_1 \frac{\sin(\omega_{\textsc{cut}}
  (t-s_1))}{\pi (t-s_1)}\frac{i  h(U^2 s_1) e^{i \varepsilon
    s_1}}{(E-\varepsilon-i\eta)}\nn
&=&0.
\eea
Here we used that $t\omega_{\textsc{cut}}$ tends to infinity in the
scaling limit. Similarly we obtain 
\bea
\Sigma_4&\rightarrow&\frac{i h(\tau)}{E-\varepsilon+i\eta}\limb
\int_{-\infty}^{t}{\rm d}s_1  e^{i \varepsilon t}
\frac{\sin(\omega_{\textsc{cut}} s_1)}{\pi s_1}e^{-i \varepsilon
  s_1}\nn
&=&0\,.
\label{Eq:filterosc}
\eea

Finally, let us consider case (ii) defined above. Focussing again
on a single contribution we now have
\begin{align}
\int_{-\infty}^{\infty}{\rm d}s_1\,e^{i E
  s_1}\frac{\sin(\omega_{\textsc{cut}} (t-s_1))}{\pi
  (t-s_1)}\int_0^{s_1}\!\!{\rm d}s_2\,  h(U^2 s_2)\,. 
\label{Eq:degeneratecase}
\end{align}
The function $\int_0^{t}\!\!{\rm d}s\,  h(U^2 s)$ does not contain
highly oscillatory contributions and thus it can not counter balance
the rapidly oscillating phase $e^{iE t}$. As a consequence the
contribution \fr{Eq:degeneratecase} vanishes in the scaling
limit. This conclusion holds true even if an appropriate
regularization of $\int_0^{t}{\rm   d}s\,  h(U^2 s)$ is considered in
order to deal with the limits $t\rightarrow\pm\infty$ (similarly to
the cases considered before).  

Putting everything together we conclude that the in the scaling limit
the smooth parts ${s}_\mu(k,\tau)$ of the mode occupation numbers
fulfil 
\begin{align}
\dot{s}_\mu(k,\tau)-\tilde I_\mu[\{s_\nu\}](k,\tau)=0\,,
\label{Eq:reduced}
\end{align}
which has precisely the same form as the Quantum Boltzmann
equation~\fr{Eq:BB}.

\section{First order EOM for the $\textrm{U}(1)$-breaking case}
\label{app:EOMnoU1}
Here we present some details regarding the first order EOM
analysis of Sec.~\ref{sec:u1b}. The first step is to move to
momentum space. We do this by working with a 2-site elementary cell
which allows us to accommodate e.g. initial states that are
invariant only under translations by two sites. We define canonical
momentum space fermion operators by
\begin{align}
&e_{k_n}=\sqrt{\frac{2}{L}}\sum_{j}e^{i {k_n} (2j)} c_{2j}~, \nn
&f_{k_n}=\sqrt{\frac{2}{L}}\sum_{j}e^{i {k_n} (2j-1)} c_{2j-1}~,
\label{c1:FT}
\end{align}
where $k_n=2\pi n/L$ with $n=1,\ldots,L/2$. The Hamiltonian
\fr{Eq:pertXY} is expressed as
\bea
{\cal H}(\gamma,h,U) &=&J 
\sum_{k>0}\Bigl[\cos( k)
  f^\dag_ke^{\phantom{\dag}}_k
+ i \gamma \sin(k) f^\dag_{k}e^\dag_{\pi-k}+\textrm{h.c.}\Bigr] \nn
&-&J h \sum_{k>0}\Bigl[f^\dag_kf^{\phantom{\dag}}_k+ e^\dag_ke^{\phantom{\dag}}_k\Bigr] \nn
&+&{4U}\sum_{{\bm k}>0} W({\bm k})e^\dag_{k_1}e^{\phantom{\dag}}_{k_2}f^\dag_{k_3}f^{\phantom{\dag}}_{k_4}~,
\eea
where 
 \begin{align}
 W({\bm k})&=\frac{1}{L}\bigl(\delta_{k_1-k_2+k_3-k_4\pm\pi,0}\nn
 &\qquad\quad+\delta_{k_1-k_2+k_3-k_4\pm\pi,0}\bigr)\cos(k_3-k_4)~.
 \end{align}
The EOMs are formulated for the fermion two point functions 
\begin{align}
\begin{split}
v_1(k;t) &\equiv \textrm{Tr}\left[{f^{\dag}_k(t) f^{\phantom{\dag}}_k(t)}\sigma_0\right],\\
v_2(k;t)&\equiv\textrm{Tr}\left[{e^{\dag}_k(t) e^{\phantom{\dag}}_k(t)}\sigma_0\right],\\
v_3(k;t)&\equiv\textrm{Tr}\left[{f^{\dag}_k(t) f^{\dag}_{\pi-k}(t)}\sigma_0\right], \\
v_4(k;t)&\equiv\textrm{Tr}\left[{e^{\dag}_k(t) e^{\dag}_{\pi-k}(t)}\sigma_0\right],\\
v_5(k;t)&\equiv\textrm{Tr}\left[{f^{\dag}_k(t) e^{\phantom{\dag}}_k(t)}\sigma_0\right],\\
v_6(k;t)&\equiv\textrm{Tr}\left[{f^{\dag}_k(t) e^{\dag}_{\pi-k}(t)}\sigma_0\right]. 
\end{split}
\label{Eq:noU1numbop}
\end{align}
As with Sec.~\ref{sec:eom}, we derive the first order EOM by writing the
Heisenberg equations for the bilinears and neglecting the
four-particle connected cumulants at all times. Defining $\bar{k}=\pi-k$
the result can be written in the form
\bea
\dot v_1(k;t) &=& 2 \cos(k) \textrm{Im}[v_5(k;t)]
+ 2 \gamma \sin(k) \textrm{Re}[v_6(k;t)]\nn
&-&4Ui\sum_{q>0}W(q,k,k,q)v_5(q;t)v^*_5(k;t)\nn
&+&4Ui\sum_{q>0}W(k,q,q,k)v_5(k;t)v^*_5(q;t)\nn
&+&4Ui\sum_{q>0}W(q,\bar{q},k,\bar{k})v_6(q;t)v^*_6(k;t)\nn
&-&4Ui\sum_{q>0}W(k,\bar{k},q,\bar{q})v_6(k;t)v^*_6(q;t),
\eea

\bea
\dot v_2(k;t) &=&  -2 \cos(k) \textrm{Im}[v_5(k;t)]+ 2 \gamma \sin(k)
\textrm{Re}[v_6(\bar{k};t)]\nn
&-&4Ui\sum_{q>0}W(k,q,q,k)v_5(k;t)v^*_5(q;t)\nn 
&+&4Ui\sum_{q>0}W(q,k,k,q)v_5(q;t)v^*_5(k;t)\nn
&+&4Ui\sum_{q>0}W(k,\bar{k},q,\bar{q})v_6(\bar{k};t)v^*_6(q;t)\nn
&-&4Ui\sum_{q>0}W(q,\bar{q},k,\bar{k})v_6(q;t)v^*_6(\bar{k};t),
\eea
\bea
\dot v_3(k;t) &=& - i \cos(k) \left(v_6(k;t)+v_6(\bar{k},t)\right)\nn
&+& \gamma \sin(k) \left(v_5(k;t)-v_5(\bar{k},t)\right)- 2 i h v_3(k;t)\nn
&+&4Ui\sum_{q>0}W(k,q,k,q)v_2(q;t)v_3(k;t)\nn
&+&4Ui\sum_{q>0}W(q,\bar{k},\bar{k},q)v_5(q;t)v_6(k;t)\nn
&-&4Ui\sum_{q>0}W(\bar{k},q,\bar{k},q)v_2(q;t)v_3(\bar{k};t)\nn
&+&4Ui\sum_{q>0}W(q,k,k,q)v_5(q;t)v_6(\bar{k};t)\nn
&-&4Ui\sum_{q>0}W(q,\bar{q},\bar{k},k)v_6(q;t)v_5(k;t)\nn
&-&4Ui\sum_{q>0}W(q,\bar{q},k,\bar{k})v_6(q;t)v_5(\bar{k};t),\ \ 
\eea
\bea
\dot v_4(k;t) &=& i \cos(k) \left(v_6(k;t)+v_6(\bar{k},t)\right)\nn
&-& \gamma \sin(k) \left(v_5(k;t)-v_5(\bar{k},t)\right) + 2 i h v_4(k;t)\nn
&-&4Ui\sum_{q>0}W(q,\bar{q},k,\bar{k})v_6(q;t)v^*_5(k;t)\nn
&+&4Ui\sum_{q>0}W(\bar{k},q,q,\bar{k})v^*_5(q;t)v_6(\bar{k};t)\nn
&-&4Ui\sum_{q>0}W(\bar{k},q,\bar{k},q)v_1(q;t)v_4(\bar{k};t)\nn
&+&4Ui\sum_{q>0}W(k,q,k,q)v_1(q;t)v_4(k;t)\nn
&-&4Ui\sum_{q>0}W(q,\bar{q},\bar{k},k)v_6(q;t)v^*_5(\bar{k};t)\nn
&+&4Ui\sum_{q>0}W(k,q,q,k)v^*_5(q;t)v_6(k;t),
\eea
\bea
\dot v_5(k;t) &=& i \cos(k)
\left(v_2(k;t)-v_1(k;t)\right)\nn
&+& \gamma \sin(k) \left(v^*_4(k;t)-v_3(k;t)\right)\nn
&-&4Ui\sum_{q>0}W(q,k,q,k)v_1(q;t)v_5(k;t)\nn
&+&4Ui\sum_{q>0}W(k,q,k,q)v_2(q;t)v_5(k;t)\nn
&+&4Ui\sum_{q>0}W(q,k,k,q)v_1(k;t)v_5(q;t)\nn
&-&4Ui\sum_{q>0}W(q,\bar{q},k,\bar{k})v^*_4(q;t)v_6(\bar{k};t)\nn
&-&4Ui\sum_{q>0}W(\bar{k},k,q,\bar{q})v_3(q;t)v^*_6(k;t)\nn
&-&4Ui\sum_{q>0}W(q,k,k,q)v_2(k;t)v_5(q;t),
\eea
\bea
\dot v_6(k;t) &=&i \cos(k) \left(v_4(k;t)-v_3(k;t)\right)\nn
&-& \gamma \sin(k) \left(v_1(k;t)+v_2(\bar{k};t)-1\right)+2ih v_6(k;t))\nn
&-&4Ui\sum_{q>0}W(q,\bar{q},k,\bar{k})v_6(q;t)v_1(k;t)\nn
&+&4Ui\sum_{q>0}W(k,q,k,q)v_2(q;t)v_6(k;t)\nn
&+&4Ui\sum_{q>0}W(\bar{k},q,\bar{k},q)v_1(q;t)v_6(k;t)\nn
&-&4Ui\sum_{q>0}W(q,k,k,q)v_5(q;t)v_4(k;t)\nn
&+&4Ui\sum_{q>0} W(q,\bar{q},k,\bar{k})v_6(q;t)(1-v_2(\bar{k};t))\nn
&+&4Ui\sum_{q>0} W(\bar{k},q,q,\bar{k})v^*_5(q;t)v_3(k;t)~.
\eea 



\begin{thebibliography}{99}
\bibitem{notefig}
We do not plot $\tau^{-1}(L/2,L/2+1)$ because large oscillations and slow decay cause very large errors in the fit.
\end{thebibliography}
\end{document}